\documentclass[10pt,aps,pra,twocolumn,showpacs,superscriptaddress,nobalancelastpage,longbibliography]{revtex4-1}

\usepackage[english]{babel}
\usepackage{graphicx,graphics,epsfig,subfigure,times,bm,bbm,amssymb,amsmath,amsfonts,mathrsfs}
\usepackage[scr=boondoxo,scrscaled=1.05]{mathalfa}
\usepackage[matrix,frame,arrow]{xypic}
\usepackage[pdfstartview=FitH]{hyperref}
\usepackage[pdftex]{color}
\usepackage{tabularx}
\usepackage{dsfont}
\usepackage[export]{adjustbox}
\usepackage[dvipsnames]{xcolor}
\usepackage{color}
\usepackage{soul}
\usepackage{wrapfig}
\usepackage{scrextend}
\newcommand{\subfigimg}[3][,]{%
  \setbox1=\hbox{\includegraphics[#1]{#3}}% Store image in box
  \leavevmode\rlap{\usebox1}% Print image
  \rlap{\hspace*{2pt}\raisebox{\dimexpr\ht1+0.2\baselineskip}{#2}}% Print label
  \phantom{\usebox1}% Insert appropriate spcing
}
\usepackage{cleveref}
\usepackage{multirow}
\usepackage[mathscr]{euscript}
\usepackage{tabularx}
\DeclareMathOperator*{\argmin}{argmin} % thin space, limits underneath in displays

\newcommand{\beq}{\begin{equation}}
\newcommand{\eneq}{\end{equation}}
\newcommand{\boleqnn}{\begin{equation*}}
\newcommand{\eneqnn}{\end{equation*}}
\newcommand{\beqy}{\begin{eqnarray}}
\newcommand{\eneqy}{\end{eqnarray}}
\newcommand{\beqynn}{\begin{eqnarray*}}
\newcommand{\eneqynn}{\end{eqnarray*}}

\newcommand{\bes} {\begin{subequations}}
\newcommand{\ees} {\end{subequations}}
	\newcommand{\bea} {\begin{eqnarray}}
	\newcommand{\eea} {\end{eqnarray}}

\newcommand{\Tr}{\mathrm{Tr}}

\newcommand{\ignore}[1]{}

\newcommand{\eq}[1]{(\ref{#1})}

\newcommand{\seq}{\hspace{1pt}{=}\hspace{1pt}}
\newcommand{\lspace}[1]{\hspace{1pt}{#1}\hspace{1pt}}
\renewcommand{\arraystretch}{1.5}

\begin{document}

\title{Frame-Based Filter-Function Formalism for Quantum  Characterization and Control}

\author{Teerawat Chalermpusitarak}\thanks{These authors contributed equally to this work.} 
\affiliation{Centre for Quantum Computation and Communication Technology (Australian Research Council), Centre for Quantum Dynamics, Griffith University, Brisbane, Queensland 4111, Australia}

\author{Behnam Tonekaboni}\thanks{These authors contributed equally to this work.}
\affiliation{Centre for Quantum Computation and Communication Technology (Australian Research Council), Centre for Quantum Dynamics, Griffith University, Brisbane, Queensland 4111, Australia}

\author{\mbox{Yuanlong Wang}}\thanks{These authors contributed equally to this work.}
\affiliation{Centre for Quantum Computation and Communication Technology (Australian Research Council), Centre for Quantum Dynamics, Griffith University, Brisbane, Queensland 4111, Australia}

\author{Leigh M. Norris}
\affiliation{\mbox{Department of Physics and Astronomy, Dartmouth College, 6127 Wilder Laboratory, 
Hanover, New Hampshire 03755, USA}}

\author{Lorenza Viola}\thanks{lorenza.viola@dartmouth.edu}
\affiliation{\mbox{Department of Physics and Astronomy, Dartmouth College, 6127 Wilder Laboratory,
 Hanover, New Hampshire 03755, USA}}

\author{Gerardo A. Paz-Silva} \thanks{g.pazsilva@griffith.edu.au}
\affiliation{Centre for Quantum Computation and Communication Technology (Australian Research Council), Centre for Quantum Dynamics, Griffith University, Brisbane, Queensland 4111, Australia}

\begin{abstract}
We introduce a theoretical framework for resource-efficient characterization and control of non-Markovian open quantum systems, which naturally allows for the integration of given, experimentally motivated, control capabilities and constraints. This is achieved by developing a transfer filter-function formalism based on the general notion of a {\em frame} and by appropriately tying the choice of frame to the available control. While recovering the standard frequency-based filter-function formalism as a special instance, this \emph{control-adapted} generalization affords intrinsic flexibility and, crucially, it permits an efficient representation of the relevant control matrix elements and dynamical integrals if an appropriate \emph{finite-size frame condition} is obeyed. Our frame-based formulation overcomes important limitations of existing approaches. In particular, we show how to implement quantum noise spectroscopy in the presence of \emph{non-stationary} noise sources, and how to effectively achieve \emph{control-driven model reduction} for noise-optimized prediction and quantum gate design. 
\end{abstract}
\date{\today}
\maketitle

%\tableofcontents

\section{Introduction}

Accurate characterization and control (C\&C) of open quantum systems coupled to realistic -- temporally correlated (``non-Markovian'') -- noise environments are vital for exploiting the full potential of quantum technologies. Open-loop control-engineering methods, based on tailored time-dependent modification of the open-system dynamics, offer a versatile and experimentally accessible approach to tackle this challenge. While techniques employing dynamical decoupling or dynamically-corrected gates~\cite{ViolaDD,DCG,DS4} are beneficial under minimal knowledge about the noise-inducing degrees of freedom, extra knowledge is instrumental for optimizing their performance and efficiency in specific noise environments \cite{Paz-Silva-DD,Wang2017}. Numerical quantum optimal control algorithms \cite{OC1,OCReview1,OCReview2} represent an extreme example: although they can in principle deliver exceptionally high fidelities, their viability in non-Markovian settings requires detailed knowledge of the time-domain noise correlation functions (or the corresponding frequency-domain spectra) \cite{OCNew}. Beside permitting noise-optimized quantum storage and gate design, accurate noise characterization is key to counter the effect of non-Markovian noise in quantum sensing and metrology applications \cite{Degen,FelixPRA}.

This has motivated the development of ``quantum noise spectroscopy'' (QNS) protocols, in which noise spectral information is inferred from the system-only reduced dynamics, under the effects of the noise and user-defined control. Despite considerable progress \cite{alvarez2011measuring,yuge2011measurement,yan2012spectroscopy,yan2013rotating,dial2013charge,Paz-Silva-NonGaussian,multiqubit,Cyw,SlepianQNS,Baugh2018Continuous,Ferrie-Bayesian-2018,qns4,multiaxis,TTN,UweSL}, existing QNS protocols suffer from several disadvantages. First, they are not applicable to important types of noise that occur in practice -- notably, \emph{non-stationary} noise, for which a frequency-domain description need not be viable \cite{Chakrabarti,ProctorDrift}. Second, they do not easily lend themselves to the identification of simplified noise models which, while providing all the required detail for optimal control to be feasible, may permit C\&C procedures to be \emph{extensible} to increasingly larger qubit networks. Finally, to the best of our knowledge no formal analysis of the ``universality'' of QNS-inferred information for control purposes has been attempted, aimed at clarifying the extent to which such information may suffice to faithfully predict the dynamics of the system under an \emph{arbitrary} control modulation, beyond what used in the QNS protocol itself.

In this paper, we overcome the above limitations by demonstrating how a ``model-reduced'' representation of the noisy dynamics of interest may be tied to the available, \emph{finite} control resources. We achieve this by integrating the language of \emph{frames}~\cite{framesoriginal,frames2,FrameIntro,FrameContinuous1,FrameContinuous2} -- already a mainstay in signal processing -- with the theory of open quantum systems. This leads to a novel transfer filter-function (FF) formalism~\cite{Kurizki2004,biercuk2011dynamical,Paz-Silva-FF2014}, in which control capabilities and constraints (dubbed $\mathscr{C}$ henceforth) play the defining role. Crucially, our framework allows the identification of the noise components relative to $\mathscr{C}$ which are, at once, accessible to estimation by a QNS protocol and sufficient to optimally predict the behavior of the system. Moreover, it provides a rigorous way to determine (and quantify) if the information inferred via QNS is useful to predict the behavior of the system under a given control sequence.

The content is organized as follows. In Sec.~\ref{section:ControlledOpenQ}, we introduce the relevant setting for controlled non-Markovian quantum dynamics, with special emphasis on highlighting the structure of the perturbative overlap integrals that enter any C\&C  protocol. Sec. \ref{section:FrameFFandReduction} provides the conceptual foundation of our approach: after providing the essential mathematical background on frames, we show how the relationship between a frame $\mathscr{F}$ and its canonical dual $\tilde{\mathscr{F}}$ naturally suggests two complementary ways for separating the dynamical overlap integrals into control-dependent and noise-dependent contributions, resulting in what we call the  \emph{standard picture} vs. the \emph{control-adapted} picture, relative to a chosen $(\mathscr{F},\tilde{\mathscr{F}})$ pair. In particular, we recover the usual frequency-domain FF formalism as a special instance associated to the use of a Fourier frame in Sec. \ref{section:SPvsCA}, whereas in Sec. \ref{section:ModelReduction} we make precise the sense in which control-driven model reduction may be achieved, provided that a suitable \emph{finite-size frame condition} is obeyed. In essence, the latter ensures that arbitrary control sequences built out of $\mathscr{C}$ may be well approximated by finite expansions over elements of $\tilde{\mathscr{F}}$.

We then proceed to showcase the added generality and flexibility of our approach by focusing, in Sec.~\ref{sec:framecontrol}, on multiqubit dynamics under simultaneous additive decoherence and multiplicative control noise. After providing a general frame construction for this setting, we specialize to the simplest paradigmatic case of single-qubit dynamics. In particular, in Sec. \ref{sub:QNS} we demonstrate how control-adapted QNS techniques that are designed to work directly in the time domain provide new capabilities over existing protocols, by allowing the characterization of non-stationary noise environments of both classical and genuinely quantum nature. A complete frame-based C\&C protocol is exemplified in Sec. \ref{CAG}, where the information about the noise correlations acquired through a first stage of control-adapted QNS is subsequently incorporated in the optimal-control search for various target unitary gates. By comparing to the optimal-control solutions obtained under access to the full dynamics, we establish that our model-reduced representation incurs no significant loss of predictive power, insofar as arbitrary controlled evolutions built out of $\mathscr{C}$ are considered. Finally, we present in Sec. \ref{sub:univ} some important remarks on the extent to which QNS-inferred information may be universal for prediction and control, before concluding. Further technical detail is included in five separate Appendices, in order to make the presentation as self-contained as possible.

\section{Controlled open quantum dynamics}
\label{section:ControlledOpenQ}

We consider a controlled $d$-dimensional open quantum system $S$ evolving in the presence of an inaccessible environment (bath) $B$. In the interaction picture associated to the free bath evolution, the joint system-bath dynamics is governed by a Hamiltonian of the form $H(t) = H_S+H_{SB}(t) + H_{\rm ctrl}(t)$, where the Hamiltonian $H_S$ determines the free evolution of $S$, $H_{SB}(t)$ couples $S$ and $B$, and $H_{\rm ctrl}(t)$ describes open-loop control modulation acting on $S$ only. Let $\{\Lambda_0\equiv I_S, \Lambda_u \}$ denote a generalized (Hermitian) Pauli basis for the operator space on $S$, with $\Tr(\Lambda_u {\Lambda_v})=\delta_{uv}$. We consider a broad class of Hamiltonians that simultaneously account for \emph{additive} (control-independent) and \emph{multiplicative} (control-dependent) noise, according to
\begin{equation}
\!H (t) = \sum_{u\ne 0} \Lambda_u \otimes {B}_u^{(a)}(t)  + \sum_{v\ne0} h_v(t) [1+ \beta_v^{(m)}(t)]\Lambda_v .
\label{IPHam}
\end{equation}
Here, the $B_u^{(a)}(t) \equiv \tilde{B}_u^{(a)}(t) + \beta^{(a)}_u(t) I_B$ describe the always-on additive ($a$) noise from quantum and classical sources, respectively, with $\tilde{B}_u^{(a)}(t)$ being bath operators (not necessarily traceless in order to account for evolution due to $H_S$) and $\beta_u^{(a)}(t)$ stochastic processes. Control is introduced via user-defined  control profiles $\{ h_v(t) \}$ which, subject to system-dependent constraints (e.g., maximum amplitude, finite time-bandwidth product), determine the control capabilities $\mathscr{C}$ in the error-free scenario. More precisely, we define the control capabilities $\mathscr{C}$ as the set of control Hamiltonians $\{ h_v(t) \Lambda_v \}_{v\neq0}$ that can be implemented in a given experiment. There are two aspects to this: the allowed control ``directions'' $\Lambda_v$ and the ``control profiles'' $h_v(t)$, usually parametrically defined by a pulse shape and a range for the specifying parameters, e.g., $h_v(t) \sim \theta
\, e^{-t^2/2 \sigma^2}$ for 
$\theta \in [\theta_\textrm{min}, \theta_\textrm{max}]$
%$\theta \in [0, \pi]$ 
and $\sigma \in [\sigma_{\textrm{min}}, \sigma_{\textrm{max}}].$  
Furthermore, multiplicative ($m$) control noise, which often arises in realistic settings \cite{dial2013charge,SlepianQNS,Didier2019}, is captured by the stochastic processes $\beta_v^{(m)}(t)$. 

Our main objects of interest are the time-dependent expectation values of system observables $O=O^\dagger$, given by
\begin{equation}
E[O(T)]_{\rho_{SB}} \lspace{=} \left\langle \Tr_{S,B} [U(T) \rho_{SB} U^\dagger(T)(O\otimes I_B)]\right\rangle_c,
\label{eq:Expectation}
\end{equation} 
where $\rho_{SB}$ is the initial state of the joint system and bath. To obtain useful expressions for them we proceed as follows.  The unitary propagator for the evolution generated by $H(t)$ is given (in units $\hbar=1$) by the time-ordered exponential $U(t) = \mathcal{T}_+ e^{-i \int_0^t ds H(s)}$, and $\langle \cdot \rangle_c$ denotes the average over realizations of the stochastic processes. We may write $H(t)\equiv H_0(t) + H_e(t)$, with $H_0(t)$ representing the intended, error-free controlled dynamics and the error component $H_e(t)$ accounting for unwanted evolution due to environmental and control noise, as well as possibly $H_S$ (e.g., $H_0(t){=} H_{\rm ctrl}(t)$ if $H_S{=}0$ and no control noise is present, $\beta_v^{(m)}(t)=0$). The evolution due to $H_e(t)$ can then be isolated by moving to the interaction (toggling) frame associated to the error-free component $H_0(t)$.  That is, we let $U(t) =  U_0(t) \widetilde{U}(t),$ where 
\[  U_0(t) = \mathcal{T}_+ e^{-i \int_0^t ds H_0(s)} , \quad 
\widetilde{U}(t) = \mathcal{T}_+ e^{-i \int_0^t ds \widetilde{H}(s)},\]
and $\widetilde{H}(t) \equiv U_0^\dagger(t) H_e(t) U_0(t)$ is the toggling-frame Hamiltonian. By construction, such Hamiltonian vanishes in the absence of noise, and can be compactly written as
\begin{equation}
\widetilde{H}(t) 
\equiv  \sum_{\alpha=a,m}\sum_{u,v} y^{(\alpha)}_{u,v}(t) \Lambda_v \otimes B^{(\alpha)}_u(t), 
\label{eq:basicHam}
\end{equation} 
where we have defined $B_u^{(m)}(t) \equiv \beta_u^{(m)}(t)I_B$ and the elements $y^{(\alpha)}_{u,v}(t)$ of the \emph{control matrix} $\mathscr{Y}^{(\alpha)}(t)$ capture the effect of the control modulation on the noise terms in $H_e(t)$. Explicitly, we have 
\begin{subequations}
\begin{align*}
y^{(a)}_{u,v}(t) &= \Tr_S [ U^\dagger_0(t) \Lambda_u U_0(t) {\Lambda_v}]/d , \\
y^{(m)}_{u,v}(t) &= h_u(t) y^{(a)}_{u,v}(t).
\end{align*}
\end{subequations}
Finally, assuming that $\rho_{SB} {=} \rho_S {\otimes} \rho_B$ (see Ref.~\cite{Bplus} for a more general treatment), and that $O$ is invertible, with $\widetilde{O}(T)\equiv U_0^\dagger (T) O U_0(T)$, we write the desired time-dependent expectation value, Eq.~\eq{eq:Expectation}, as
\begin{equation}
E[O(T)]_{\rho_S \otimes \rho_B}= \Tr_S [V_O(T) \rho_{S} \widetilde{O}(T)],
\label{expec}
\end{equation}
where the system operator 
\begin{equation}
\label{eq:VO}
V_O(T)\equiv \langle \Tr_B  [\widetilde{O}^{-1} (T)\widetilde{U}^\dagger(T) \widetilde{O} (T)\widetilde{U}(T) \rho_B] \rangle_c
\end{equation}
contains all the unwanted effects due to $H_e(t)$ up to time $T$; that is, $V_O(T) = I_S$ represents ideal, noise-free dynamics. 

The operator $V_O(T)$ may be computed perturbatively, for instance through a Dyson or cumulant expansion~\cite{Kubo, multiaxis}. Regardless of what technique is chosen it can be shown that, due to time ordering, the dynamics depends only on certain linear combinations, say, $\mathcal{L}_{\vec{v}} (\vec{t}~)$, of the multi-time noise correlation functions $\{\langle B_{v_1}^{(\alpha_1)}(t_{\mu(1)}) \cdots B_{v_k}^{(\alpha_k)}(t_{\mu(k)}) \rangle\}$ with respect to the combined quantum and classical averages, $\langle\cdot\rangle\equiv \langle \Tr_B [\cdot \rho_B] \rangle_c$, with $\alpha_j \in \{a, m\}$ and $\mu$ being an arbitrary permutation of $k$ elements, $k\in {\mathbb N}$. While the specifics of the linear combinations which enter the expectations in Eq. \eqref{expec} depends on both $O$ and the details of the system and control setting, the possible contributions are determined by overlap integrals of the form (see Appendix \ref{App:OpenDynamics})
\begin{equation}
\label{eq:kIntTime}
{\cal I}^{(k)}_{\vec{\alpha};\vec{u},\vec{v}} (T)= \int_0^T d_> \vec{t}_{[k]}  \bigg[\!\prod_{j=1}^k y^{(\alpha_j)}_{u_j,v_j}(t_j) \bigg]
\mathcal{L}_{\vec{\alpha};\vec{v}} (\vec{t}~),
\end{equation}
with $\int_0^T d_> \vec{t}_{[k]} \equiv \int_0^T dt_1 \int_0^{t_1} dt_2 \cdots \int_0^{t_{k-1}} d t_k $.  

These integrals are key to understanding the effect of the noise on the system. The conventional approach to their analysis involves rewriting them as multi-dimensional overlap integrals in frequency space via an appropriate Fourier transform. This leads to the standard FF formalism~\cite{Kurizki2004,biercuk2011dynamical,Paz-Silva-FF2014}, where frequency-domain FFs and, for general non-Gaussian noise, higher-order (poly-)spectra capture the effect of the control and noise, respectively. On the one hand, one expects that, by (experimentally) obtaining the value of any such integral for various FFs, it may be in principle possible to deconvolve it, and thus infer relevant information about the noise correlation functions (or their Fourier transform). This is indeed how QNS protocols work. On the other hand, one also sees that mitigating the effect of the noise is akin to minimizing the value of such integrals, which is the working principle behind existing decoherence suppression and optimal control protocols. The frequency- (or time-)domain representations, however, are agnostic to the control capabilities or constraints $\mathscr{C}$, which are unavoidable in any realistic setting, since the $\{h_v(t)\}$ cannot be arbitrarily chosen. Indeed, these constraints translate into limitations to the information that can be inferred about the noise (e.g., in comb-based QNS protocols the frequency-domain spectra are sampled via a discrete grid), or superfluous information used for synthesizing optimal control (e.g., rather than full knowledge of the noise correlation functions, only knowledge the overlap between noise and the admissible controls should be necessary). This motivates the search for a space -- or a mathematical language -- in which the relevant overlap integrals can be more efficiently represented by incorporating information about $\mathscr{C}$ from the outset.

\section{Frame-based filter functions and control-driven model reduction}
\label{section:FrameFFandReduction}

In order to meet the challenge identified in the previous section, instead of moving to the frequency domain we use the more general language of {\it frames} \cite{framesoriginal,frames2,FrameIntro,FrameContinuous1,FrameContinuous2}. Frames have a long tradition in signal processing, thanks to the flexibility they afford as compared to bases (e.g., in generalizing time-to-frequency-domain transforms) as well as to various properties particularly advantageous to signal reconstruction (e.g., robustness to noise~\cite{noiseframes}). Moreover, they are already successfully exploited in different quantum applications~\cite{Bplus,FrameTomo,QuasiProb}. Leveraging the frame language in our context will first and foremost afford extra flexibility, as a frame can be chosen in a way that matches -- in a sense that will be made precise later -- the available control. In turn, aided by a change in point of view, this will allow us to {\it efficiently} represent each of the overlap integrals $\mathcal{I}^{(k)}$s as a sum over a finite domain, thereby achieving control-driven model reduction.

\subsection{Basics on frames}
\label{section:FrameBasics}

The mathematical theory of frames is quite sophisticated and an exhaustive review of the topic is beyond our scope. While we refer the interested reader to the extensive literature for a more complete and rigorous account~\cite{frames2,FrameIntro,FrameContinuous2}, we summarize here the basic definitions needed for the exposition of our result. We further discuss in Appendix \ref{App:Frames} illustrative examples directly relevant to the control scenarios we analyze.

Let $\mathcal{H}$ be a complex (finite-dimensional or separable) Hilbert space, consisting of functions $f(t)$, $t \in [0,T]$, with inner product and norm given, respectively, by 
$$(f,g) \equiv \frac{1}{T} \int_0^T dt f(t) g^*(t), \quad \Vert f \Vert^2 \equiv (f,f).$$ 
A \emph{discrete frame} for $\mathcal{H}$ is an (at most) countable sequence $\mathscr{F}\equiv \{ \phi_n\}_{n}^{},$ with $\phi_n \in \mathcal{H}$ and $n\in {\mathbb Z}$, satisfying the \emph{frame condition}, that is, 
\begin{equation}
\label{eq:FrameCond}
A \Vert f \Vert^2 \leq \sum_{n} |(f,\phi_n)|^2 \leq B \Vert f \Vert^2,  \,\quad \forall f \in \mathcal {H},
\end{equation}
with $0<A \leq B< \infty$ being the lower and upper frame bounds, respectively. Of particular interest are {\em tight frames}, for which $A=B$, and {\em Parseval frames}, for which, in addition, $A=B=1$. Notably, an orthonormal basis in ${\mathcal H}$ is a Parseval frame, and indeed one sees that in this case the frame condition is equivalent to Parseval's identity.

Frames may be seen as redundant (linearly-dependent) spanning sets for $\mathcal{H}$. More precisely, any $f \in \mathcal{H}$ can be expanded as $f(t) = \sum_n c_n \phi_n(t),$ $t \in [0,T]$, 
where the coefficients are given by the \emph{reconstruction formula},
\begin{equation}
\label{eq:ReconstructionFormula}
f(t) = \sum_n (f,\tilde{\phi}_n)\, \phi_n(t) = \sum_n (f,{\phi}_n)\, \tilde{\phi}_n(t),
\end{equation}
and $\{\tilde{\phi}_n\}_n$ are elements of a frame \emph{dual} to $\mathscr{F}$. While a frame $\mathscr{F}$ may in general admit infinitely many dual frames, there exists a {\em canonical dual frame} $\tilde{\mathscr{F}}$ which is special, in the sense that it minimizes the norm of the expansion 
$$\sum_n | (f,\tilde{\phi}_n ) |^2 \le \sum_n |c_n|^2,$$
for any sequence $\{c_n\}_n \in \ell_2(\mathbb{Z})$ that satisfies $f=\sum_n c_n \phi_n$, with equality holding if and only if $c_n =(f, \tilde{\phi}_n)$ for all $n$. Such a dual frame is determined by $\tilde{\mathscr{F}}=\{\tilde{\phi}_n\} = \{\mathbf{S}^{-1}{\phi}_n\},$ in terms of the positive {\it frame operator}  $\mathbf{S}: \mathcal{H} \rightarrow \mathcal{H},$  given by
$$\mathbf{S} f \equiv \sum_n (f,\phi_n) \phi_n.$$ Note that $\mathbf{S}$ is a multiple of the identity if the frame is tight and it coincides with the identity for a basis, that is, an orthonormal basis is self-dual. Of special relevance to this work will be 
\emph{finite frames}, $\{\phi_n\}_n$, $n=1,\ldots, N_\# <\infty$.

More generally, {\it continuous frames} for which the labelling index $ n \mapsto x\in X$ changes continuously can also be constructed. Given a measure space $(X,\mathscr{W},\mu)$, a family $\mathscr{F}\equiv \{ \phi_x\}_{x}$ is a \emph{continuous frame} if (i) for all $f \in \mathcal{H}$,  $(f,\phi_x)$ is $\mathscr{W}$-measurable in $X$; and (ii) there exist constants $0<A \leq B <\infty $ such that
\begin{equation}
 \label{eq:FrameCont}
A \Vert f \Vert^2 \leq \int_X |(f,\phi_x) |^2 d\mu(x) \le B \Vert f \Vert^2, \, \quad \forall \, f \in \mathcal{H}.
\end{equation}
From the above it follows that $\overline{\rm span} \{(\phi_x)\}_{x \in X} = \mathcal{H},$ and an appropriate notion of dual frame and the reconstruction formula can be introduced by essentially replacing sums -- as in Eq.~\eqref{eq:ReconstructionFormula} -- with integrals of the form $\int_X d\mu(x)$. In this way, the special case of a discrete frame is recovered when $\mu$ is a counting measure and $X= \mathbb{Z}$ (or $\mathbb{N}$). Prominent examples of continuous frame expansions are expansions into canonical and generalized coherent states and the continuous short-time Fourier (or G\'{a}bor) transform \cite{FrameContinuous2}, whereas Fourier series are a special case of expansions into discrete exponential (or Fourier) frames \cite{frames2,FrameIntro}.

\noindent
\begin{table*}
\setlength{\fboxrule}{1pt}
\fbox{
\begin{minipage}[t][10.5cm]{0.45\linewidth}
\centering
        \textbf{(i)} \emph{Standard-Picture} (SP)\\
        \rule{7cm}{0.5pt}
\begin{subequations}
        \begin{flalign*}
    \mathcal{L}_{\vec{\alpha},\vec{v}}(\vec{t}~) &= \sum_{\vec{n}}  \Big(\mathcal{L}_{\vec{\alpha},\vec{v}}(\vec{s}~)
    , \prod_i {\phi}_{n_i}^{(\alpha_j)} (s_i)^{{\ast}}
    \Big)   \prod_j \tilde{\phi}_{n_j}^{(\alpha_j)} (t_j)^\ast \notag& \\ & \equiv \sum_{\vec{n}} S_{\vec{\alpha};\vec{v}}^{\!(k)}(\vec{n}) \prod_j \tilde{\phi}_{n_j}^{(\alpha_j)} (t_j)^\ast
          %\label{subeq:BathSP}
          &\\
          %--------
        F_{\vec{\alpha};\vec{u},\vec{v}}^{(k)}(\vec{n}) &= \int d_>\vec{t}_{[k]} \prod_j y^{(\alpha_j)}_{u_j,v_j}(t_j) \tilde{\phi}_{n_j}^{(\alpha_j)} (t_j)^\ast
        %\label{subeq:FilterFrameSP}
        &\\
         %--------
         S_{\vec{\alpha};\vec{v}}^{\!(k)}(\vec{n})&=\!\int\! d\vec{t}_{[k]}  \mathcal{L}_{\vec{\alpha};\vec{v}} (\vec{t}~) \prod_i \phi_{n_i}^{(\alpha_i)}(t_i)
         %\label{subeq:SpectraFrameSP}
         &\\
         %----------
        \mathcal{I}^{(k)}_{\vec{\alpha};\vec{u},\vec{v}}
        &= \sum_{\vec{n}}  F_{\vec{\alpha};\vec{u},\vec{v}}^{(k)}(\vec{n})  S_{\vec{\alpha};\vec{v}}^{(k)}(\vec{n})&
        %\label{subeq:DynamicalIntegralFrameSP}
        \end{flalign*}
\end{subequations}
%-----------------------------------------------------
\centering
\vspace{-1pt}
\rule{8cm}{1pt}
\textbf{(iii)} \emph{Frequency-domain} (SP) \\
\rule{7cm}{0.5pt}
\begin{subequations}
\begin{align*}
F^{(k)}_{\vec{\alpha};\vec{u},\vec{v}}(\vec{\omega}) &= \int_0^T \!\! d_>\vec{t}_{[k]} \,  \prod_{j=1}^k y^{(\alpha_j)}_{u_j,v_j}(t_j) e^{i  \vec{\omega} \cdot \vec{t}}\\
S^{(k)}_{\vec{\alpha};\vec{v}}(\vec{\omega}) &= \int_{-\infty}^\infty d\vec{t}_{[k]} ~ \mathcal{L}_{\vec{\alpha},\vec{v}}(\vec{t}~) e^{- i \vec{\omega} \cdot \vec{t}} \\
\mathcal{I}^{(k)}_{\vec{\alpha};\vec{u},\vec{v}} &=  \int_{-\infty}^\infty  d\vec{\omega} F^{(k)}_{\vec{\alpha};\vec{u},\vec{v}} (\vec{\omega}) S^{(k)}_{\vec{\alpha};\vec{v}}(\vec{\omega})
\label{eq:kIntFreq}
\end{align*}
\end{subequations}
\end{minipage}
}
%-------------------------------------
\fbox{
\begin{minipage}[t][10.5cm]{0.45\linewidth}
\centering
\hspace{1cm} \textbf{(ii)} \emph{Control-Adapted} (CA)\\
\hspace{10mm}\rule{7cm}{0.5pt}
\vspace{-3.5mm}
\begin{subequations}
       \label{CA}
       \begin{flalign*}
              \prod_j y^{(\alpha_j)}_{u_j,v_j}(t_j)& = \! \prod_j\sum_{\vec{n}} \!\big(  y^{(\alpha_j)}_{u_j,v_j}(s_j) ,\tilde{\phi}_{n_{{j'}}}^{(\alpha_{j})} (s_{{j}}) \big) \phi_{n_{j'}}^{(\alpha_j)} (t_{j}) \notag&\\
              & \equiv \! \prod_{j}\sum_{\vec{n}}  F_{\alpha_j;u_j,v_j}^{(1)}\!(n_{j'}) \phi_{n_{j'}}^{(\alpha_j)} (t_{j})&\\
        %-----------
        F_{\alpha_j;u_j,v_j}^{(1)}\!(n_j) &= \!\!\int \!dt\, y^{(\alpha_j)}_{u_j,v_j}(t)  \tilde{\phi}_{n_j}^{(\alpha_j)} (t)^\ast
        %\label{subeq:FilterFrameCA}
        &\\
        %-----------
        \bar{S}_{\vec{\alpha};\vec{v}}^{(k)}(\vec{n}) &= \int d_>\vec{t}_{[k]} \mathcal{L}_{\vec{\alpha},\vec{v}}(\vec{t}~) \prod_i {\phi^{(\alpha_i)}_{n_i} (t_i)}
        %\label{subeq:SpectraFrameCA}
        &\\
       %------------
        \mathcal{I}^{(k)}_{\vec{\alpha};\vec{u},\vec{v}}
       &= \sum_{\vec{n}} \prod_j F^{(1)}_{\alpha_j;u_j,v_j}({n_j}) \bar{S}_{\vec{\alpha};\vec{v}}^{\!(k)}(\vec{n})&
       %\label{subeq:DynamicalIntegralFrameCA}
       \end{flalign*}
\end{subequations}
\centering 
\vspace{0.3mm}
\rule{8.9cm}{1pt}\\
\hspace{1cm}\textbf{(iv)} \emph{Frequency-domain} (CA) \\
\hspace{1cm}\rule{7.cm}{0.5pt}
\vspace{-2.5mm}
\begin{subequations}
\begin{align*}
    F_{\alpha_j;u_j,v_j}^{(1)}\!(\omega) &= \int_0^T dt\, y^{(\alpha_j)}_{u_j,v_j}(t) e^{i \omega t}\\
    \bar{S}_{\vec{\alpha};\vec{v}}^{(k)}(\vec{\omega}) &= \int_{-\infty}^{\infty} d_>\vec{t}_{[k]} \mathcal{L}_{\vec{\alpha},\vec{v}}(\vec{t}~) e^{-i \vec{\omega}\cdot\vec{t}}\\
    \mathcal{I}^{(k)}_{\vec{\alpha};\vec{u},\vec{v}} &=\!\!\int_{-\infty}^\infty \! d \vec{\omega} \Big( \prod_j F_{\alpha_j;u_j,v_j}^{(1)}\!(\omega_j) \Big) \,\bar{S}_{\vec{\alpha};\vec{v}}^{(k)}(\vec{\omega})
\end{align*}
\end{subequations}
\end{minipage}
}
\caption{Summary of the defining relationships for standard (i) vs. control-adapted (ii) pictures. The corresponding frequency-domain SP and CA representations are also included in (iii) and (iv) for comparison.} 
\label{tab:SPvsCA}
\end{table*}

\subsection{Standard-picture vs. control-adapted filters and spectra}
\label{section:SPvsCA} 

To apply the frame formalism to our problem,
we start by noting that there are two ways of representing any of the time-ordered integrals of Eq.~\eq{eq:kIntTime}. On the one hand, we can write the noise correlation function in the chosen frame and rewrite the remaining factors as a FF associated to that frame. We dub this the ``standard picture'' (SP). On the other hand, we may expand the control matrix elements in the frame and have the remaining factors be the equivalent of a noise spectra in the frame language -- resulting in what we dub the ``control adapted'' (CA) picture. Mathematically, these two approaches lead, respectively, to
\begin{subequations}
\begin{eqnarray}
{\cal I}^{(k)}_{\vec{\alpha};\vec{u},\vec{v}}(T)
& \!= \!& \!\sum_{\vec{n}} {F}_{\vec{\alpha};\vec{u},\vec{v}}^{(k)}(\vec{n}, T) \, 
{S}_{\vec{\alpha};\vec{v}}^{(k)}(\vec{n}) \label{eq:standard} \\
& \!= & \!\sum_{\vec{n}} \Big[\prod_{j=1}^k {F}_{\alpha_j;u_j,v_j}^{(1)}(n_j, T) \Big]    
\bar{S}_{\vec{\alpha};\vec{v}}^{(k)}(\vec{n}).
\label{eq:tconvo} 
\end{eqnarray}
\end{subequations}
Eq.~\eq{eq:standard} represents the direct generalization to the (discrete) frame language of the standard frequency-domain representation: the $k$th-order frame-based FFs and frame-based power spectra are given, respectively, by
\begin{subequations}
\begin{align}
    F_{\vec{\alpha};\vec{u},\vec{v}}^{(k)}(\vec{n}, T) &\equiv
        \int d_>\vec{t}_{[k]} \prod_j y^{(\alpha_j)}_{u_j,v_j}(t_j)\tilde {\phi}_{n_j}^{(\alpha_j)}(t_j)^\ast, \\
    S_{\vec{\alpha};\vec{v}}^{(k)}(\vec{n}) &= \int\! d \vec{t}_{[k]} \, \mathcal{L}_{\vec{\alpha};\vec{v}} (\vec{t}) \prod_{i=1}^k {\phi}^{(\alpha_i)}_{n_i} (t_i),
\end{align}	
\end{subequations}
where we allow for different frames $\mathscr{F}^{(\alpha)}$ for $\alpha \in\{a,m\}$. Indeed, as shown explicitly in Appendix \ref{App:Fourier}, the standard frequency-domain FF formalism \cite{Kurizki2004,Paz-Silva-FF2014} is recovered when one uses a Fourier frame \cite{ortega2002fourier}. In contrast, Eq.~\eq{eq:tconvo} does not only represent a generalization to the frame language; but, importantly for our purposes, it also provides a change in viewpoint -- a dual representation in which 
\begin{subequations}
\begin{align}
{F}_{\alpha_j;u_j,v_j}^{(1)}(n_j, T) &\equiv (y^{(\alpha_j)}_{u_j,v_j} {\displaystyle ,} ~\tilde{\phi}_{n_j}^{(\alpha_j)}) \notag \\ 
&= \int_0^T dt \, y^{(\alpha_j)}_{u_j,v_j}(t)  \tilde{\phi}_{n_j}^{(\alpha_j)} (t)^*  , \label{FFF}\\
\bar{S}_{\vec{\alpha};\vec{v}}^{(k)}(\vec{n}) &= \int d_>\vec{t}_{[k]} \,\mathcal{L}_{\vec{\alpha};\vec{v}} (\vec{t}~) 
\prod_{i=1}^k {\phi^{(\alpha_i)}_{n_i} (t_i)}, \label{CAS}
\end{align}
\end{subequations}
are the {\it frame-based fundamental FFs} and the {\it frame-based CA-spectra}, respectively. Note that, in the CA representation, the time-ordering is moved from the FFs to the CA-spectra, with a twofold implication: on the one hand, as in the SP setting, the CA-spectra encode all the information about noise correlations that influence the system dynamics, as we will explicitly demonstrate in Sec. \ref{sec:framecontrol}. On the other hand, the fact that the CA-FFs are all one-dimensional integrals without any time ordering makes their use more advantageous to both theoretical analysis and numerical implementation of C\&C.

We also note that there is a degree of arbitrariness in the above definitions since, given a frame $\mathscr{F}= \{ \phi_n\}$ and its dual $\tilde{\mathscr{F}}= \{ \tilde{\phi}_n\}$, their corresponding complex conjugates (say, ${\mathscr{F}}^\ast$ and $\tilde{\mathscr{F}}^\ast$) are also frames. In order to maintain a certain symmetry in our expressions, in this paper we will choose to expand in $\tilde{\mathscr{F}}^\ast$ and ${\mathscr{F}}$ in order to generate, respectively, the SP- and CA-formalism. For completeness, we summarize the resulting expressions in Tab.~\ref{tab:SPvsCA}.

\subsection{Frame-based control-driven model reduction}
\label{section:ModelReduction}

The next key step in our approach is to observe that, as far as the effect of the noise on the system is concerned (captured by the operator $V_O$ in Eq. \eqref{eq:VO}), what matters are not the control inputs $\{ h_u(t)\}$ themselves, but rather their control matrix elements $\{ y_{u,v}^{(\alpha)}(t) \},$ which are related to the $\{ h_u(t) \}$ via conjugation under the known map $U_0(t)$. Moreover, any constraints on $h_u(t)$ (say, limited bandwidth, a minimum time between two consecutive pulses, or a finite time resolution) necessarily translate into limitations on the possible form that the $y_{u,v}^{(\alpha)}(t)$ can take.  
 
These observations, the flexibility of frames, and the CA representation of $\mathcal{I}^{(k)}_{u,v}$ come together as follows: (i) for given control capabilities $\mathscr{C}$, the possible $\{y^{(\alpha)}_{u_,v}(t)\}$ are known; (ii) it is in principle possible to tailor the choice of frame $\mathscr{F}^{(\alpha)}$ so that it ``efficiently'' represents such $\{y^{(\alpha)}_{u_,v}(t)\}$; (iii) this, in turn, leads to an ``efficient'' representation of the integrals in $\mathcal{I}^{(k)}_{u,v}$. More formally, we introduce the following condition:

\medskip

{\bf Definition [Finite-size frame (FSF) condition].} {\em Let $\mathscr{C}$ specify fixed control capabilities, which determine a (possibly infinite) set of control matrix elements,  $y_{u,v}^{(\alpha)}(t) \in L^2([0,T])$, $\alpha\in\{a,m\}$. We say that the FSF condition holds if one can build \emph{finite-size} frames $\mathscr{F}^{(\alpha)}_\# \equiv \{\phi^{(\alpha)}_n\}$, ${n=1,\dots,N^{(\alpha)}_\#},$ and dual frame $\tilde{\mathscr{F}}^{(\alpha)}_\#$, such that for all $y_{u,v}^{(\alpha)}(t)$ allowed by \vspace*{-2mm}$\mathscr{C}$, 
\begin{equation}
\label{CS}
y^{(\alpha)}_{u,v}(t) \!= \!\!\sum_{n=1}^{N^{(\alpha)}_\#} {F}^{(1)}_{\alpha; u,v}(n,t) \phi^{(\alpha)}_{n}(t).
\vspace*{-0.1mm}
\end{equation} 
We say that the FSF condition is satisfied to tolerance $\varepsilon \geq0$ over $[0,T]$ if the above equality can be approximately obeyed with error no larger than $\varepsilon$ (in the $L^2$ norm).}

\medskip

\noindent 
If the FSF condition holds, the $\{y^{(\alpha)}_{u_,v}(t)\}$ are represented efficiently in the sense that they are well approximated by a finite expansion over the elements of $\tilde{\mathscr{F}}^{(\alpha)}_\#$. It then follows that 
\begin{equation} 
\label{FSFconsequence}
{\cal I}^{(k)}_{\vec{\alpha};\vec{u},\vec{v}}(T)
 \simeq  \! \sum_{n_i=1}^{N^{(\alpha_j)}_\#} \Big( \prod_j F^{(1)}_{\alpha_j;u_j,v_j}\!(n_j) \Big) \bar{S}_{\vec{\alpha};\vec{v}}^{(k)}(\vec{n}),
\end{equation}
that is, each integral can be efficiently represented by a finite sum up to an error which scales as $\mathcal{O}(\varepsilon k).$ A key consequence of the above is that it allows us to identify the components of the noise that are relevant to the dynamics allowed by $\mathscr{C}$: namely, the {\it finite set} of CA-spectra, 
\begin{equation}
\mathscr{\bar{S}}\vert_\mathscr{C} = \{\bar{S}_{\vec{\alpha};\vec{v}}^{(k)}(\vec{n})\}, \quad {n_j\in [1, N_\#^{(\alpha_j)}]},
\label{caspectra}
\end{equation} 
(or specific combinations thereof) that contribute to the expectations of observables as in Eq. \eqref{expec}. Thus,  $\mathscr{\bar{S}}\vert_\mathscr{C}$ represents {\it both} the information that can be extracted from the reduced system dynamics, and what suffices to optimally control it, under the resource constraints $\mathscr{C}$, i.e., a {\it  model-reduced} description of the noisy dynamics. Generally, there will be a trade-off between the model-reduction properties of the frame and the accuracy: a larger frame will lead to a smaller $\varepsilon$, which however necessarily implies that each $\mathcal{I}^{(k)}_{\vec{\alpha};\vec{u},\vec{v}}(T)$ is represented by a sum over a larger domain.

It is clear then why there is a need for a flexible language: one must design $\mathscr{F}$ according to $\mathscr{C}$. The frame language provides a constructive and relatively straightforward mechanism to do so. For instance, one can choose as frame elements a subset of the possible $y_{u,v}^{(\alpha)}(t)$, say $\mathscr{C}_0$, and expand every control matrix element in terms of this subset via Eq. \eqref{eq:ReconstructionFormula}. If $\mathscr{C}_0$ is chosen adequately, the error $\epsilon$ in the reconstruction can be made small, and an accurate model reduction is achieved. This ($\mathcal{F} = \{ y(t)\vert_{\mathcal{C}_0}\}$) will be essentially how we build our frames in this paper, as it greatly simplifies the CA-QNS problem: if one chooses as control for a QNS protocol an element of $\mathscr{C}_0$, the sum in Eq.~\eqref{FSFconsequence} can reduce to a single term. Now, while a full CA-QNS protocol is not as simplistic as that, this observation gives a glimpse into how much the ``correct'' language can simplify the task. Notice that such a simplification would not be achievable in general if one insists on considering only expansions in orthonormal bases, as the $y_{u,v}^{(\alpha)}(t)$ are typically not orthogonal to each other.

Clearly, our choice for $\mathscr{F}$ is not unique and other ways of building a convenient frame may be possible, depending on the task at hand. In general, this problem is related to that of building a {\it parsimonious model}  -- in  the language of the model-reduction literature \cite{Obinata} -- which in our context characterizes the ability of the chosen frame to approximate the elements of the control matrix. We leave it to future work to explore what frame, for fixed control capabilities $\mathscr{C}$, allows for maximum parsimony while retaining sufficient predictive power. 
We stress, however, that the key is tying the choice of $\mathcal{F}$ to the available $\mathscr{C}$ and the tolerance $\varepsilon.$ This 
%of course 
follows the reasoning that if $\mathscr{C}$ changes, then the components of the noise that affect the quantum system also change. Indeed, given a change $\mathscr{C} \mapsto \mathscr{C}'$ the first step should be verifying that $\mathscr{F}$ still satisfies the FSF condition to an acceptable $\varepsilon.$ Now, the change in $\mathscr{F}$ can take many forms depending on the change in $\mathscr{C}$. It can be as radical as completely changing the functional form of the frame elements or as simple as adding more elements to the original frame. The former is necessary when, for example, the accessible control profiles change from $h_v(t) \sim e^{-(t-\tau)^2/2 {\sigma}^2}\vert_{\sigma \in [\sigma_{\rm min},\sigma_{\rm max}]}$ to 
$h'_v(t) \sim e^{-|t-\tau|/2 \sigma^2}\vert_{\sigma \in [\sigma_{\rm min},\sigma_{\rm max}]}.$ In contrast, the latter type of frame change would be sufficient, for example, when the shape of the profiles $h_v(t)$ remains the same but the range of the defining parameters changes, e.g., to $\sigma \in [{\sigma}'_{\rm min},{\sigma}'_{\rm max}]$.

\section{Frame-based approach to characterization and control}
\label{sec:framecontrol}

\subsection{From noise and control assumptions to frame construction}
\label{sec:CAframeconstruction}

We are now ready to deploy our tools. To do so, we first introduce a multiqubit system control problem and show that the same frame can be used irrespective of the number of qubits, as long as the control constraints are homogenous. We will exemplify the frame construction for both instantaneous and non-instantaneous control settings and use these in the two key applications we anticipated: QNS beyond the standard frequency domain and noise-tailored optimized gate design.

\subsubsection{The multiqubit model system}

To demonstrate the usefulness of our formalism, we now specialize our system to $N$-qubits, and correspondingly take $\{\Lambda_u \}\equiv \{\Sigma_u\}$ to be the usual Pauli product-operator basis, with $u\in \{0, \dots,4^N-1 \}$ and $\Sigma_0 = \mathbb{I}^{\otimes N}.$ The qubits are exposed to additive noise, described as before by $B_u^{(a)}(t) \lspace{=} \tilde{B}_u^{(a)}(t) +\beta_u^{(a)}(t)I_B$, along with multiplicative control noise which, for simplicity, we assume to be isotropic, $B^{(m)}_u(t) \lspace{\equiv} \beta^{(m)}(t)I_B$, for all $u$. We focus on the paradigmatic scenario in which $\mathscr{C}$ comprises $M$ non-overlapping pulses of duration $\tau\lspace{\equiv} T/M$ applied over $[0,T]$, implemented by 
\begin{equation}
\label{eq:MultiQubitCtrl}
H_{\rm ctrl}(t) = \left(1 + \beta^{(m)}(t)\right)\sum_{j=1}^M \theta_j h(t_j,t) \, {\vec{n}^{(j)}\!\cdot\vec{\Sigma}/2}, 
\end{equation}
where $\vec{\Sigma}$ excludes $\Sigma_0$, $\vec{n}^{(j)} \in {\mathbb R}^{4^N-1}$, $\vert\vert \vec{n}^{(j)} \vert\vert=1$ and $\theta_j \in [0, 2\pi]$ specifies the $j$th pulse,
described by a fixed (normalized) control profile $h(t_j,t)$ centered around  $t_j\lspace{\equiv}(j{-}1/2)\tau$   and proportional to a window function $W_{j, \tau}(t)$ defined via
 \begin{equation}
 \label{window}
 	W_{j, \tau}(t) \equiv
 		\begin{cases}
 			1 & (j{-}1) \tau \le t < j\tau,\\
 			0 & \text{otherwise}.
 		\end{cases}
 \end{equation} 
 The above form of $H_{\rm ctrl}(t)$ is general enough to support single- or multi-body Hamiltonian controls, although in a realistic system one would typically be limited to at most two-body Hamiltonians. To simplify our calculations, however, we demand that the possible control inputs $h_u(t)$ are the same regardless of $\Sigma_u,$ that is, the same control constraints apply to any of the possible $H_{\rm ctrl}(t)$.

Under the above assumptions, one realizes that the form of the frame is independent of the number of qubits, as we will show. Mathematically, the finite support of $h(t_j,t)$ implies that one can write $U_0(t)$ as a piece-wise function given by
\begin{equation*}
\label{uctrl}
U_0(t) = \mathcal{U}_j \prod_{\ell=1}^{j-1} U_{\ell}, \quad t \in \left[(j-1) \tau, j \tau \right],
\end{equation*}
where 
\begin{subequations}
\begin{align*}
	\mathcal{U}_j &= e^{-i \frac{\theta_{j}}{2} \psi_t^{({j})} \vec{n}^{({j})}\cdot \vec{\Sigma}}, 
	\quad 
	%\\
	U_\ell = e^{ -i   \frac{\theta_{\ell}}{2} ~ \vec{n}^{({\ell})}\!\cdot \vec{\Sigma} }, \\
	\text{with} \quad
	\psi^{(j)}_t &\equiv \int_{(j-1) \tau}^{t} \!\!\!ds \, h(t_j,s). 
\end{align*}
\end{subequations}
In turn, this leads to a toggling-frame Hamiltonian (Eq.~\eqref{eq:basicHam}) of the form 
\begin{equation}
\label{Hactual}
\widetilde{H}(t) = \sum_{u,v} \left[ y_{u,v}^{(a)}(t) \Sigma_v \otimes B_u^{(a)} (t) + y_{v}^{(m)}(t) \beta^{(m)}(t) \Sigma_v  \right]. 
\end{equation} 
Here, the additive control  matrix elements are given by
\begin{align}
y^{(a)}_{u,v}(t) &= \frac{1}{2^N}\Tr[ U_0(t)^\dagger \Sigma_u U_0(t) \Sigma_v] \notag \\ 
& = \frac{1}{2^N}\sum_{w=0}^{4^{^N}\!\!\!-1} c^{(j-1)}_{v,w}  \Tr\left[ \mathcal{U}_j^\dagger \Sigma_u \mathcal{U}_j \Sigma_w \right],
\label{CtrlElementTr} 
\end{align}
for $ t \in [(j-1) \tau, j \tau]$, where we have used that
\begin{align*}
\big(\prod_{\ell=1}^{j-1} U_{\ell}\big) \Sigma_v \big(\prod_{\ell=1}^{j-1} U_{\ell}\big)^{\!\dagger} & \!=\! \sum_w \Tr \Big[ \big(\prod_{\ell} U_{\ell}\big) \Sigma_v \big(\prod U_{\ell} \big)^{\!\dagger} \Sigma_w \Big]\!\frac{\Sigma_w}{2^N} \\ 
& \equiv \! \sum_w c^{(j-1)}_{v,w} \frac{\Sigma_w}{2^N}.
\end{align*}
This ultimately leads to
\begin{align} 
y^{(a)}_{u,v}(t) =& \sum_{w=0}^{4^{^m}\!\!\!-1} c^{(j-1)}_{v,w} \Big(  
				k^{(0)}_{j,l;u,w}\cos[ \theta_{j} \psi^{({j})}_t  p_l(\vec{n}^{(j)})] \notag \\
				+ &k^{(1)}_{j,l;u,w} \sin[ \theta_{j} \psi^{({j})}_t p_l(\vec{n}^{(j)})] +k^{(2)}_{j,l;u,w} \Big),
\label{ctrl(a)}
\end{align}
where $\{k^{(s)}_{j,l;u,w}\}$ and $p_l(\vec{n}^{(j)})$ are polynomials of the elements of $\vec{n}^{(j)}$, whose values can be numerically calculated (from 
%the last two lines of 
Eq.~\eq{CtrlElementTr}) given a fixed number of qubits $N$. 

The multiplicative control matrix elements can be similarly calculated, by noting that $h(t_j,t) $ is assumed to be compactly supported in $[(j-1) \tau, j \tau]$.  Thus, 
\begin{align*}
 U_0(t)^\dagger  \big( h(t_j,t) ~& \vec{n}^{(j)} \cdot \vec{\Sigma} \big) U_0(t) = \\
 & h(t_j,t) \big(\prod_{\ell=1}^{j-1} U_{\ell} \big)^{\!\dagger} \!\big( \vec{n}^{(j)} \cdot \vec{\Sigma} \big) \!\big(\prod_{\ell=1}^{j-1} U_{\ell} \big),
\end{align*}
for $t \in [(j-1) \tau, j \tau]$, since $\mathcal{U}_j ( \vec{n}^{(j)} \!\cdot \vec{\Sigma}) \,\mathcal{U}_j^\dagger =\vec{n}^{(j)} \!\cdot \vec{\Sigma}$. It follows that
\begin{equation}
	y^{(m)}_{v}(t) =\Big( \sum_{w} n^{(j)}_w {c_{v,w}^{(j-1)}}\Big) h(t_j,t){\theta_j/2}.
	\label{ctrl(m)}
\end{equation} 
% We note that $y_{u}^{(a)}(t)$ is a linear combination of $\{W_{j, \tau}(t) \cos [{ \theta_j \psi^{(j)}_t}], W_{j, \tau}(t) \sin[ {\theta_j \psi^{(j)}_t}]\},$ with 
%$$\psi^{(j)}_t \equiv \int_{(j-1) \tau}^{t} ds h(t_j,s)$$ and $y_{u}^{(m)}(t) {\propto} \,h(t_j,t)$.

\subsubsection{The frame}
\label{subsection:frame}

The important consequence of the above calculations is that $y^{(a)}_{u,v}(t)$ is {spanned} by the functions $\{W_{j,\tau}(t) \sin[ \eta \psi^{(j)}_t] ,W_{j,\tau}(t) \cos[ \eta \psi^{(j)}_t] \}$, for some $\eta$ and $W_{j,\tau}(t)$, while the $y_{u,v}^{(m)}(t)$ are spanned by the $h(t_j,t)$. Given this, it is then natural to use as frames 
\begin{align}
\label{eq:frame}
\mathscr{F}^{(a)}_\#  &\equiv \{W_{j, \tau}\cos[\eta \psi^{(j)}_t] , W_{j, \tau}\sin[\eta \psi^{(j)}_t]\}, \notag \\
\mathscr{F}^{(m)}_\#  &\equiv \{h(t_j,t)\}, \\
j &\in [1,M] , \quad \eta =  2 \pi k/\tilde{N}_\#,  \quad k \in [0,\tilde{N}_\#], \notag
\end{align}
where $\tilde{N}_\#$ is a free parameter which determines the size of the frame and the tolerance $\varepsilon$, i.e., how well the FSF condition is satisfied. We note that while a finite $\tilde{N}_\#$ implies a non-zero $\varepsilon$, the latter decreases as $\tilde{N}_\#$ grows: if $\mathscr{F}^{(\alpha)}_\#$ contains {\em all} admissible $y_{u,v}^{(\alpha)}(t)$, the FSF is exactly satisfied  ($\varepsilon=0$). Moreover, the non-overlapping nature of the pulses implies that the error $\varepsilon\vert_{M}$ for an $M$-pulse control matrix grows only linearly with $M$, i.e., $\varepsilon\vert_{M} \sim \mathcal{O}(M\,  \varepsilon\vert_{1})$ for a frame with $N_\# = M (2 \tilde{N}_\#+1)$ elements. A larger $\tilde{N}_\#$ can thus be chosen so that  $\varepsilon\vert_{M}$ is below a user-defined error tolerance as $M$ increases. 

The remaining missing element in our description of the model are the control capabilities $\mathscr{C}$. This is where the frame construction shows its flexibility in dealing with various types of control. For illustration, we will consider two scenarios: 

\smallskip

$\bullet$ First, we address the important limiting case in which control is enacted via perfect, instantaneous pulses. In this case, $\beta^{(m)}(t) \lspace{=} 0$ and $h(t_j,t) = \delta(t-j \tau)$, leading to piecewise-constant ``switching functions'' $\{y_{u,v}^{(a)}(t)\}$. Given this,  $\mathscr{F}^{(a)}_\#$ reduces to a collection of window functions $\{ W_{j,\tau}\}_{j=1}^M$, for which the FSF holds exactly, with $N_\#=M$. Thus, in this case, any \emph{digital basis} suffices as a finite $\mathscr{F}^{(a)}_\#$, and an especially compelling choice is provided by the Walsh functions, thanks to their potential for minimizing sequencing complexity \cite{walsh1923closed,WalshLV,WalshCooper}. 

\smallskip

$\bullet$ Second, we consider a windowed Gaussian control profile, $h(t_j, t) = W_{j,\tau} e^{- {(t-t_j)^2}/{ 2\sigma^2}}$, with $\sigma= 1 \,\mu$s, $\tau=10\,\mu$s, although other possibilities, such as square or Slepian pulses \cite{SlepianQNS}, can be easily accommodated. With $M=1$ and $\mathscr{F}^{(a,m)}_\#$ as above, one finds that ${\varepsilon \vert_{\tilde{N}_\#=2}} = 2.4\cdot 10^{-5}$ i.e., a $~1.1\%$ relative error \cite{Relative}, whereas ${\varepsilon \vert_{\tilde{N}_\#=4}} = 2.8\cdot 10^{-8}$, i.e., a $~0.0014\%$ relative error. That is, a modest-size frame ensures the FSF condition is basically satisfied. Exemplifying the $M$ and $\tilde{N}_\#$ interplay, for $M=100$ pulses, a value of $\tilde{N}_\# =4$, i.e., an ${N}_\# =900$ elements frame, ensures an overall error  $\sim 0.15\%$.   

\smallskip

The above highlights the  aforementioned trade-off between model-reduction and accuracy, highlighting the importance of building a (maximally) parsimonious frame --  our choice here need not be optimal as is only meant as a proof-of-principle. Moreover, we note that the dynamics of the system generally depends on linear combinations of the $\mathcal{I}^{(k)}_{\vec{\alpha},\vec{u},\vec{v}}$ ($k \leq k_{\rm max}$), 
where the number of terms is typically a function of the range of the values the indices $\{u,v,\alpha\}$ can take.
%, i.e., of the dimension of the system. 
Accordingly, one must ensure that the error in each $\mathcal{I}^{(k)}_{\vec{\alpha},\vec{u},\vec{v}}$ and also in the relevant linear combinations is small. Ultimately, this implies that while the form of the frame is the same regardless of the dimension of the system $d$, the number of non-overlapping pulses $M$, or the largest order of perturbation being considered $k_{\rm max}$, the tolerance $\varepsilon$ must be small enough -- and thus $\tilde{N}_\#$ must be large enough -- so that the overall error is below a user-defined tolerance for a given $d$, $M$, and $k_{\rm max}$.

\subsubsection{Case study: Single-qubit reduced dynamics}
\label{sub:reduced}

The final step is to obtain the expectation values of system's observables, $E[O(t)]_{\rho_S \otimes \rho_B}$. To simplify our expressions and the analysis in the illustrative applications, we shall specialize in what follows to the simplest paradigmatic setting of a single-qubit dephasing model. Thus, we consider additive noise only along one direction, $B_u(t)\equiv B_z(t)$, whereas the multiplicative noise will be present, as in Eq.~\eqref{Hactual}, whenever the control is assumed to be imperfect. Noting that in this scenario the $\Sigma_u$ reduce to the single-qubit Pauli operators $\sigma_u$, $u \in \{0,x,y,z\},$ Eq.~\eq{ctrl(a)} simplifies to 
\begin{align}
\label{ctrl(a)single}
&y^{(a)}_{u}(t) \equiv  y^{(a)}_{z,u}(t) =\\
&\sum_{v} \! c^{\small{(j-1)}}_{u,v} \left(  k^{(0)}_{j,v}\cos[ \theta_{j} \psi^{({j})}_t] + k^{(1)}_{j,v} \sin[ \theta_{j} \psi^{({j})}_t] {+k^{(2)}_{j,v}}\right). \notag
\end{align}
Further, we enforce in our model that the multiplicative and additive noise sources are uncorrelated, i.e., $\langle B^{(m)}(t_1) B^{(a)}(t_2) \rangle \lspace = \langle B^{(m)}(t_1)\rangle \langle B^{(a)}(t_2) \rangle \lspace {=}~0$. We do \emph{not}, however, require stationarity nor Gaussianity.

In a suitable weak-coupling regime where, formally, $\text{max}_{\{t_i\}} \langle B^{(\alpha)}(t_1)\cdots B^{(\alpha)}(t_k)\rangle T^2 \ll 1$, Eq.~\eq{expec} becomes 
$$E[O(T)]_{\rho_{S}\otimes \rho_B} \approx \langle \Tr_S[ (I_S \lspace{-} D^{(2)}_O(T)) \rho_S \widetilde{O}(T)]\rangle,$$ where the second-order Dyson term $D^{(2)}_O(T)$ can be written as a functional of a reduced set of CA-spectra. Specifically, the components of the spectra relevant to $\mathscr{C}$ are found to be (see Appendix \ref{oneQ} for full detail)
$$\mathscr{\bar{S}}\vert_\mathscr{C}= \{\bar{S}^{(+)}_\alpha(n,n'), (\bar{S}^{(-)}_\alpha(n,n')-\bar{S}^{(-)}_\alpha(n',n))\},$$ 
for $n \in [1,N_\#]$ and $\alpha \in \{a,m\}$, where $\bar{S}^{(\pm)}_\alpha (n,n')$ are associated to the ``classical'' ($+$) and ``quantum'' ($-$) two-point bath correlation functions \cite{multiqubit}, 
$$C^{(\alpha)}_\pm(t_1,t_2) \lspace \equiv \langle [B^{({\alpha})}(t_1),B^{({\alpha})}(t_2)]_{\pm}\rangle,$$
with $[A,A']_\pm\equiv AA'\pm A'A$. Since only FFs $\{{F}^{(1)}_{\alpha;u}(n,T)\}$ allowed by $\mathscr{C}$ can be generated, only the above noise parameters can be inferred from the reduced dynamics via CA-QNS. Still, such finite information suffices for the prediction - and eventual optimization - of the qubit dynamics at time $T$ under \emph{any} of the (infinite) control sequences allowed by $\mathscr{C}$.

\subsection{QNS beyond frequency domain} 
\label{sub:QNS}

Regardless of whether multi-pulse or continuous control modulation is employed (such as, respectively, in comb-based \cite{Cyw} or Slepian- \cite{SlepianQNS} and spin-locking-based \cite{yan2013rotating,Baugh2018Continuous} protocols), existing QNS methods largely rely on the possibility to describe the noise properties in the frequency domain.  This prevents applicability to non-stationary noise \cite{Chakrabarti,ProctorDrift} as well as noise with singular correlation functions \cite{Uhrig}, which must be described in the time domain.

To illustrate how such limitations are overcome in our frame-based approach, in this section we implement CA-QNS via instantaneous perfect pulses applied at separate uniform intervals over a total time $T$. We thus specialize the control Hamiltonian in Eqs.~\eq{eq:MultiQubitCtrl}-\eq{window} to 
\begin{equation*}
\label{eq:SingleQubitCtrl}
H_\mathrm{ctrl}(t) = \sum_{j=1}^{M}  \delta(t-j\tau)\theta_j \vec{n}^{(j)} \cdot \vec{\sigma}/2,
\end{equation*}
where $M$ is now the number of intervals and $\tau=T/M$ is the minimum separation time between pulses (that is, each pulse is applied at a non-zero multiple of the minimum ``switching-time'' $\tau>0$ \cite{Limits}). A direct specialization of Eq.~\eq{ctrl(a)single} reveals that the control matrix elements (i) are necessarily linear combinations of $\{ W_{t_j,\tau} \cos [\theta_j\psi^{({j})}_t],W_{t_j,\tau} \sin [\theta_j\psi^{({j})}_t]\}$; and (ii) satisfy the constraints $y^{(a)}_{u}(t) {\in} [-1,1]$; and $\sum_{u} |y^{(a)}_{u}(t)|^2 {=}1$. As discussed in Sec.~\ref{sec:CAframeconstruction}, a suitable (self-dual) frame in this case is the Walsh basis, which obeys the FSF condition exactly ($\varepsilon=0$). 
In our CA-QNS protocol, we chose $\theta_{j} \in \{ 0,\pi/2,\pi\}$ and $\vec{n}^{(j)} = (0,1,0)$, for all $j$. In contrast with previous Walsh-based characterization methods~\cite{WalshCooper}, the use of non-$\pi$ pulses now makes it possible to generate control matrix elements $\{y_u^{(a)}(t)\}$ that are linear combinations of Walsh functions with $M$ switches over the time range $[0,T]$. This leads to the ability to infer the Walsh-basis CA-spectra $\bar{S}^{(\pm)}_\alpha(n,n')$ for all $n,n'$, not only generalizing the approach of \cite{WalshCooper} beyond reconstruction of signals with a finite number of frequency components, but allowing  reconstruction of non-stationary noise. 

In particular, we apply the CA-QNS protocol to two distinct settings: (i) a classical non-stationary Wiener process; and (ii) a genuinely quantum (bosonic) non-stationary environment. Our task will be to show that the protocol provides sufficient information about the correlation functions, which in turns allows one to infer the parameters describing the noise model under consideration. Full detail about the sequences we used is included in Appendix~\ref{App:CA-QNS}.

\begin{figure}[t!]
 \begin{tabular}{@{}p{0.9\columnwidth}@{}p{0.08\columnwidth}}
 \subfigimg[width=0.9 \columnwidth]{(a)}{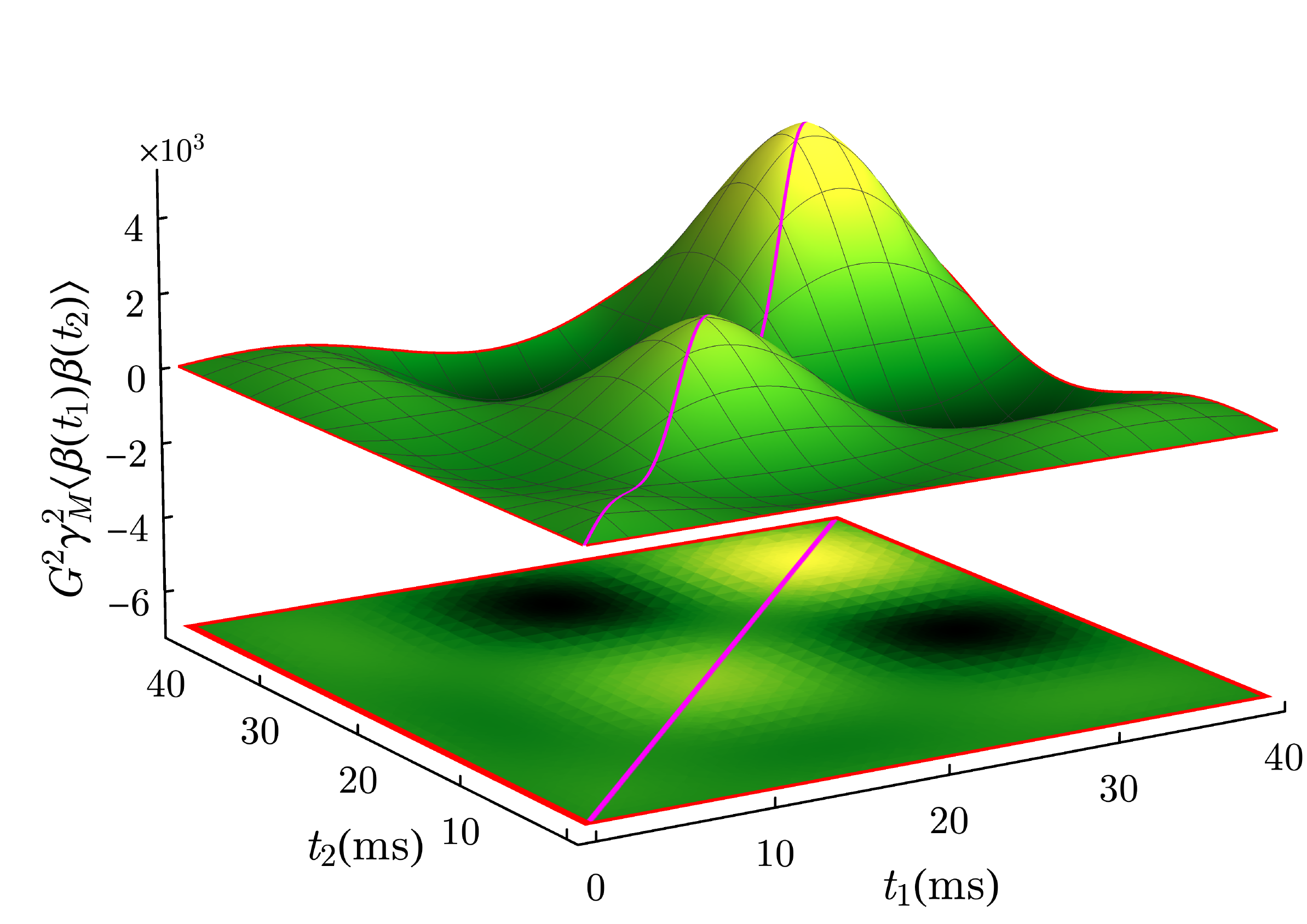} &\subfigimg[width= 0.08\columnwidth]{}{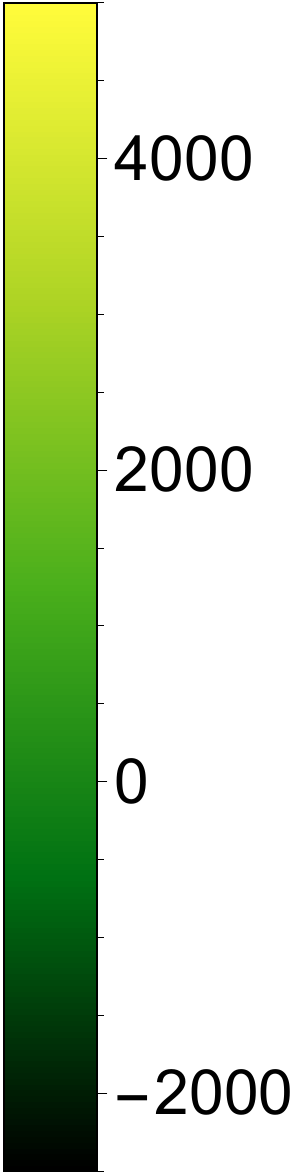} \\
   \subfigimg[width=0.9\columnwidth]{(b)}{diffusion.pdf} &\subfigimg[width= 0.08\columnwidth]{}{bar.pdf}
\end{tabular}
\vspace{-1em}
\caption{(Color online) (a) Theoretical vs. (b) digitally-reconstructed version of the non-stationary correlation function $\langle \beta(t_1) \beta(t_2) \rangle$ in unit of $\rm{rad}^2\rm{s}^{-2}$, for a time resolution of $T/M$, $T=  40~ \rm{ms}$ and $M=16$.  The solid magenta curve in (a) highlights where $\langle \beta(t_1) \beta(t_2) \rangle$ is not
differentiable.}
\label{fig:diffusion}
\end{figure}

\begin{figure*}[th]
 \begin{tabular}{@{}p{0.485\linewidth}@{}p{0.485\linewidth}@{}}
   \subfigimg[width=\linewidth]{(a)}{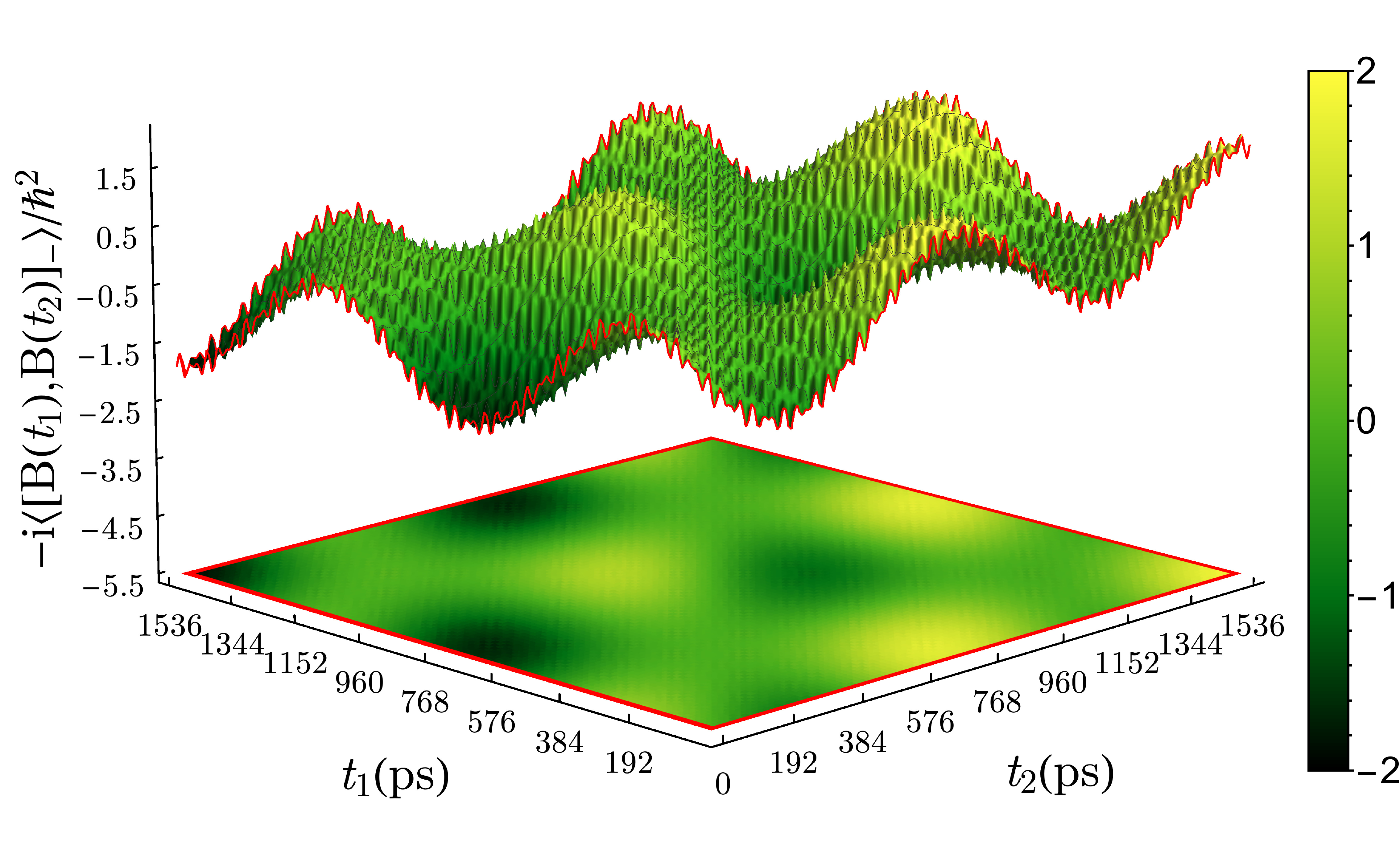} &
   \subfigimg[width=\linewidth]{(c)}{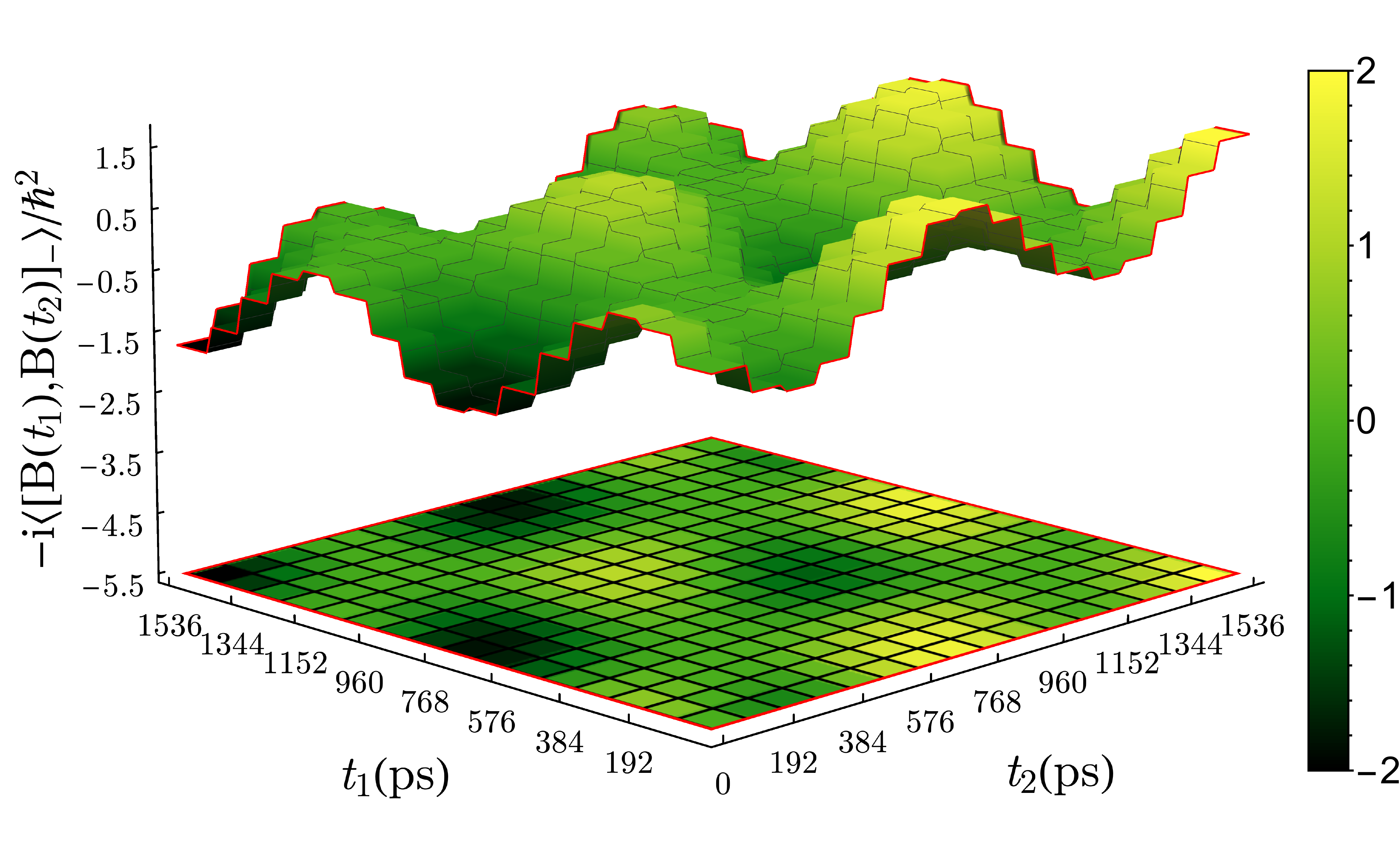} \\
   \subfigimg[width=\linewidth]{(b)}{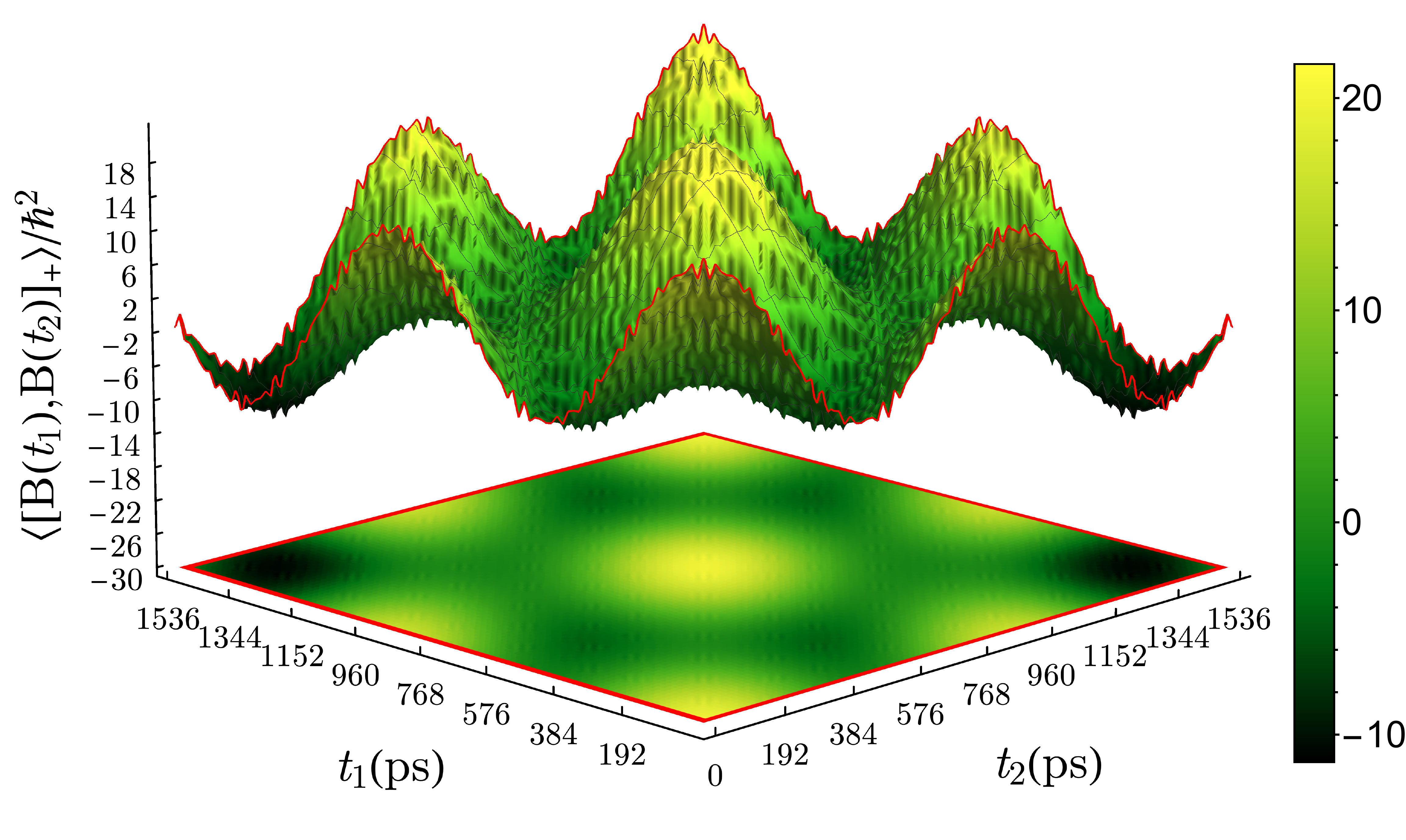}&
   \subfigimg[width=\linewidth]{(d)}{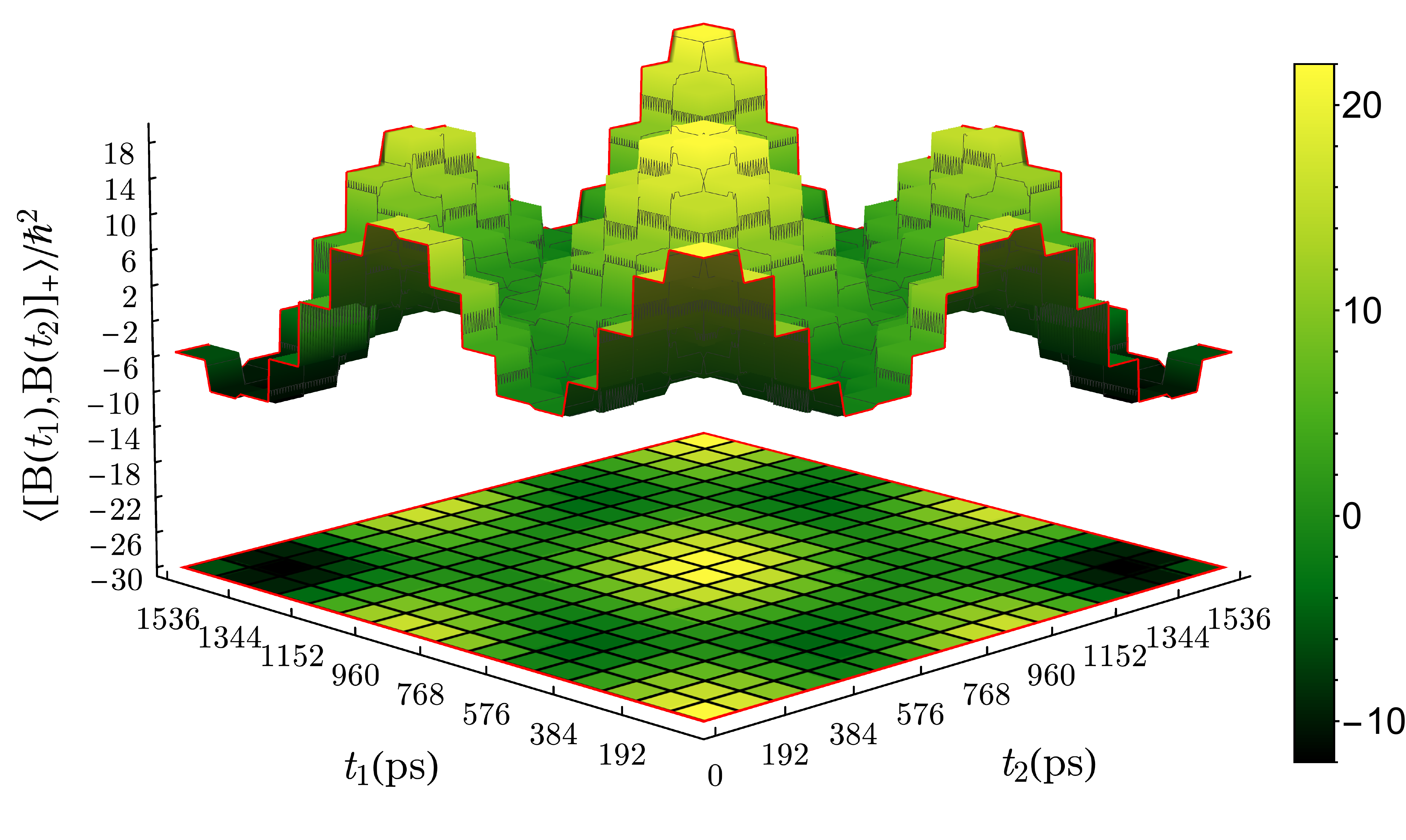}
  \end{tabular}
\vspace{-1em}
\caption{(Color online) (a) Actual anti-symmetric (quantum), (b) symmetric (classical) components of the correlation function, and their corresponding digital reconstructions ((c) and (d), respectively) in units of $10^{18}\rm{Hz}^2$, for $T= 1536 ~\rm{ps}$ and $M = 16$. In our numerical simulations we choose: $\rm{w}_1=125\pi/96~\rm{GHz}$, $\rm{w}_2=7.5\rm{w}_1$, $\Omega_{2}=60\Omega_{1}$, $\Omega_{1}=125\pi/128~\rm{GHz}$, couplings $\bar{g}_1 / \hbar =976.56~\rm{MHz}$, $\bar{g}_2/ \hbar=345.27~\rm{MHz}$, and inverse temperatures are $\hbar \beta_1=61.44~\rm{ps}$, $\hbar\beta_2=2.05~\rm{ps}$ (corresponding to $\rm{T}_1=0.12~{\rm{K}}$ and $\rm{T}_2=3.73~{\rm{K}}$). The reconstruction with the chosen resolution is capable of capturing the slow oscillations, but not the fast oscillations. }
\label{fig:plot}
\end{figure*}

\subsubsection*{Example 1. Non-stationary noise from a classical time-dependent diffusion process}
\label{diffusion}

\begin{table}[h!]
\setlength{\tabcolsep}{5mm}{\begin{tabular}{ |c|c|c| }
 \hline
  & $D$ $(\mathrm{m}^2\mathrm{s}^{-1})$ & $\nu $ $\mathrm{(Hz)}$ \\
 \hline
 Actual & $2.3\cdot 10^{-9}$ & $60\pi \approx 188.496$  \\
 \hline
 Estimated &  $2.3\cdot 10^{-9}$ & $188.496$ \\
 \hline
\end{tabular}}
\caption{Actual and estimated parameters for the classical
oscillatory-diffusive model of Example 1.}
\label{Wiener}
\end{table}

Motivated by the physical setting described in \cite{Chakrabarti}, we consider noise induced by a random walk of molecules in solution, resulting in translational diffusion in the presence of an external magnetic field -- a process ubiquitously encountered in liquid-state NMR and beyond. The relevant dephasing Hamiltonian may be written as 
$$H(t)  = \sigma_z \cdot \hbar \gamma_M  {\rm{G}}  \beta(t) \equiv \sigma_z B(t),$$
where $\gamma_M=2.67 \cdot 10^8 \textrm{ rad } \textrm{s}^{-1} \textrm{T}^{-1}$ is the gyromagnetic ratio of protons, $\rm{G}=0.0214 \, \textrm{T }\textrm{m}^{-1}$ is a constant magnetic gradient along $z$, and $\beta(t)$ is a Gaussian stochastic process representing the Brownian excursions of the molecule. That is, 
$$\langle\beta(t)\rangle=0, \quad C(t_1, t_2)\equiv \langle \beta(t_1)\beta(t_2)\rangle = D \min(t_1,t_2),$$ 
with $D$ being the molecular diffusion constant. While the non-stationary nature of the process is evident -- $C(t_1,t_2)$ is \emph{not} invariant under an arbitrary shift in time -- we further assume here that the diffusion constant varies periodically in time according to $D \mapsto D \cos(\nu t_1) \cos(\nu t_2)$, where $\nu$ is an unknown angular frequency (see Fig.~\ref{fig:diffusion}a). This renders the modified process $\beta(t)$ \emph{second-order cyclostationary} \cite{Napolitano}, with 
$$C\left(t_1 +  {2\pi\,n}/{\nu}, t_2 + \, {2\pi\,n}/{\nu}\right) = C(t_1,t_2), \; \forall t_1,t_2, n \in {\mathbb Z}. $$
We further assume that, for each of the control experiments required by a QNS protocol, the system is re-initialized, i.e., the diffusion process is also effectively reset.

Applying the CA-QNS protocol, we estimate the CA-spectra, $\bar{S}^{(+)}(n,n')$ (notice that in this case  $\bar{S}^{(-)}(n,n')=0$) and, from there, the digitized correlation function $\langle [B(t_1), B(t_2)]_+ \rangle \vert_{\mathscr{C}}$
by using the general relation
\begin{align}
&\notag \langle [B(t_1), B(t_2)]_\pm \rangle \vert_{\mathscr{C}} \\
 &= \sum_{n,n'} \big(\bar{S}^{(\pm)}(n,n')\pm\bar{S}^{(\pm)}(n',n)\big) \tilde{\phi}_n (t_1)\tilde{\phi}_{n'} (t_2). \label{eq:timedomainreco} 
\end{align}
The result of the digital reconstruction is shown in  Fig.~\ref{fig:diffusion}. In addition, by leveraging the knowledge of the physical origin of the noise, we infer the parameters in the model Hamiltonian, $P\equiv \{ D, \nu\}$ (see Tab.~\ref{Wiener}), by following an approach we outline in more detail in Example 2 below. It should be noted that the accuracy in the estimation of the parameters relies on the control capabilities. While in our example the given $T= 40~{\rm ms}$ and $M= 16$ suffice to achieve a good estimation, this need not be the case in general, as we exemplify next.

\subsubsection*{Example 2. Non-stationary noise from a quantum time-dependent \\ 
bosonic environment}

Consider now a scenario where the qubit couples to a two-mode bosonic environment via a periodically varying coupling operator, that is, 
\begin{align*}
	B(t) = \sum_{\ell=1}^2 g_\ell(t) (e^{ i \Omega_\ell t} a^\dagger_\ell + \text{H.c.}), \;\,
g_\ell(t) = \bar{g}_\ell \cos(  {\rm w}_\ell t). \notag
\end{align*}
Again, the time dependence in the couplings makes the noise non-stationary, as $\langle B(t_1)B(t_2)\rangle$ is manifestly a function of both $t_1+t_2$ and $t_1-t_2$, although periodic in the former. Also, we assume that the initial state of the bath is thermal, 
$\rho_B \propto e^{-\sum_{\ell=1}^2 \beta_\ell \hbar\Omega_{\ell}a^{\dagger}_{\ell}a_{\ell}},$
with $\beta_{\ell}\equiv 1/k_B\rm{T}_{\ell}$, so that $\langle B(t) \rangle_c =0$, and the symmetric and anti-symmetric part of the correlation function can be written, respectively, as
\begin{widetext}
\begin{subequations}
\begin{align}
\langle [B(t_1),B(t_2)]_+ \rangle_{c} &= \sum_{\ell} |\bar{g}_{\ell}|^2\Big(\cos{(\textrm{w}_\ell(t_1+t_2))}+\cos{(\textrm{w}_\ell(t_1-t_2))}\Big) \cos(\Omega_{\ell}(t_1-t_2))\coth(\hbar\beta_\ell \Omega_{\ell}/2), 
\label{eq:anticommutequantumbath}\\
\langle[B(t_1),B(t_2)]_-\rangle_{c}&= -i \sum_{\ell}|\bar{g}_{\ell}|^2\Big(\cos{(\textrm{w}_\ell(t_1+t_2)}+\cos{(\textrm{w}_\ell(t_1-t_2))}\Big) \sin(\Omega_{\ell}(t_1-t_2)).
\label{eq:commutequabtumbath}
\end{align}
\end{subequations}
\end{widetext}

\begin{figure}[h]
 \begin{tabular}{c}%@{}p{0.485\linewidth}@{}}
    \subfigimg[width=0.95\columnwidth]{(a)}{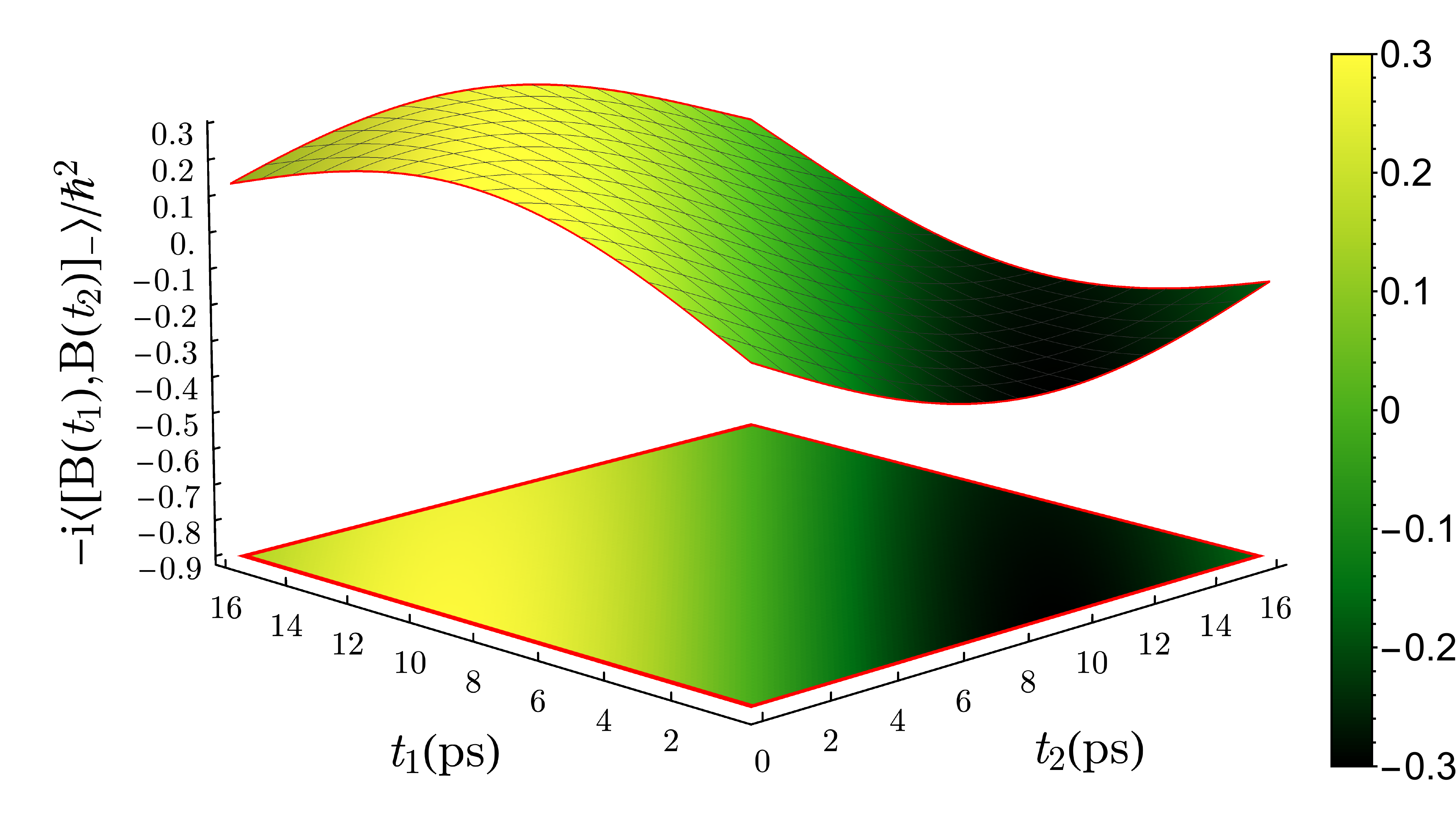}  \\
   \subfigimg[width=0.95\columnwidth]{(b)}{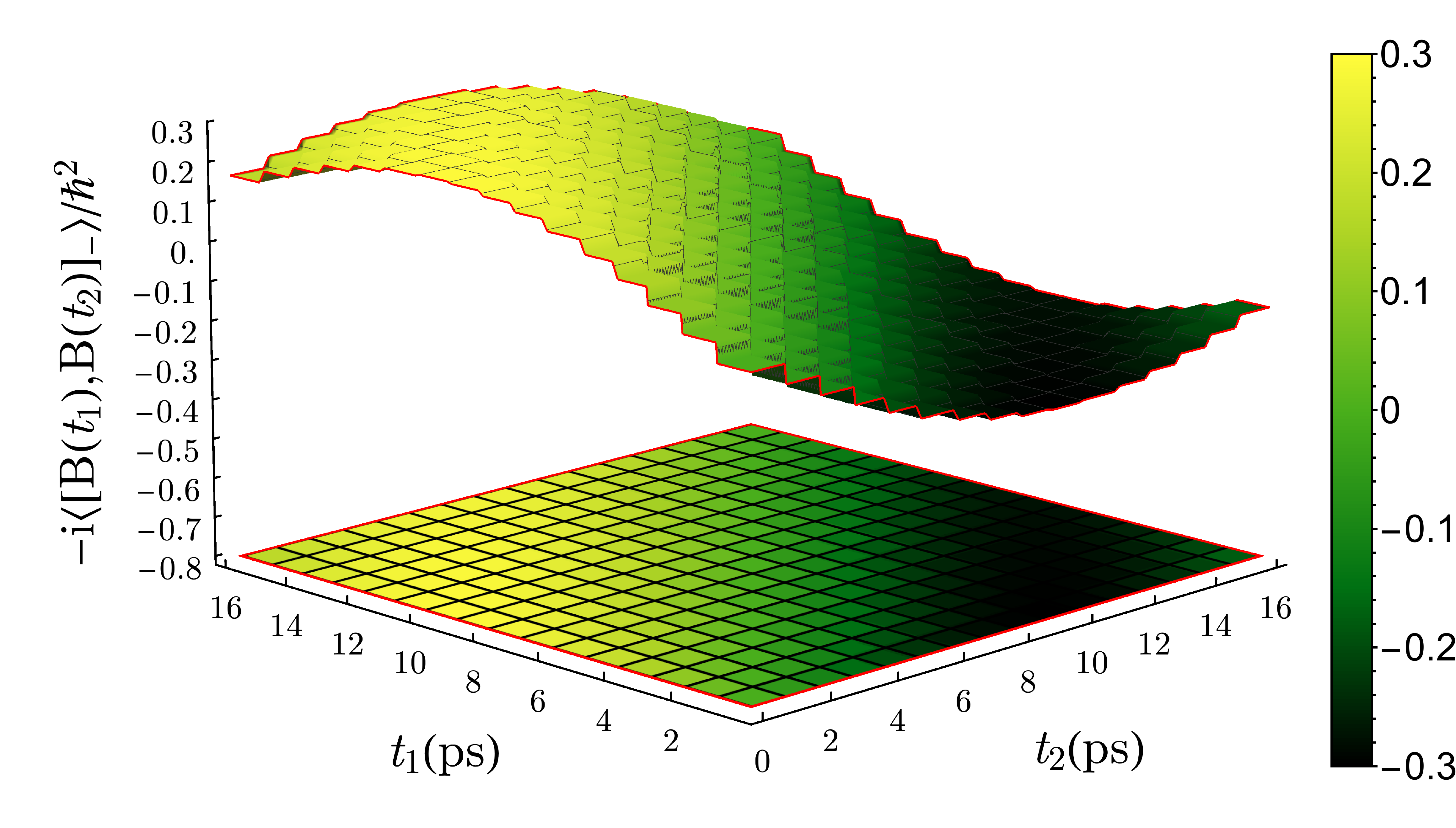} 
   \end{tabular}
\vspace{-3mm}
\caption{(Color online)
(a) Actual anti-symmetric (quantum), component of the correlation function, and (b) their corresponding digital reconstruction, in units of $(10^9 \rm{GHz})^2$ for $T= 16 ~\rm{ps}$ and $M = 16.$ While the resolution is $\tau = 1 ~\rm{ps} = 
\frac{16 ~\rm{ps}}{16} = \frac{1536 ~\rm{ps}}{1536}$, equivalent to performing a digital reconstruction on a total time of $1536~ \rm{ps}$ with a $1536 \times 1536$ grid, notice that in the reconstruction with $T =16 ~\rm{ps}$, the slow oscillations cannot be appreciated.}
\label{fig:zoom}
\end{figure}

Applying the CA-QNS protocol described earlier, one can infer $\bar{S}^{(+)}(n,n')$, $\bar{S}^{(-)}(n,n')-\bar{S}^{(-)}(n',n)$ and, from there, obtain a digital reconstruction (by using Eq.~(\ref{eq:timedomainreco})) of both the classical and quantum components of the correlation function. While this information is also crucial for control, in this section we focus only on the open-system characterization aspect of our problem, i.e., leveraging the information CA-QNS provides and knowledge of the noise model to estimate the relevant parameters.
We execute the protocol for two resolutions $\tau = T/M$, namely, for $T = 1536~\rm{ps}$, $M=16$ and $T = 16~\rm{ps}$, $M=16$. The resulting reconstructions are presented in Fig.~\ref{fig:plot} and Fig.~\ref{fig:zoom}, respectively, which reveal the impact of the time resolution. As Fig.~\ref{fig:plot} demonstrates, the coarse resolution reconstruction does not detect the effect of the fast oscillations. Equipped only with this information, it is not possible to infer the value of comparatively large frequencies with high accuracy.
In contrast, the high-resolution reconstruction -- consistent with a minimum inter-pulse timing of $1 ~\rm{ps}$ (see Fig.~\ref{fig:zoom}) --  can detect the fast oscillations in our model, and allows us to accurately estimate all the model parameters.
\begin{table*}[ht]
\centering
\setlength\tabcolsep{6pt}
\begin{tabular}{ |c|c|c|c|c|c|c|c|c| }
 \hline
  & $\bar{g}_1/\hbar~\mathrm{(MHz)}$ & $\bar{g}_2/\hbar~\mathrm{(MHz)}$& $\Omega_1 \mathrm{(GHz)}$ & $\Omega_2 \mathrm{(GHz)}$ & $\textrm{w}_1 \mathrm{(GHz)}$ & $\textrm{w}_2 \mathrm{(GHz)}$ & $\frac{\hbar}{k_B T_1}$ $(\mathrm{ps})$ & $\frac{\hbar}{k_B T_2}$ $(\mathrm{ps})$ \\
 \hline
 Actual & 976.56 & 345.27  & 3.07 & 184.08 & 4.09 & 30.68 & 61.44 & 2.05 \\
 \hline
 Coarse & 976.55 & 317.80 & 3.07 & 179.63 & 4.09 & 25.93 & 61.44 & 2.03\\
 \hline
 Coarse and Fine & 976.56 & 345.27  & 3.07 & 184.08 & 4.09 & 30.68 & 61.44 & 2.05 \\
 \hline
\end{tabular}
\caption{Actual and estimated physical parameters in two-mode bosonic model described by Eq.~\eqref{eq:anticommutequantumbath} and Eq.~\eqref{eq:commutequabtumbath}. The second row shows the actual parameters of the model. The third and the forth row show the estimated parameter using CA-spectra from coarse reconstruction only and both coarse and fine reconstruction, respectively.}
\label{table:tableboson}
\end{table*}
By using both the low and high resolution, we infer the physical parameters as follows. The parameters of interest are the set 
$$P \equiv \{{\rm w}_1,{\rm w}_2,\Omega_1,\Omega_2,\bar{g}_1,\bar{g}_2,\rm{T}_1,\rm{T}_2\}.$$ 
Assuming knowledge of the model, we estimate them by minimizing a cost function
\begin{equation}
    \mathcal{C}_\$ (P) \equiv \sum _{\mu=\pm}\sum_{n,n'=1}^{N_{\#}} \Big({S}^{(\mu)}(n,n')\vert_P-\hat{{S}}^{(\mu)}(n,n')\Big)^2,
\end{equation}
where the ${S}^{(\pm)}(n,n')\vert_P$ is calculated from the assumed model for a given set of parameters $P$, and 
$\hat{{S}}^{(\pm)}(n,n')$ is estimated as $\hat{{S}}^{(\pm)}(n,n')={\bar{S}}^{(\pm)}(n,n') \pm {\bar{S}}^{(\pm)}(n',n)$, with the input spectra calculated from the $\overline{\mathscr{S}}\vert_\mathscr{C}$.

We perform the optimization, $ \argmin_P \mathcal{C}_\$(P),$ in two settings: (i) with only the low (or coarse) resolution  $\overline{\mathscr{S}}\vert_\mathscr{C}$; and (ii) combining both the low- and high-resolution information. As expected, the optimization in the first approach only accurately estimates the parameters corresponding to the slow frequencies, but not the high frequency generating the fast oscillations in the correlation functions. In contrast, in the second more powerful approach, we estimate all the parameters of interest with high accuracy (assuming no other source of error but the digitization of the reconstruction induced by the available control). We summarize our estimation results in Tab.~\ref{table:tableboson}. We highlight that the example above shows that it is possible to perform ``local bath thermometry'' using a single qubit probe in a ``short-time'' regime, in contrast with existing approaches for stationary noise, which require either a steady-state regime~\cite{quintana2017} or multiple probes~\cite{multiqubit}.

\subsection{Control-adapted noise-tailored optimized gate design} 
\label{CAG}

Beyond the task of bath characterization, and perhaps more relevant to the implementation of quantum technologies, one can leverage the information QNS provides to achieve high-fidelity operations, by tailoring the control to the noise affecting the qubit, via numerical optimal control algorithms~\cite{OCReview2,OCNew} or geometric techniques \cite{Barnes}. While the details are method-dependent, and the underlying non-Markovian dynamics may be modeled in different forms (i.e., via a master equation or a Dyson expansion), a common feature is that -- {in the absence of extra assumptions on the bath} -- information about the noise correlation functions $\{\langle B_{v_1}^{(\alpha_1)}(t_{\mu(1)}) \cdots B_{v_k}^{(\alpha_k)}(t_{\mu(k)}) \rangle\}$ is needed as an input.

The issue, however, is that given limited control capabilities $\mathscr{C}$, one cannot characterize such correlation functions {\it in full}, but rather only the portion of them which is relevant to $\mathscr{C}$. While the intuition rings true, namely, one can only infer what the control allows one to see, this is also evident from our previous developments. Take for example the instantaneous pulse case, for which $\varepsilon =0$, hence there is no loss of information due to model reduction. Given a minimum-switching time constraint, we showed that one can only recover a digitized version of the true correlation functions, $\langle[B(t_1),B(t_2)]_{\pm}\vert_\mathscr{\bar{\mathscr{F}}}\rangle$. The key point, however, is that while $[B(t_1),B(t_2)]_\pm\vert_{\bar{\mathscr{F}}}$ is only a ``control-reduced'' version of $[B(t_1),B(t_2)]$, it is exactly what is needed to predict, and eventually optimize, the dynamics of any pulse sequence resulting from $\mathscr{C}.$ 

To demonstrate this, we consider the task of executing a target quantum gate $G$ with the highest possible fidelity. Importantly, we will assume no {\em a priori} knowledge of the noise correlation functions and, in contrast to the previous subsection, we will consider the realistic setting of {noisy bounded-strength (non-instantaneous) control}. Specifically, we will restrict our control capabilities $\mathscr{C}$ to the scenario where $h_u (t)$ in each of the $M=2$ pulses has a Gaussian shape, and a total execution time ${T= 10 \mu s}$. In the non-Markovian setting, achieving a high-quality operation implies minimizing an appropriate cost function, which is a functional of the overlap integrals $\mathcal{ I}^{(k)}_{\vec{\alpha};\vec{u},\vec{v}} (T)$ and whose explicit form depends on the perturbative expansion of choice. While many choices are possible, we define our as follows. Noting that a single-qubit gate $G$ can be specified by the expectations 
$$E[\sigma_u]_{\sigma_v}= \Tr_S[ G \sigma_v G^\dagger \sigma_u ] \equiv E_{u,v; {\rm G}},\quad  {u,v\in\{0,x,y,z\}},$$ 
we define the cost function for executing $G$ over time $T$ as 
\begin{equation}
\mathscr{E}^G_\$ (P;T) \lspace{\equiv}  \sum_{u,v} |E_{u,v; {G}}-e_{u,v} (P;T)|^2,
\label{costG}
\end{equation}
where $e_{u,v}(P;T)$ is a fixed-order (here, $k=2$) perturbative expansion of $E[\sigma_u(T)]_{\sigma_{v}\otimes \rho_B}$ corresponding to a control parameter set $P=\{\theta_i,\vec{n}^{(i)}\}$, calculated using full knowledge of $\langle [B(t_1), B(t_2)]_+\rangle$ (as would be the case in numerical optimal control routines). In contrast, when we specialize our equations to the model-reduced representation associated to $\mathscr{F}$ of the integrals $\mathcal{ I}^{(k)}_{\vec{\alpha};\vec{u},\vec{v}} (T)$ (such as e.g., Eq. (\ref{FSFconsequence})), we will write the cost function as 
\begin{equation}
\mathscr{E}^G_\$ (P;T)\vert_{\mathscr{F}} \lspace{\equiv}  \sum_{u,v} |E_{u,v; {G}}-e_{u,v} (P;T)\vert_\mathscr{F}|^2. \label{costPractical}
\end{equation}
In each case the optimization, given $\mathscr{C}$, consists in finding the set $P$ such that the corresponding cost function is minimized. Our objective will be to show that the model-reduced and full-knowledge optimal solutions, namely, 
$$P^\ast\vert _\mathscr{F}\equiv\argmin_{P}\mathscr{E}^G_\$ (P;T)\vert_{\mathscr{F}}, \;\text{vs. }\; P^\ast\equiv \argmin_{P}\mathscr{E}^G_\$ (P;T),$$
yield similar performances, in the sense that $\mathscr{E}^G_\$ (P^\ast\vert_{\mathscr{F}};T) \approx \mathscr{E}^G_\$ (P^\ast;T).$ If this indeed happens, it follows that there is no significant loss of information and effective model reduction has been achieved, and we will show this is the case below.

\begin{figure*}[th]
 \begin{tabular}{@{}p{0.33\linewidth}@{}p{0.33\linewidth}@{}p{0.33\linewidth}
}
\begin{minipage}[t][9cm]{0.94\linewidth}
\centering
\vspace{1mm}
\subfigimg[width=0.98\linewidth]{{\;\;\;\;\;\;\;\;\;\;\;\;\;\;Full knowledge}}{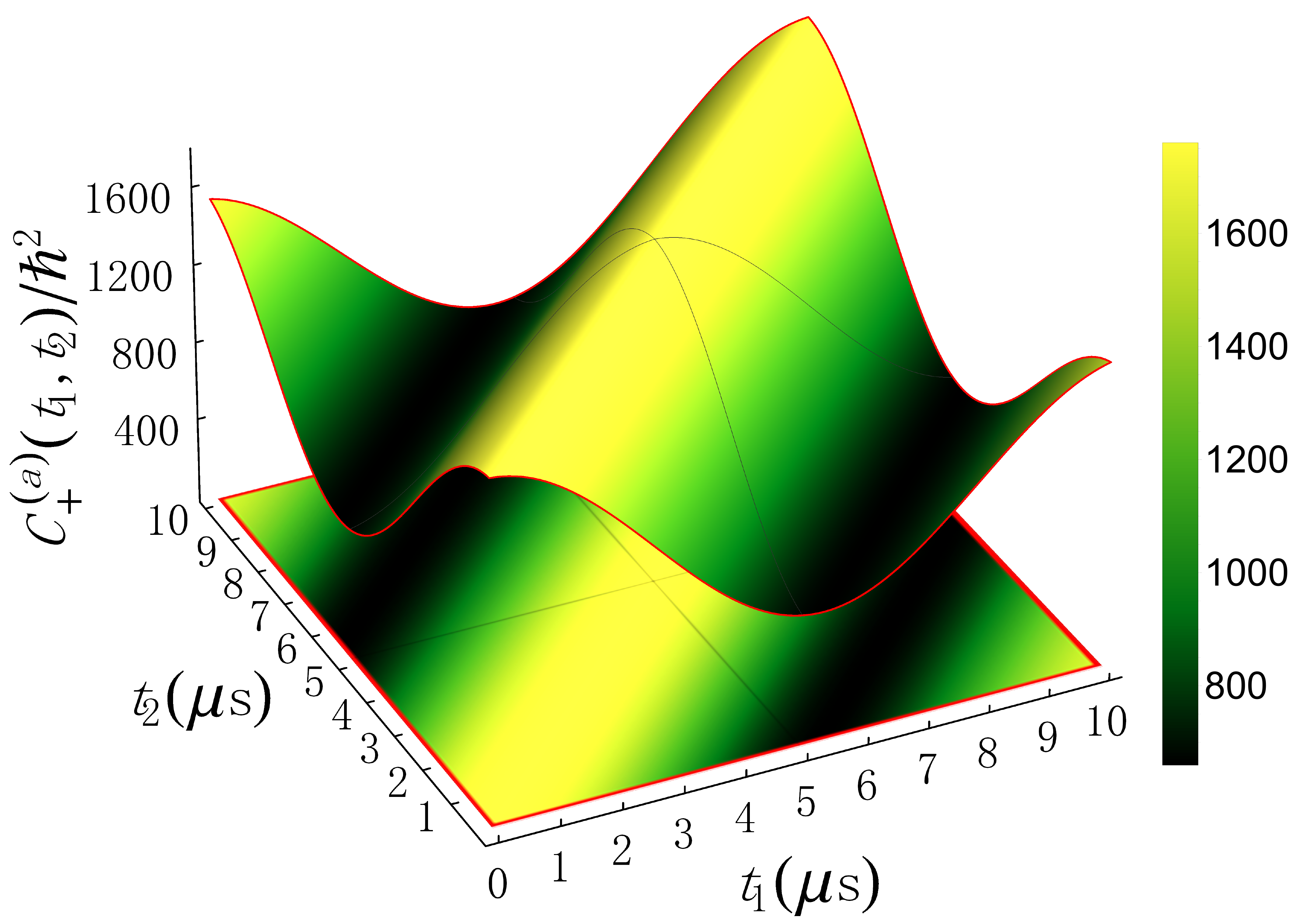}\\
\vspace{1.5mm}
\subfigimg[width=0.98\linewidth]{{\;\;\;\;\;\;\;\;\;\;\;\;\;Error in prediction}}{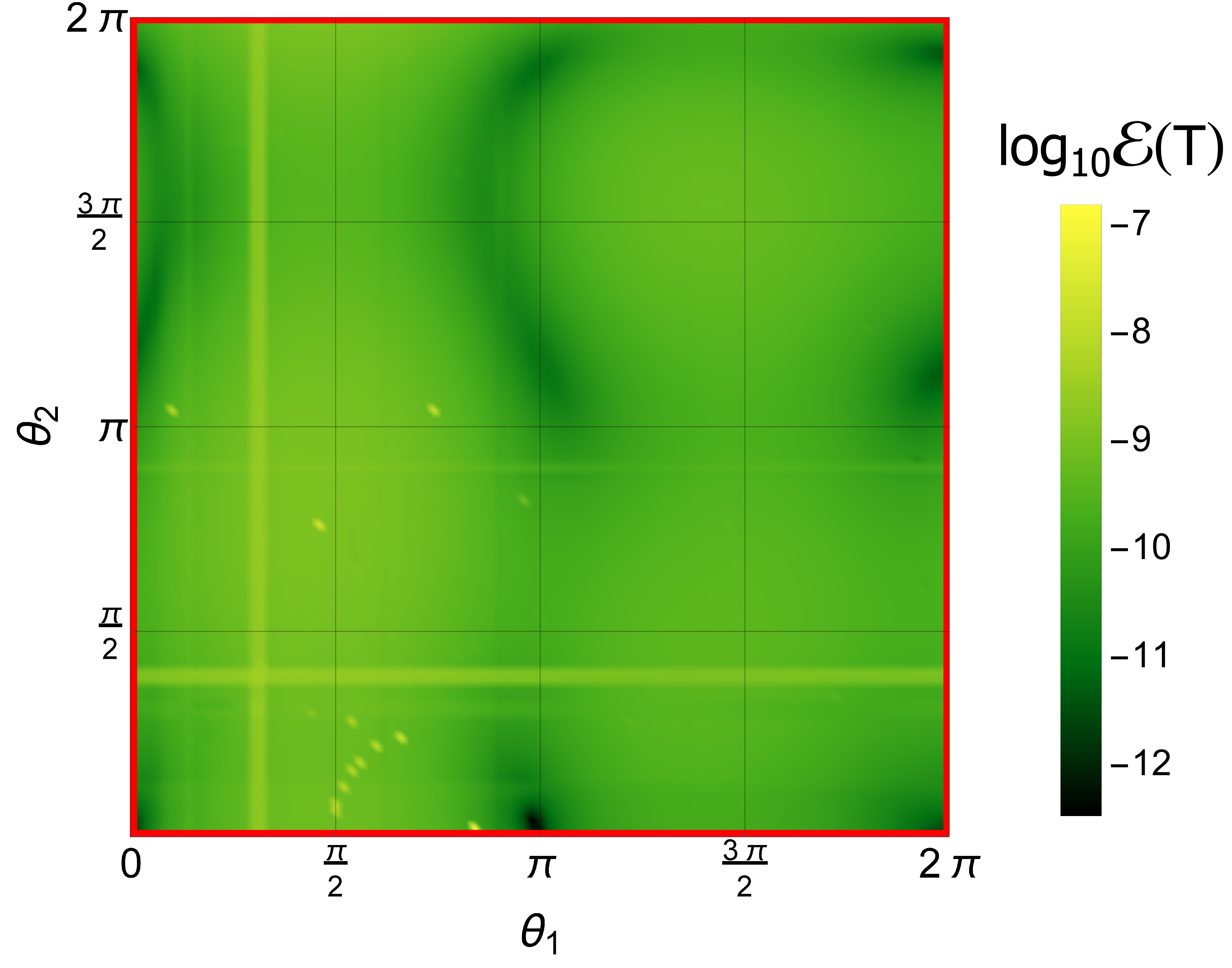}
\end{minipage}
&
{\setlength{\fboxsep}{0pt}\fbox{
\begin{minipage}[t][8.9cm]{0.95\linewidth}
\centering
\vspace{2.75mm}
 \subfigimg[width=0.98\linewidth]{{\;\;\;\;\;\;\;\;\;\;\;\;\;\;\;\;\;\;\;\;\;\;\;\;\;\;\;Ideal}}{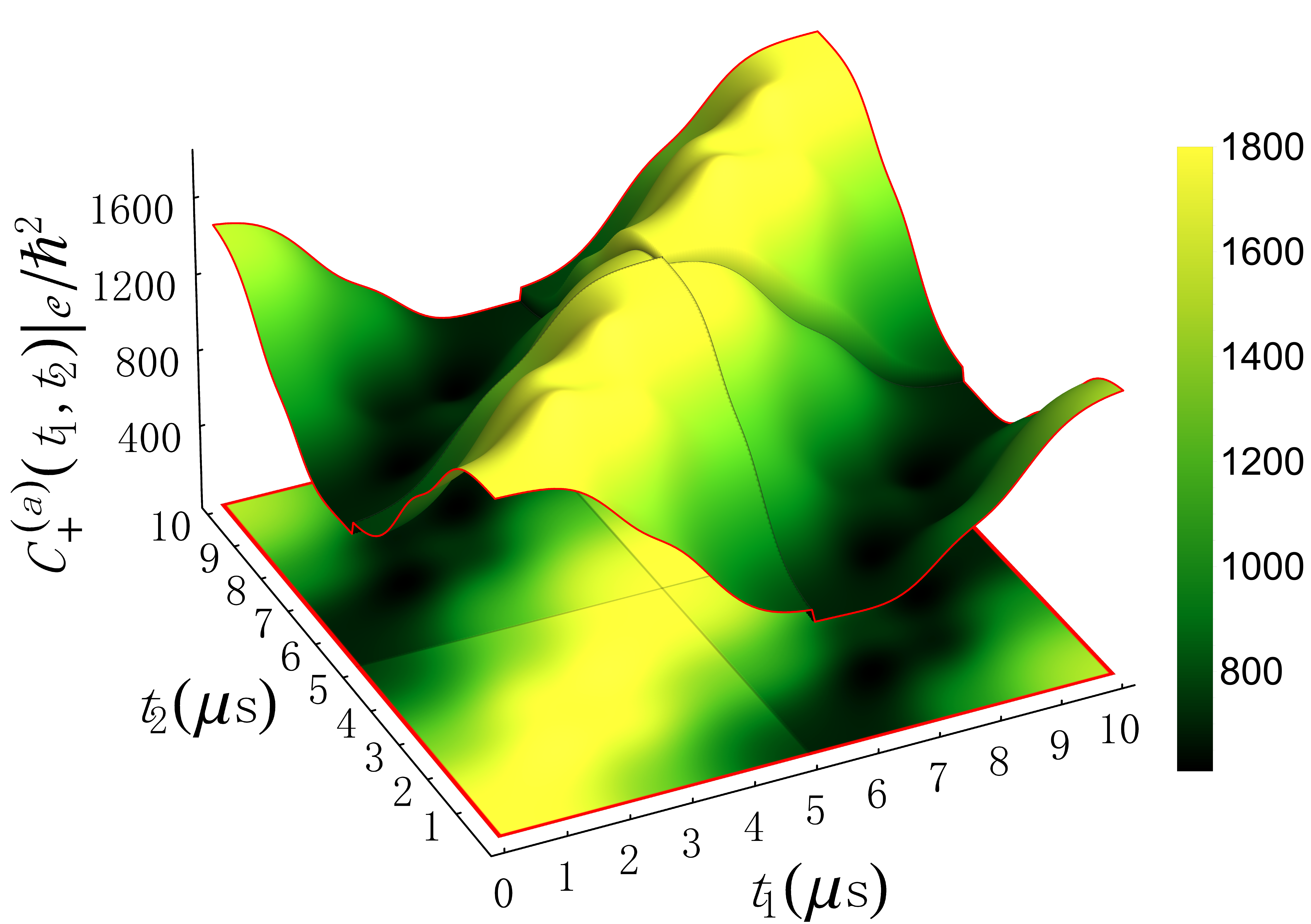}\\
 \vspace{1mm}\subfigimg[width=0.99\linewidth]{{}}{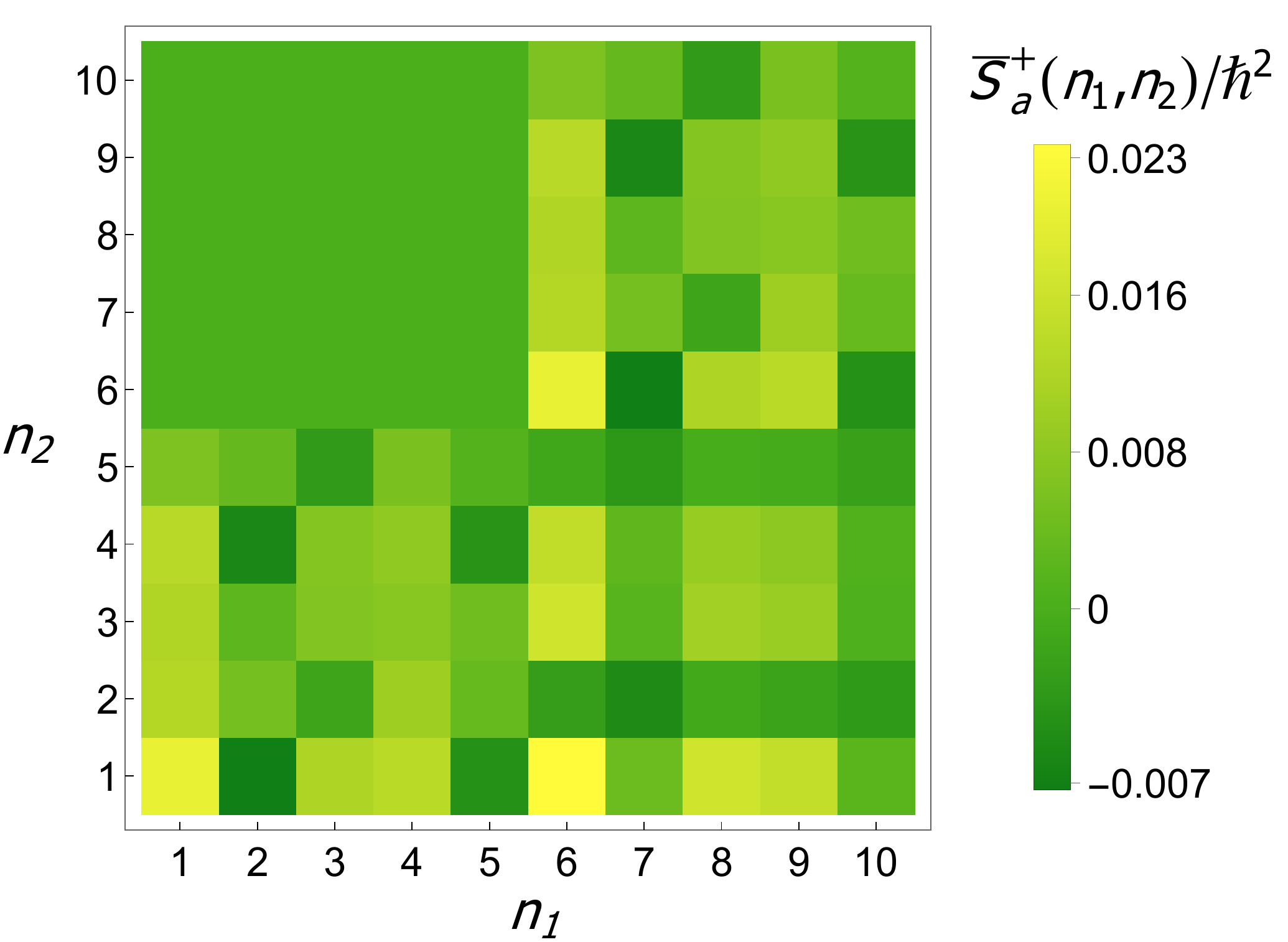}
\end{minipage}
}}&
{\setlength{\fboxsep}{0pt}\fbox{
\begin{minipage}[t][8.9cm]{0.95\linewidth}
\centering
\vspace{3mm}
\subfigimg[width=0.98\linewidth]{{\;\;\;\;\;\;\;\;\;\;\;\;\;\;\;\;\;\;\;\; QNS-inferred }}{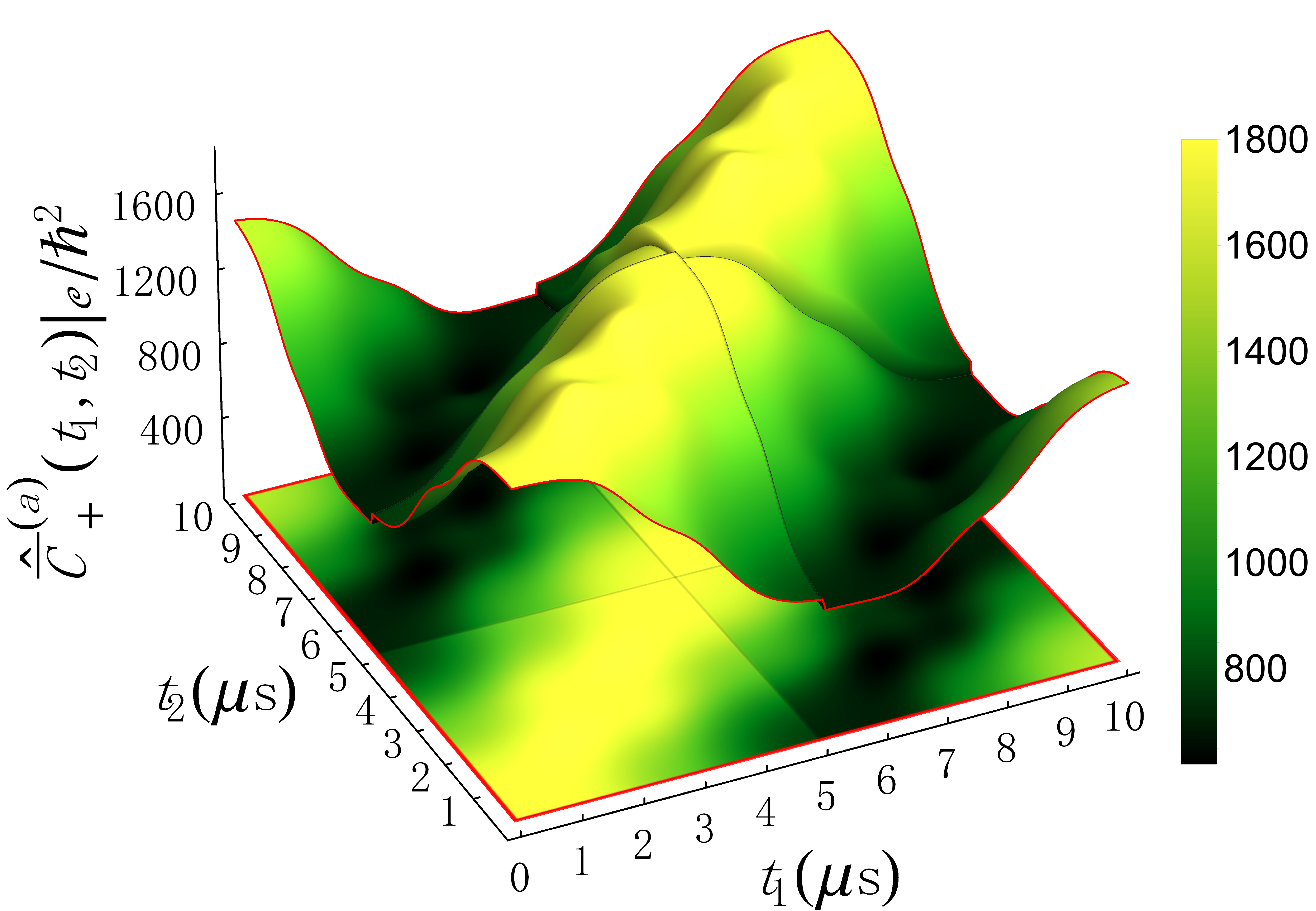} 
    \\
    \vspace{2mm}
    \subfigimg[width=0.99\linewidth]{{}}{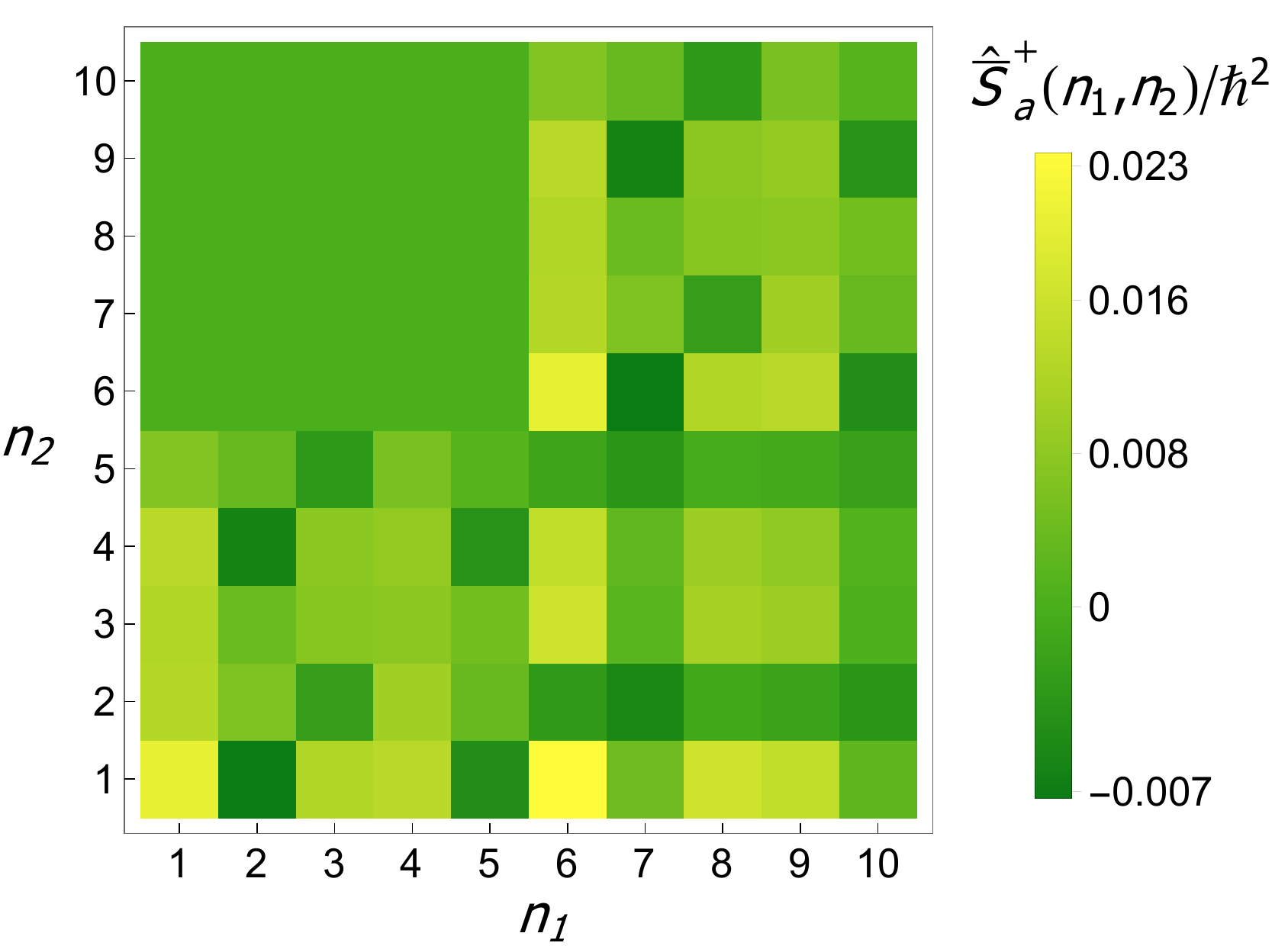}
\end{minipage}
}}
  \end{tabular}
\caption{
(Color online) The model assumes a correlation function (upper left) $C^{(a)}_+(t_1\space{-}t_2) =[B^{(a)}(t_1),B^{(a)}(t_2)]_+$ unknown to the experimenter. Given $\mathscr{C}$ (hence $\mathscr{F}$), one can calculate the ideal frame projection $\bar{S}_a^{+}(n_1,n_2)$ (bottom-middle) and the corresponding time-domain representation $[B^{(a)}(t_1),B^{(a)}(t_2)]_+\vert_{\mathscr{F}}$ (top-middle). A CA-QNS protocol provides an accurate estimate $\hat{\bar{S}}_a^{+}(n_1,n_2)/\hbar^2$ (bottom-right) of $\bar{S}_a^{+}(n_1,n_2)/\hbar^2$, with a maximum absolute error of $5.38  \cdot 10^{-4}$.  For multiplicative noise, our estimates are $\hat{\bar{S}}_{m}^+(1,1)/\hbar^2=1.125 \cdot 10^{-3}\ 
{\rm Hz^2}$ and $\hat{\bar{S}}_{m}^+(2,1)/\hbar^2=2.251 \cdot 10^{-3}\ 
{\rm{Hz^2}}$, with relative errors less than $0.001\%$. The magnitude of these errors depends on a non-zero $\varepsilon$. The distance ${\mathscr{E}(P;T)}$ between the actual dynamics -- calculated using $[B^{(\alpha)}(t_1),B^{(\alpha)}(t_2)]_+$ -- and the predicted dynamics -- using $\hat{\bar{S}}_\alpha^{+}(n_1,n_2)$ -- at the time $T$ is small. We showcase the prediction error by randomly selecting pulse directions (here we use $\vec{n}^{(1)}=(0.658,0.751,-0.052)$ and $\vec{n}^{(2)}=(-0.411,0.000,0.912)$) and sweeping over all pulse angles allowed by $\mathscr{C}$ (bottom-left).  All figures are given in units of ${\rm{kHz}^2}$.
}
\label{fig:bigone}
\end{figure*}

\subsubsection{Noise characterization}\label{section:Characterization}

Since we stipulated that no knowledge of the noise correlation functions is available, we first need to characterize the open quantum system to the best of our ability, i.e., within the limits allowed by $\mathscr{C}$. As in Sec. \ref{sub:reduced}, 
we assume that the qubit is evolving in the presence of uncorrelated additive and multiplicative noise sources (both zero-mean), 
which we take here to be stationary and characterized by correlation functions $C^{(\alpha)}_+(t_1\space{-}t_2)$, $\alpha \lspace{=} a,m$. While unknown to the experimenter, for demonstration we choose the latter to be the inverse Fourier transforms of 
$$S^{(a)}(\omega)= \frac{b^{(a)}_{0}}{ {1 + c^{(a)}_0 \omega^2}} + \frac{b^{(a)}_{1}}{{1 + c^{(a)}_1 (\omega -\omega^{(a)}_{1})^2}}, $$ 
$$S^{(m)}(\omega)=b_0^{(m)}e^{- {(\omega-\omega_0^{(m)})^2}/{ 2 {c_0^{(m)}}^2}},$$
with parameters $b^{(a)}_{0}/\hbar =400\ {\rm kHz}$, $c_0^{(a)}=0.004\ {\rm ms}^2$, $b^{(a)}_1/\hbar= 10^6\ {\rm kHz}$, $c^{(a)}_1=0.64\ {\rm s}^2$, $\omega_1^{(a)}=600\ {\rm kHz}$, $b^{(m)}_{0}/\hbar=2\ {\rm mHz}$, $c^{(m)}_0/\hbar=5\sqrt{2\pi} \ {\rm Hz}$ and $\omega_0^{(m)}=50\ {\rm Hz}$. The total time $T$ is chosen so that the weak-coupling approximation is valid, allowing the cumulant or Dyson expansion to be truncated at order two. {We also assume that we can prepare any Pauli eigenstate at $t=0$ and measure any Pauli observable at times $t=T/2$ and $t=T$.}

Recalling our declared $\mathscr{C}$, i.e., $M=2$ pulses and $\tilde{N}_\#=2$, we start by building the relevant frames. Namely, 
\begin{align*}
\mathscr{F}^{(a)}_\# =\{\phi^{(a)}_{n}\} \equiv
\{&W_{1,T/2},W_{2,T/2},\\
&W_{1,T/2}\cos[\pi \psi_t^{(1)}],W_{1,T/2}\sin[\pi \psi_t^{(1)}],\\
&W_{2,T/2}\cos[\pi \psi_t^{(2)}],W_{2,T/2}\sin[\pi \psi_t^{(2)}],\\
&W_{1,T/2}\cos[2\pi \psi_t^{(1)}],W_{1,T/2}\sin[2\pi \psi_t^{(1)}],\\
&W_{2,T/2}\cos[2\pi \psi_t^{(2)}],W_{2,T/2}\sin[2\pi \psi_t^{(2)}] \},
\end{align*}
(see also Fig.~\ref{frame1} in Appendix~\ref{AppSymmetry}), and 
$$\mathscr{F}^{(m)}_\# =\{\phi^{(m)}_{n}\} \equiv \{h(t_1,t),h(t_2,t)\},$$
where $W_{j,\tau}$ is given in Eq. \eq{window} and $h(t_j,t)$ is a Gaussian profile centered at $t_j$. The canonical dual frames $\tilde{\mathscr{F}}^{(a,m)}_\#$ are built as described in Appendix~\ref{custom}, and are essentially the two-pulse extension of the dual frame shown in Fig.~\ref{fig:dualFrame}(a) therein.

We then perform CA-QNS via control sequences obtained from Eq. \eq{eq:MultiQubitCtrl} in the single-qubit case, with pulse directions and angles given, respectively, by $\vec{n}^{(1)}=\vec{n}^{(2)} \equiv \vec{n}$, 
$$ \vec{n} \!\in\! \Big\{ (0,1,0),\Big(\frac{1}{\sqrt{2}},\frac{1}{\sqrt{2}},0\Big) ,\Big(\frac{1}{\sqrt{2}},0,\frac{1}{\sqrt{2}}\Big),\Big(0,\frac{1}{\sqrt{3}},\frac{\sqrt{2}}{\sqrt{3}}\Big) \Big\},$$  
$$ \theta_1, \theta_2 \in \Big\{0, \pi, \frac{64 \pi}{35}, \frac{17 \pi}{10}, 2\pi \Big\}.$$ 
While in principle any ``sufficiently large'' set of directions and angles works for our purposes, the above set is sufficient to estimate all the necessary CA-spectra components while leading to a well-defined and stable estimation problem. Notice that in principle one would need to infer the $10 \times 10 + 2 \times 2=104$ parameters describing $\overline{\mathscr{S}}|_{\mathscr{C}}$. However, given a chosen frame there are necessarily symmetries in the CA-spectra. To account for them, we systematically classify these symmetries by a kernel analysis method, as detailed in Appendix \ref{AppSymmetry}. In our example, this implies that the total number of parameters to infer reduces to $30 +2 = 32$. Our estimation of these parameters, i.e., of $\overline{\mathscr{S}}|_{\mathscr{C}},$ is summarized in Fig. \ref{fig:bigone} (right panels).

\begin{table*}[th] 
\huge
 \centering
 \resizebox{\textwidth}{!}{
\begin{tabular}{c|cccc|ccc}
{\Huge{Gate}}& \multicolumn{4}{c}{\Huge{Model-reduced, $P^\ast\vert_{\mathscr{F}}=\argmin_P \mathscr{E}^{G}_\$(P;T)\vert_\mathscr{F}$}} &  \multicolumn{3}{c}{\Huge{Full knowledge, $P^\ast=\argmin_P \mathscr{E}^{G}_\$(P;T)$}}\\
\hline 
\multirow{2}{*}{$G$} & \multirow{2}{*}{$(\theta^\ast_1, {\vec{n}^{(1)\ast}})$} & \multirow{2}{*}{$(\theta^\ast_2, {\vec{n}^{(2)\ast}})$} & $\mathscr{E}^{G}_\$(P^\ast\vert_{\mathscr{F}};T)\vert_\mathscr{F}$ \vspace{-4mm}& $\mathscr{E}^{G}_\$(P^\ast\vert_{\mathscr{F}};T)$ & \multirow{2}{*}{$(\theta^\ast_1, {\vec{n}^{(1)\ast}})$} & \multirow{2}{*}{$(\theta^\ast_2, {\vec{n}^{(2)\ast}})$} & $\mathscr{E}^{G}_\$(P^\ast;T)$\\ & & & $(\cdot 10^{-3})$ & $(\cdot 10^{-3})$ & & & $(\cdot 10^{-3})$ \\ \hline
$I$ & $(1.56,\{-1,0,0\})$& $(1.56,\{-1,0,0\})$& $10.8$ & $11.1$& $(1.56,\{1,0,0\})$& $(1.56,\{1,0,0\})$& $11.1$\\
$X$ & $(1.56,\{-1,0,0\})$& $(3.14,\{1,0,0\})$& $3.48$ &$3.51$ &  $(1.56,\{-1,0,0\})$& $(3.14,\{1,0,0 \})$& $3.51$\\
\multirow{2}{*}{$Z$} & $(1.74,$ \vspace{-3mm}& $(1.74,$& \multirow{2}{*}{$6.99$} & \multirow{2}{*}{$7.13$}&  $(1.74,$& $(1.74,$& \multirow{2}{*}{$7.12$}\\
& $\{-0.56,0.80,-0.19\})$& $\{0.80,0.57,-0.19\})$& & & $\{-0.70,0.70,-0.18\})$ & $\{0.69,0.70,-0.19\})$ & \\
$e^{i \pi \sigma_x/8}$
& $(1.36,\{-1,0,0\})$& $(1.36,\{-1,0,0\})$& $9.47$ & $9.71$ &  $(1.37,\{-1,0,0\})$ & $(1.36,\{-1,0,0\})$& $9.71$\\
\multirow{2}{*}{$e^{i\pi\sigma_z/8}$}
& $(2.24,$ \vspace{-4mm}& $(2.24,$& \multirow{2}{*}{$2.22$} &\multirow{2}{*}{$2.25$} & $(2.24,$ & $(2.24,$& \multirow{2}{*}{$2.25$}\\
& $\{-0.98,0.13,-0.16\})$ & $\{0.95,0.26,-0.16\})$ & & & $\{-0.99,0,-0.16\})$ & $\{0.91,0.38,-0.16\})$ & \\
\multirow{2}{*}{$H$} & $(1.35,$ \vspace{-4mm}& $(2.18,$& \multirow{2}{*}{$6.69$} & \multirow{2}{*}{$6.82$}& $(1.35,$ & $(2.18,$& \multirow{2}{*}{$6.82$}\\
& $\{0.82,0.58,0.02\})$ & $\{-0.66,0.68,0.31\})$ & & & $\{0.82,0.57,0.01\})$ & $\{-0.65,0.69,0.31\})$ & 
\end{tabular}}
\caption{
Optimal control parameters $P^\ast =\{ \theta_i^\ast, {\vec{n}^{(i)\ast}}\}$ (all rounded to two decimals) found by minimizing the model-reduced and full knowledge cost functions, respectively.
A fair and experimentally relevant comparison is made by evaluating $\mathscr{E}^G_\$(P;T)$ at the corresponding optimal values $P=P^\ast\vert_{\mathscr{F}}$ (fifth column) and comparing them with the minimal values of the full-knowledge cost function  rightmost column). We find that there is virtually little difference in doing so, indicating that there is no significant loss of information due to model reduction so long as the controls are in $\mathscr{C}.$ Notably, across multiple examples our numerical routine was less likely to be trapped in a local minimum when optimizing $\mathscr{E}^G_\$\vert_{\mathscr{F}}$ as compared to optimizing $\mathscr{E}^G_\$,$ suggesting a potential additional benefit of model reduction. Optimal gate-design results for the same noise model with different noise parameters are also included in Appendix~\ref{appendNoise}; similar to dynamically corrected gates or composite pulses \cite{DCG,DS4}, these results show that doing a faster gate might lead to a larger cost function than the optimal gate that takes longer time. }
\label{table1}
\end{table*}

\subsubsection{Prediction and optimization} 

In order to test the ability of our model-reduced representation to predict the dynamics of the system at time $T$, 
we calculate the distance 
$$ {\mathscr{E}(P;T)} \equiv \sum_{u,v} | e_{u,v}(P;T) -e_{u,v}(P;T)\vert_{\mathscr{F}}|^2,$$
between the model-reduced and full-knowledge predictions for 1000 randomly chosen configurations $P=\{\theta_1,\theta_2, \vec{n}^{(1)},\vec{n}^{(2)}\}$. We find that the average ${\mathscr{E}}(P;T)_{\textrm{avg}}= 1.18\cdot 10^{-9},$ with a worst case ${\mathscr{E}(P;T)}_{\textrm{worst}} =   8.17\cdot 10^{-8}$. Further, choosing a pair of random pulse directions $\{\vec{n}^{(1)},\vec{n}^{(2)}\}$, we evaluate $\mathscr{E} (P;T)$ for $\{\theta_1, \theta_2\}$ in $[0,2\pi]$. The results are shown in Fig.~\ref{fig:bigone} (bottom-left).

Prediction is the precursor to optimization, and we can thus demonstrate the benefits that our model-reduced representation brings to the problem of optimally executing a desired gate $G$ given $\mathscr{C}$. Using standard Nelder-Mead numerical routines, for several representative choices of $G$, we search for the optimal parameters $P^\ast=\{\theta_i^\ast, \vec{n}^{(i)\ast}\}$ that minimize: (i) the cost function $\mathscr{E}^G_\$(P;T)\vert_\mathscr{F}$, by using the information $\overline{\mathscr{S}}|_\mathscr{C}$ inferred in the above characterization stage; or (ii) the cost function $\mathscr{E}^G_\$(P;T)$, by assuming full knowledge of the noise model, with access to the full-model time-domain equations. The scenario (ii) is a drastic idealization as such a knowledge is never available in practice, and no QNS protocol can provide such information unless one assumes arbitrary control capabilities. Nevertheless, it is a useful benchmark, as our objective is to show that our model-reduced representation of the dynamics captures all the relevant information, as dictated by $\mathscr{C}$, to a very good approximation.

The results are presented in Tab. \ref{table1}. They demonstrate the model reduction capabilities of the formalism, as there is little to no predictive power lost. Note that $\mathscr{E}^G_\$ (P^\ast\vert_{\mathscr{F}};T) \sim \mathscr{E}^G_\$ (P^\ast;T)$, as desired. Finally, for completeness and to highlight the benefits of control, we calculate the value of the cost function $\mathscr{E}^G_\$(P_0{~=~}\{\theta_i{~=~}0\};~t)$ for $G=I$ using full information and in the absence of control, i.e., the effect of the natural decoherence of the system. One finds that in the absence of control $\mathscr{E}^I_\$ (P_0;T/2) =9.77\cdot 10^{-3}$  and $\mathscr{E}^I_\$(P_0;T) =9.09\cdot 10^{-2}$, which should be contrasted, for example, with the optimal control solution over time $T$, namely  $\mathscr{E}^I_\$ (P^\ast;T) = 11.1\cdot 10^{-3}.$

\subsection{On the universality of QNS-inferred information \\ for control purposes} 
\label{sub:univ}

Fundamental to the paradigm of C\&C of open quantum systems is the assumption that QNS-inferred information is sufficient to implement high-accuracy operations. This, however, is not guaranteed. QNS protocols infer information about the noise by measuring the response of the system to a \emph{fixed} set of control sequences, say, $\mathscr{C}_0\subseteq \mathscr{C}$, so by design the information they access is only the one these sequences can sense, namely, $\mathscr{S}\vert_{\mathscr{C}_0}$. The question is whether this information is universal, that is, whether it can be used to accurately predict the dynamics of the system under a sequence not in $\mathscr{C}_0$.

\begin{figure}[t]
    \centering
    \includegraphics[width=0.9\columnwidth]{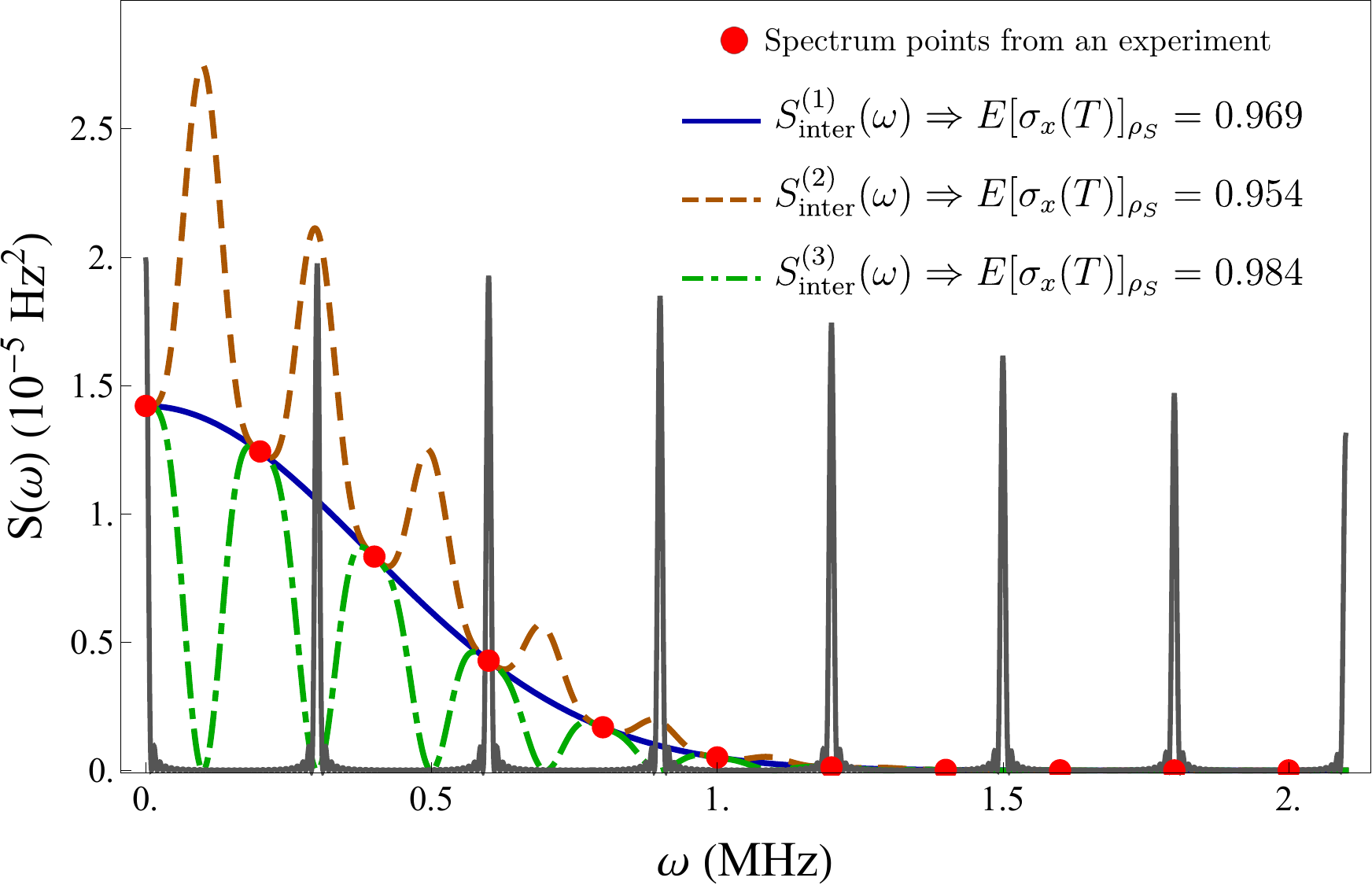}
    \caption{(Color online) Illustrative example of the possible issues from incompatibility between a control sequence of interest and QNS-inferred information. Red dots represent sampled points from comb-based SP-QNS for a representative Gaussian power spectrum, $S(\omega) = \exp{[-\omega^2/(3\cdot 10^{11})]}/7042.$ Shown are also three possible interpolations ${S}_{\rm inter}^{(i)}(\omega)$ consistent with them. Namely: (i) ${S}_{\rm inter}^{(1)}(\omega)=S(\omega)$ (blue line); (ii) ${S}_{\rm inter}^{(2)}(\omega)=S(\omega) \left(1+\sin^4[\frac{\pi}{2}\omega\cdot 10^{-5}]\right)$ (brown dashed line); and (iii) ${S}_{\rm inter}^{(3)}(\omega)=S(\omega) \left(1-\sin^4[\frac{\pi}{2}\omega\cdot 10^{-5}]\right)$ (green dash-dotted line). Given in the legend are the predicted expectation values of $\sigma_x$, obtained by using the initial state $\rho_S = (\sigma_0 + \sigma_x)/{2}$ and the expression $E[\sigma_x(T)]_{\rho_S} = \exp[{- \int d\omega |F^{(1)}(\omega,T)|^2 {S}_{\rm inter}^{(i)}(\omega)}].$  The predicted value can vary significantly depending on the interpolation being considered 
if the control sequence being studied is not ``of the form'' of those used in the QNS protocol. In this case, this means ensuring that $|F^{(1)}(\omega,T)|^2$ (in grey) has peaks out of phase with the interpolation points. An example of such filter~\cite{alvarez2011measuring} can be generated by $M\gg1$ repetitions of a $\pi$-pulse sequence with cycle time $T_c= (20\pi)/{3}\ \mu {\rm s}.$ Shown here is the filter for $M=10$ repetitions of a Hahn-echo sequence given by $[{T_c}/{2} - \sigma_x - {T_c}/{2} - \sigma_x].$ 
}
    \label{fig:interpolation}
\end{figure}

To gain a concrete feeling about this problem, note that standard comb-based QNS protocols~\cite{Cyw}, but also Slepian-based~\cite{SlepianQNS} or spin-locking \cite{yan2013rotating} protocols, sample the leading noise power spectra in frequency domain at a finite set of points. For control purposes, however, one is interested in overlap integrals of the form given in Tab.~\ref{tab:SPvsCA} (panel (iii)), hence the information provided by the sampling is necessarily incomplete. Therefore, it is necessary to complement it with additional assumptions, by interpolating between the sampled points. The assumptions that are more or less implicitly made in this completion step -- e.g., in choosing a particular interpolation method -- can be highly arbitrary and user-defined, and yet they can decisively influence our ability to predict the dynamics accurately. For instance, given a sampling set, there are in principle infinitely many possible interpolations consistent with it, and it is easy to build a control sequence for which the details of the interpolation are crucial: a simple example demonstrating how the latter can directly impact observable expectation values is given in Fig.~\ref{fig:interpolation}. That is, $\mathscr{S}\vert_{\mathscr{C}_0}$ is \emph{not} universal in general. Consequently, obtaining rigorous criteria to characterize the control sequences whose effect on the system can be accurately predicted given such information is not only desirable but also imperative. Of course, this is not a problem exclusive to spectral estimation in either the classical~\cite{Madsen,Byrnes-Georgiou-Lindquist,Minda-Barbinta-Cillich} or quantum settings, and indeed the task of picking a good interpolation given sampled data is a mainstay in applied mathematics, with various possible criteria available \cite{caruso-interpolation-1998}.

The frame-based approach we proposed in Sec.~\ref{section:FrameFFandReduction} and demonstrated in Sec. \ref{sec:framecontrol} provide a way around the above problem. On the one hand, by construction, if the finite frame $\mathscr{F}_\#$ exists, the FSF condition guarantees that for an appropriately chosen $\mathscr{C}_0$, i.e., a well-designed QNS-protocol, we have ${\mathscr{S}}\vert_{\mathscr{C}_0} = \bar{\mathscr{S}}\vert_\mathscr{C}.$ Accordingly, the finite frame provides a space in which the (finite) sampling is precisely what is needed to accurately (up to an error scaling with $\varepsilon$) predict the behavior of the system under an arbitrary control modulation allowed by $\mathscr{C}$. This naturally obviates the need to complement the sampled noise information, for instance, via interpolation. In the SP frequency-domain example shown in Fig.~\ref{fig:interpolation}, this would be akin to demanding that all the sequences we are interested in were such that $\sqrt{\int d\omega |F
(\omega,T) - \sum_{k} F(\omega,T) \delta (\omega- k \omega_0)|^2} \leq \varepsilon,$ with $\omega_0 = 2\pi/T_c$. Clearly, such a condition would severely constraint the control sequences we can accurately predict the behavior of. The frame-based approach can achieve the desired effect while overcoming this limitation. Likewise, given an \emph{arbitrary} control sequence (not necessarily in $\mathscr{C}$), one can estimate the prediction error one would incur in by using QNS-inferred information $\bar{\mathscr{S}}\vert_\mathscr{C}$. Namely, by calculating the distance between the control matrix elements associated to such sequence and their projection on the frame $\mathscr{F}_\#$ associated to the QNS protocol, one essentially obtains an effective $\varepsilon$, and it is then up to the user to determine if such error is acceptable.

In closing, we note that the question of universality is vital when model-reduced representations of complex systems are introduced (not necessarily related to control). For example, recent work \cite{Szankowski} has addressed similar questions but related to the ``objectivity" of classical noise representations of quantum baths. In our setting, if one thinks of the \emph{context} \cite{Szankowski} to be defined by $\mathscr{C}$, then universality of the QNS-inferred information based on $\mathscr{C}_0$ may be taken to signify a degree of objectivity of the noise representation for control purposes. Moving forward, it would be interesting to better understand how different model simplifications interact and can lead to more comprehensive model-reduced representations (relative to the combined contexts, for example).

%\vfill

\section{Conclusion and outlook} 

We have introduced a framework for constructing a model-reduced representation of open quantum dynamics \emph{relative to given control capabilities}, which both mathematically formalizes and substantially simplifies the problem of C\&C for general non-Markovian noise environments. While we have exemplified our analysis in two paradigmatic applications -- QNS of non-stationary noise and model-reduced C\&C of a single qubit -- our results also formally justify the success of the machine-learning enhanced approach for noise discrimination proposed in  \cite{Akram2020}.

Our framework lends itself to several generalizations. On the one hand, a natural and important next step is to develop explicit frame-based protocols applicable to multiqubit C\&C tasks in the presence of more general noise models, including multiaxis and non-Gaussian noise. As we mentioned, of special significance in this context will be to understand how, for fixed control capabilities $\mathscr{C}$, a model-reduced description with maximum parsimony may be obtained without sacrificing accuracy \cite{Obinata}. On the other hand, the use of frame-based optimization need not be restricted to the synthesis of unitary target gates; one could imagine leveraging the natural decoherence of the system in the presence of applied control to optimally implement a reachable completely positive trace-preserving map, possibly in connection with ideas from \cite{QScience}. Ultimately, we believe that the use of frames will ease the integration of signal processing tools into quantum control and prove instrumental to develop efficient model-reduced approaches to C\&C of realistic open quantum systems of growing complexity, as needed by both realistic NISQ-era devices and full-fledged fault-tolerant architectures.

\section*{Acknowledgements}

It is a pleasure to thank Howard M. Wiseman, Michael J.W. Hall, and Areeya Chantasri for valuable comments. Work at Griffith was supported by funding from the Australian Government via the AUSMURI grant AUSMURI000002. G.P.S. is pleased to acknowledge support from the DECRA fellowship DE170100088. Work at Dartmouth was supported in part by the U.S. Army Research Office through the U.S. MURI Grant No. W911NF1810218.

\appendix

\section{Time-dependent expectation values} 
\label{App:OpenDynamics}

As discussed in the main text, in order to capture the system dynamics under the simultaneous effect of the noise and the applied time-dependent, open-loop control we consider the expectation value of a system-only (invertible) observable $O$, given in the physical frame by $E[O(T)]_{\rho_S\otimes\rho_B} = \left \langle \Tr\left[U(T) (\rho_S\otimes\rho_B) U^\dagger(T)O\right] \right \rangle_c$. The latter may be rewritten in the form
\begin{align*}
    E[O(T)]_{\rho_S\otimes\rho_B} &= \left \langle \Tr[\widetilde{U}(T) (\rho_S\otimes\rho_B) \widetilde{U}^\dagger(T)\widetilde{O}(T)] \right \rangle_c \\
& \hspace*{-9mm}
= \Tr_S\left[ \left\langle \widetilde{O}(T)^{-1} \widetilde{U}^\dagger(T) \widetilde{O}(T) \widetilde{U}(T) \right\rangle \rho_S \widetilde{O}(T)\right] \notag, \\
    & \hspace*{-9mm}
\equiv \Tr\left[V_O(T) \rho_s \widetilde{O}(T)\right],
\end{align*}
where $\langle \cdot \rangle_c$ represents averaging over realizations of the stochastic process, $\langle \cdot \rangle = \langle \Tr_B[\cdot ~ \rho_B] \rangle_c$ is the joint classical-quantum average, and $ \widetilde{O}(T) = U_0^\dagger(T) O U_0(T)$. Following a similar line of reasoning as   in \cite{Paz-Silva-MultiQubit}, we write $V_O(T)$ as a time-ordered exponential, 
$V_O(T) = \langle \mathcal{T}_+ e^{-i\int_{-T}^T ds H_O(s)} \rangle$, with
\begin{align}
    H_O(t) =
    \begin{cases}
    \overline{H}(T-t) & t \in [0,T],\\
    \widetilde{H}(T+t) & t \in [-T,0],
    \end{cases} 
    \end{align}
$\overline{H}(t) \equiv -\widetilde{O}(T)^{-1} \widetilde{H}(t) \widetilde{O}(T)$, and $ \widetilde{H}(t) $ being the toggling-frame Hamiltonian given in Eq. \eq{eq:basicHam} of the main text. In turn, this allows us to expand $V_O(T)$ via a cumulant or Dyson expansion,
\begin{align*}
    \left\langle \mathcal{T}_+ e^{-i \int_{-T}^T H_O(t)dt} \right\rangle & =
    e^{\sum_{k=1}^\infty (-i)^k {\mathcal{C}_O^{(k)}(T)}/{k!}}  \\
     & =1 + \sum_{k=1}^\infty \frac{\mathcal{D}_O^{(k)}(T)}{k!},
\end{align*}
where $\mathcal{C}_O^{(k)}$ is  the generalized cumulant defined implicitly as a function of the Dyson-like terms
%\begin{widetext}
\begin{align}
\frac{\mathcal{D}^{(k)}_O(T)}{k!} &= (-i)^k \int_{-T}^T d_>\vec{t}_{[k]} \left\langle H_O(t_1) \cdots H_O(t_k) \right\rangle,\\
 \label{changeofV} 
 &= (-i)^k \sum_{\ell=0}^k \sum_{\pi\in \Pi_{\ell;k}} \int_0^T \!\!\!d_> \vec{t}_{[k]} 
 \mathbf{C}_{\ell;k}, 
 %\left\langle \prod_{j=1}^\ell \overline{H}(t_{\pi(j)}) \prod_{j'=\ell+1}^k \widetilde{H}(t_{\pi(j')}) \right\rangle,
 %\left\langle \overline{H}(t_{\pi(1)}) \cdots  \overline{H}(t_{\pi(\ell)})  \, \widetilde{H}(t_{\pi(\ell+1)}) \cdots \widetilde{H}(t_{\pi(k)})\right\rangle,
\end{align}
with correlators
\begin{align*}
 \mathbf{C}_{\ell;k} \equiv \bigg\langle \prod_{j=1}^\ell \overline{H}(t_{\pi(j)}) \prod_{j'=\ell+1}^k \widetilde{H}(t_{\pi(j')}) \bigg\rangle,
\end{align*}
%\end{widetext}
and $d_>\vec{t}_{[k]}$ represents time-ordered integration, i.e., such that $t_1 \geq \cdots \geq t_k$.
In Eq.~\eq{changeofV}, we have performed a change of variables, allowing us to change the integration domain, which leads to the sum over the set $\Pi_{\ell;k}$, containing the permutations of $\{1,\ldots,k\}$ such that $t_{\pi(1)} \leq\cdots  \leq t_{\pi(\ell)}$ and $t_{\pi(\ell+1)} \geq\cdots \geq  t_{\pi(k)}$.
Expanding each of the correlators, we find
\begin{align*}
\mathbf{C}_{\ell;k} \notag &=  \\
 (-1)^\ell & \sum_{\vec{\alpha},\vec{u},\vec{v}} f^O_{\vec{v}|_\ell} \bigg\langle \prod_{j=1}^k y_{u_j,v_j}^{(\alpha_j)}(t_{\pi(j)}) \Lambda_{v_j} \otimes B_{u_j}^{(\alpha_j)}(t_{\pi(j)}) \bigg\rangle = \notag \\
 (-1)^\ell &\sum_{\vec{\alpha},\vec{u},\vec{v}} \! f^O_{\vec{v}|_\ell} \kappa_{\vec{v}}\hat{\Lambda}_{\vec{v}} \prod_{j=1}^k y_{u_j,v_j}^{(\alpha_j)}(t_{\pi(j)}) \bigg\langle  \prod_{j=1}^k B_{u_j}^{(\alpha_j)}(t_{\pi(j)}) \bigg\rangle , \notag
\end{align*}
where $f^O_{\vec{v}|_\ell} \equiv \frac{1}{d} \Tr[ O^{-1}\Lambda_{v_{1}} \cdots \Lambda_{v_{\ell}} O \big(\Lambda_{v_{1}} \cdots \Lambda_{v_{\ell}}\big)^{-1} \hat{\Lambda}_{\vec{v}}]$ and we have assumed for simplicity that the chosen operator basis is such that $ \prod_{j=1}^k \Lambda_{v_j} \equiv   \kappa_{\vec{v}} \hat{\Lambda}_{\vec{v}},$ for $\kappa_{\vec{v}} \in \mathbb{C}$ and $\hat{\Lambda}_{\vec{v}}$ invariant under permutations of $\vec{v}.$ Finally, this implies that we can write
\begin{widetext}
\begin{align}
    \mathcal{D}_O^{(k)}(T) & = k!(-i)^k \sum_\ell \sum_\pi \sum_{\vec{\alpha},\vec{u},\vec{v}} (-1)^\ell  f^O_{\vec{v}|_\ell} \kappa_{\vec{v}} \hat{\Lambda}_{\vec{v}} \int_0^T d_>\vec{t}_{[k]} \prod_{j=1}^k y_{u_j, v_j}^{(\alpha_j)}(t_{\pi(j)}) \left\langle \hat{B}_{\vec{u}}^{(\vec{\alpha})}(\pi(\vec{t})) \right\rangle, \notag \\
& = k!(-i)^k \sum_\ell (-1)^\ell \sum_\pi \sum_{\vec{\alpha},\vec{u},\vec{v}}
\kappa_{\vec{v}} \hat{\Lambda}_{\vec{v}}
\int_0^T d_>\vec{t}_{[k]} \prod_{j=1}^k y_{u_{\pi^{-1}(j)},v_{\pi^{-1}(j)}}^{(\alpha_{\pi^{-1}(j)})}(t_{j})
\Big(
f^O_{\vec{v}|_\ell}   \left\langle \hat{B}_{\vec{u}}^{(\vec{\alpha})}(\pi(\vec{t}))  \right\rangle\Big), \notag \\
& = k!(-i)^k \sum_\ell (-1)^\ell \sum_\pi \sum_{\vec{\alpha},\vec{u},\vec{v}}
\hat{\Lambda}_{\vec{v}}
\int_0^T d_>\vec{t}_{[k]} \prod_{j=1}^k y_{u_{j},v_{j}}^{(\alpha_{j})}(t_{j})
\Big(
\kappa_{\pi(\vec{v})} f^O_{\pi(\vec{v})|_\ell}   \left\langle \hat{B}_{\pi(\vec{u})}^{(\pi(\vec{\alpha}))}(\pi(\vec{t}))  \right\rangle\Big). \label{eq:finaldyson}
\end{align}
\end{widetext}
From Eq. \eqref{eq:finaldyson}, obtained by an adequate relabeling of the indices and observing that the sum is over all $\vec{u},\vec{v},\vec{\alpha}$, one deduces that for each configuration of $\alpha_j,u_j,v_j$ in $\{x,y,z\}$, the term $\hat{\Lambda}_{\vec{v}} \prod y^{(\alpha_j)}_{u_j v_j} (t_j)$ appears modulating a linear combination of bath correlation functions, $\left\langle \hat{B}_{\pi(\vec{u})}^{(\pi(\vec{\alpha}))}(\pi(\vec{t})) \right\rangle,$ i.e., a linear combination $\mathcal{L}_{\vec{\alpha};\vec{v}} (\vec{t})$. That is, each of the relevant overlap integrals $\mathcal{I}^{(k)}_{\vec{\alpha};\vec{u},\vec{v}}(T) $ appearing in the sum have the structure claimed in Eq. \eq{eq:kIntTime} of the main text. We also highlight that any other perturbative expansion, e.g., a cumulant expansion, can be written in term of structurally similar overlap integrals. Moreover, any function of the reduced dynamics, e.g., the fidelity, can be expanded in terms of the above integrals and resulting filter function representation. Which function is chosen is then a matter of convenience given a task at hand.

\section{Illustrative frame examples}
\label{App:Frames}

\subsection{Fourier frames and frequency-domain FFs revisited}
\label{App:Fourier}

The frame formalism encompasses both the Fourier series and the short-time Fourier transform, by relating them to expansions in terms of appropriate discrete Fourier frames \cite{ortega2002fourier} or, respectively, discrete and continuous Gabor frames \cite{FrameContinuous2,Grochenig}. Specifically, the frame of complex exponentials,
$$\mathscr{F}_{\rm FS} \equiv \{\phi_n(t) = e^{-i n \frac{2 \pi}{|\Lambda|} t }, n \in \mathbb{Z} \},$$ is a discrete, self-dual frame for functions $f \in L^2(\Lambda)$, where $\Lambda$ is a closed interval on ${\mathbb R}$ (e.g., $\Lambda = [0,1]$) and the inner product $(a,b) = \int_\Lambda \frac{dt}{|\Lambda|} a(t) b(t)^*$. The resulting frame expansion corresponds to the usual Fourier series on $\Lambda$, namely, $f(t) =  \sum_{n} (f,\phi_n ) e^{-i n \frac{2 \pi}{|\Lambda|} t }$, for $t \in \Lambda$.

Likewise, given $f \in L^2 ({\mathbb R})$, recall that the short-time Fourier transform (STFT, also known as the ``windowed'' FT or the Weyl-Heisenberg transform) with respect to a window function $g \in L^2 ({\mathbb R})$ is given by 
$$F_g (\omega, \tau)\equiv \int_{-\infty}^\infty dt f(t) {\phi}^*_{\omega,\tau}(t), \;\, \phi_{\omega,\tau}(t)= g(t-\tau) e^{-i\omega\cdot t},$$ 
where $\phi_{\omega,\tau}(t)$, $\omega,\tau \in {\mathbb R},$ are elements of a (continuous) Gabor frame~\cite{Grochenig, CGabor2}. That is, a two-dimensional representation of the signal is obtained by taking the FT of $f$ as the window function (e.g., a Gaussian or Hanh function centered around zero) is slid along the time axis. Conversely, the inverse STFT is given by $f(t) = \int_{-\infty}^\infty d\omega F_g(\omega,\tau) {\tilde{\phi}^*}_{\omega,\tau}(t),$ where the functions $\{\tilde{\phi}_{\omega,\tau}(t)\}$ are dual to  $\{\phi_{\omega,\tau}(t)\}$.

\begin{figure*}[t]
\begin{center}
\includegraphics[width=0.85\textwidth]{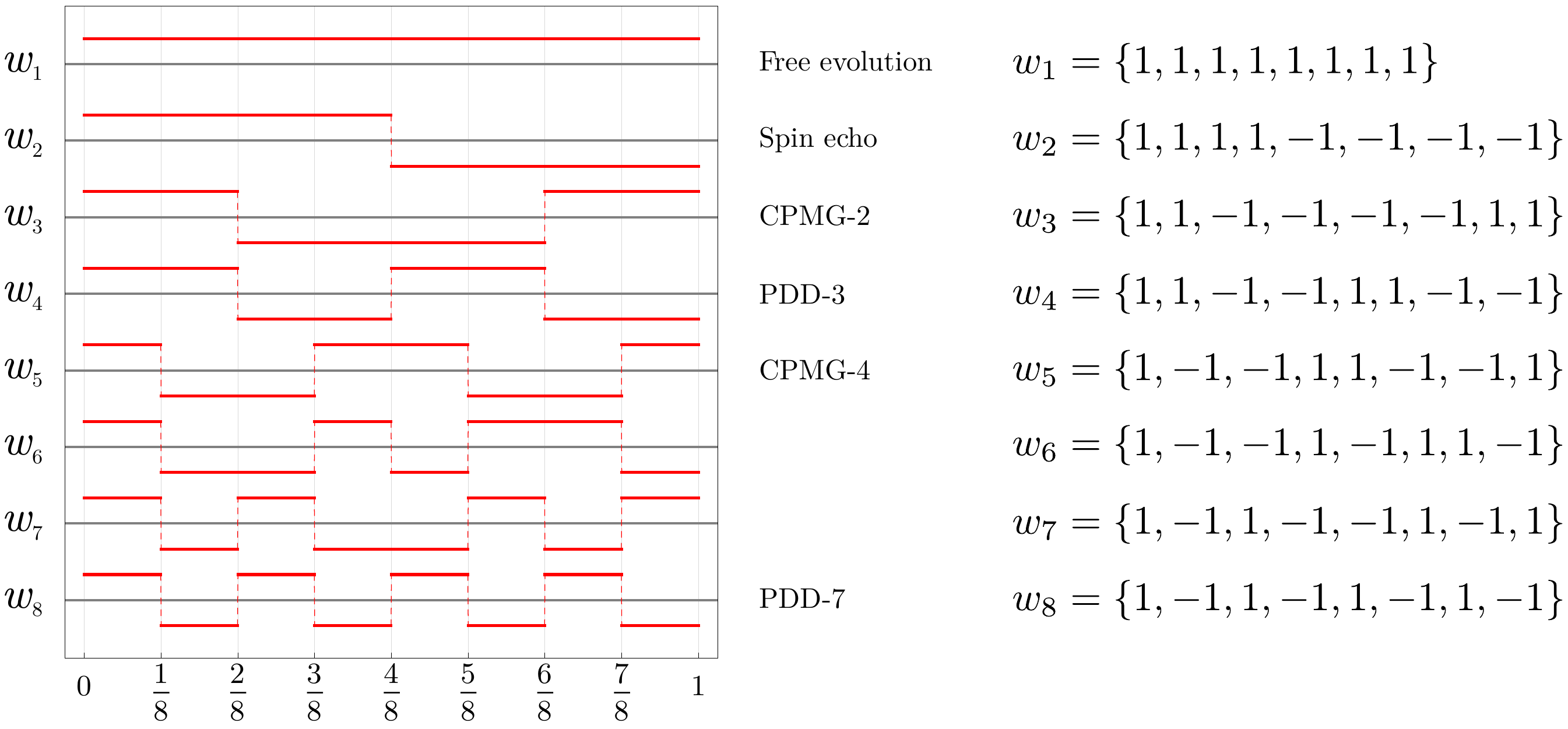}
\end{center}
\vspace*{-5mm}
\caption{(Color online) Left: the first eight Walsh functions in sequency ordering. The gray solid lines represent zero for each function. The name of the traditional $\pi$-pulse sequences \cite{WalshLV,WalshCooper} are written next to the plot. Right: each Walsh function is represented with its values in intervals.}
\label{fig:WalshBases}
\end{figure*}

Given the above formalism, one can now see how the standard frequency-domain FF formalism~\cite{Kurizki2004,Paz-Silva-FF2014} constitutes a particular limit of our construction. At an intuitive level, this is possible by considering the FT as an appropriate limit of the Fourier series, in the context of eigenfunction expansions (see e.g. Sec. 5.7 in \cite{Byron}). More formally in our context, the key observation is that any function corresponding to a physically admissible control (including free evolution) is necessarily time-limited, i.e., $t_i \leq \bar{T}$ for some finite $\bar{T}$ in our overlap integrals. Further, one has for the SP that
\begin{align*}\mathcal{I}^{(k)}_{\vec{\alpha};\vec{u},\vec{v}}
&= \int_{-\infty}^\infty d\vec{\tau}_{[k]}\int_{-\infty}^\infty d\vec{\omega}_{[k]}\,  F^{(k)}_{\vec{\alpha};\vec{u},\vec{v}} (\vec{\omega}) S^{(k)}_{\vec{\alpha};\vec{v}}(\vec{\omega},\vec{\tau}),
\end{align*}
where $S^{(k)}_{\vec{\alpha};\vec{v}}(\vec{\omega},\vec{\tau}) = \int_{-\infty}^\infty d\vec{t}_{[k]} ~ \mathcal{L}_{\vec{\alpha},\vec{v}}(\vec{t}~) g(\vec{t}-\vec{\tau}) e^{- i \vec{\omega} \cdot \vec{t}}$ is the FTFT associated to a sliding-window function $g(\vec{t})$~\cite{Grochenig}. By noticing that for any $\bar{T}$, there is a $\bar{\tau}$ such that  $ g(\vec{t}-\vec{\tau})$ and $\prod_{j=1}^k y_{u_j,v_j}^{(\alpha_j)}(t_j)$ have negligible overlap when $|\vec{\tau}| > \bar{\tau}$, one finds that
$$\mathcal{I}^{(k)}_{\vec{\alpha};\vec{u},\vec{v}} \simeq \int_{ |\vec{\tau}| \leq \bar{\tau}} d\vec{\tau}_{[k]}\int_{-\infty}^\infty d\vec{\omega}_{[k]} F^{(k)}_{\vec{\alpha};\vec{u},\vec{v}} (\vec{\omega}) S^{(k)}_{\vec{\alpha};\vec{v}}(\vec{\omega},\vec{\tau}). $$

In other words, the standard frequency FF formalism is effectively an expansion on the frame given by $\{\phi_{\vec{\omega},\vec{\tau}}\}$, for $|\vec{\tau}| < \bar{\tau}.$ A similar reasoning can be applied to the CA representation. The two representations are 
summarized in Tab. \ref{tab:SPvsCA} (iii)-(iv) of the main text.

\subsection{Digital frames for instantaneous pulses}
\label{app:digital}

Beyond Fourier and Gabor frames, the frame formalism allows for considerable flexibility. Consider the scenario in which the control matrix elements, $y_{{u},v}(t)$, are piece-wise constant in time. As we show in Appendix~\ref{oneQ}, this is relevant, for instance, when one considers $M$ (equidistant) instantaneous pulses over a time $T$, implemented by control profiles $h(t_j,t)=\delta(t- (t_j+ \tau/2))$, for $j\in[1,M]$ and $\tau = T/M$. Such control matrix elements are naturally spanned by the sequence $\mathscr{F}_W = \{ W_{j,\tau}\},$ where the window function $W_{j,\tau}$ is defined in Eq. \eqref{window} in the main text. As it turns out, $\mathscr{F}_W$ is not only a frame but also a basis, and for the above scenario the FSF condition is exactly satisfied ($\varepsilon =0$). What is more, {\em any} digital basis suffices, among which the Walsh functions provide a compelling  choice~\cite{WalshLV,WalshCooper}.

\emph{Walsh functions}~\cite{walsh1923closed} $w_n(t)$ are a complete set of orthogonal functions in an interval $[0,T]$ with the inner product $(a,b) = \int_0^T \frac{dt}{T} a(t) b(t)^\ast, $ which form a basis for piece-wise constant functions in $[0,T]$ with $2^N$ intervals ($N \in \mathbb{N}$). They are digital, taking values in $\{-1,1\}$, and can be defined via the rows of the Hadamard matrix $H_{2^N}.$ We choose the so-called \textit{sequency} ordering \cite{walsh1923closed,WalshLV,WalshCooper} for their labeling for convenience, such that the sequence $\mathscr{F}_{{\rm Walsh}_N}=\{w_j(t)\}_{j=1}^{2^N}$ is a basis for $2^N$-interval piecewise constant functions. The first eight Walsh functions are illustrated in  Fig.~\ref{fig:WalshBases}.

\subsection{Custom-built frames for arbitrary pulse profiles}
\label{custom}

\begin{figure*}[ht!]
 \begin{tabular}{@{}p{\linewidth}@{}}
 \centering
    \subfigimg[width=0.46\linewidth]{{(a)}}{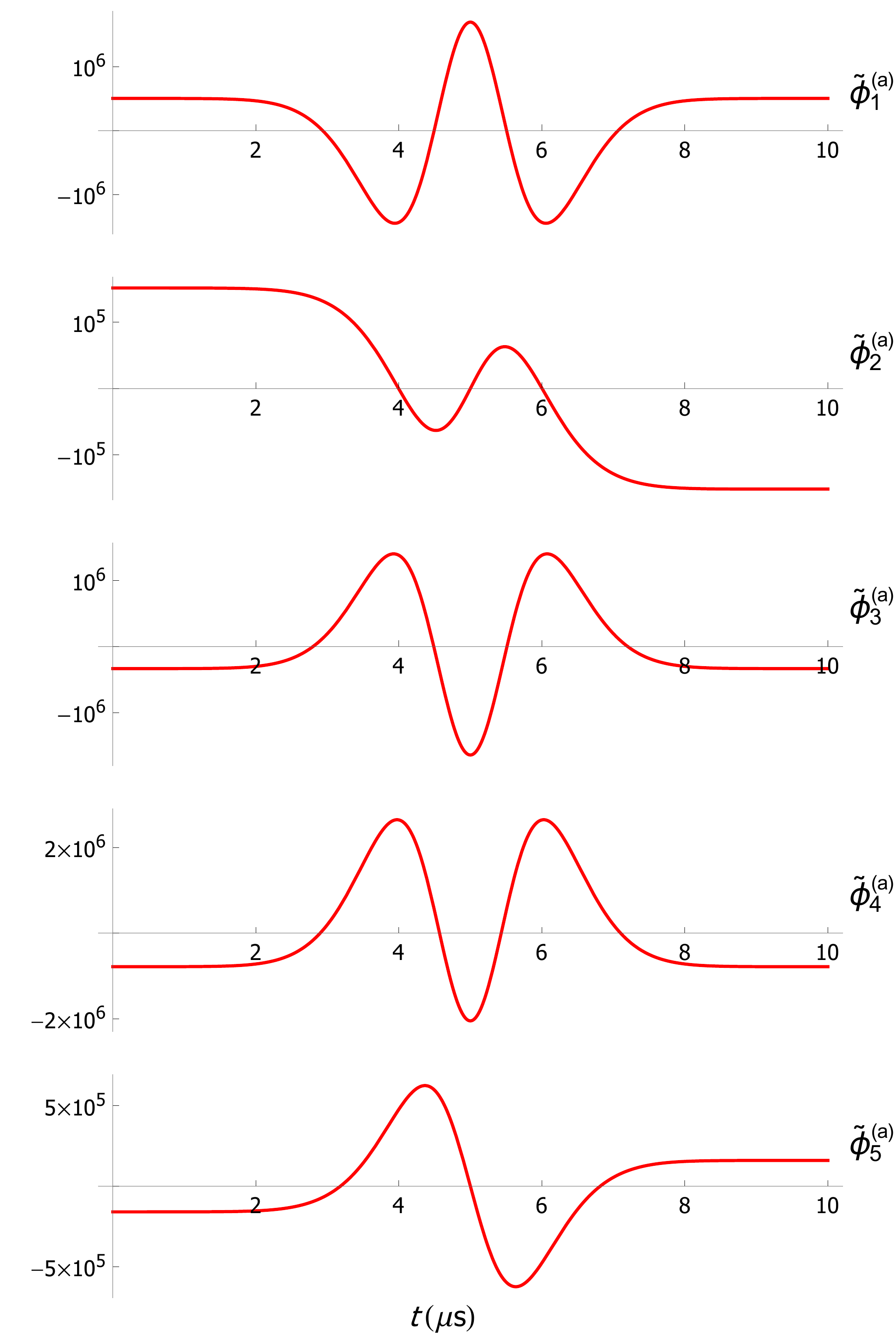}
    \hspace{6mm}
    \subfigimg[width=0.434\linewidth]{{(b)}}{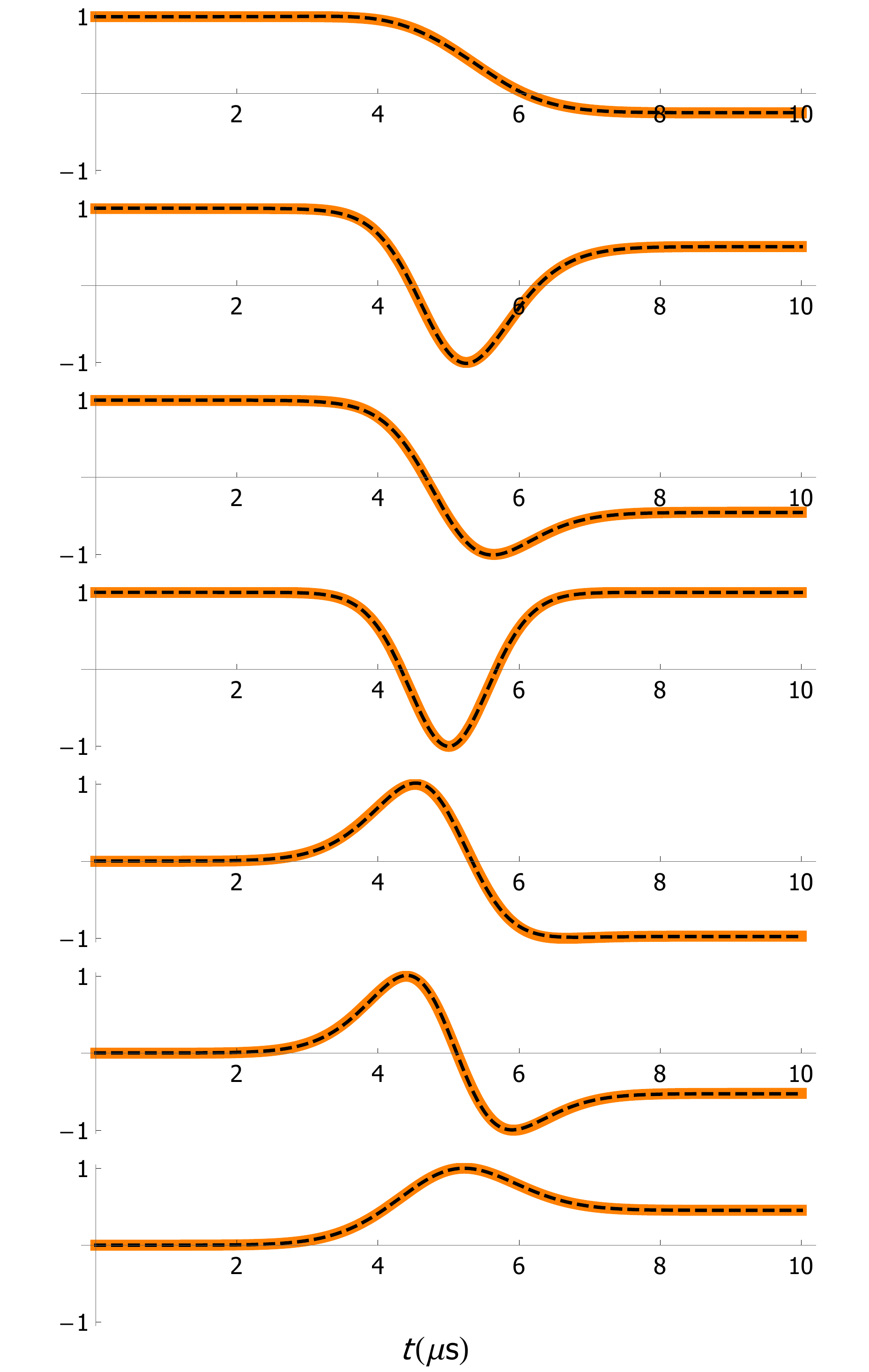}
  \end{tabular}
\vspace{-1em}
\caption{(Color online) (a) The five functions in the dual frame ${\tilde{\mathscr{F}}}^{(a)}$, obtained via Moore-Penrose pseudo-inverse. (b) Seven examples (orange thick) of $\cos(\theta_1\psi_t^{(1)})$ and $\sin(\theta_1\psi_t^{(1)})$ with randomly generated $\theta_1$ and their approximated frame representation (black dashed) {with ${\tilde{N}_\#=2}$}.}
\label{fig:dualFrame}
\end{figure*}

For general non-instantaneous pulses, corresponding to arbitrary control profiles, one often encounters the situation where the allowed control matrix $\mathscr{Y}$ has a particular structure, e.g., its components are linear combinations of specific functions of the available  control profile, and its elements belong to a particular Hilbert space $\mathcal{H}$. Two notable examples, typical of the unitary control scenario resulting from $M$ non-overlapping pulses in time $T$ we describe in the main text (and in detail in Appendix~\ref{oneQ}), are
\begin{align}
\mathscr{Y}^{(a)} &\in \mathcal{H}_{\mathscr{Y}^{(a)}}={\rm span} \Big\{W_{j,\tau}\cos[ \theta_j \psi _t^{(j)}],W_{j,\tau}\sin[ \theta_j \psi _t^{(j)}\Big\}, \nonumber \\
& \,\,\, {\rm with     } \,\,\, \psi _t^{(j)} = \int_{(j-1)\tau}^t  h(t_j,s)ds , \label{addFrame}\\
\mathscr{Y}^{(m)} &\in \mathcal{H}_{\mathscr{Y}^{(m)}}={\rm span} \Big\{ \theta_j h(t_j,t)\Big\}.
\end{align}

Consider a generic case where 
$$\mathcal{H} = {\rm span}_{\mathbb C} \Big\{  f_\ell (\{  \theta_j\},  \{b_j(t)\}) \: | \: {\ell\in[1,L],j\in[1,M]}\Big\},$$ for some set of functions $\{b_j(t)\}$, e.g., $b_j(t)= h(t_j,t)$ as  above or where each $b_j(t)$ is an element of a convenient (truncated) basis in which control profiles can be expanded. One can build a frame/dual-frame pair as follows. Imagine that each $\theta_j\in[0,2\pi]$ takes values among integer multiples of $2\pi/N_{\rm ctrl}$, for certain $N_{\rm ctrl}\in \mathbb{Z}$. Then, the ${N}_\#={N}_\#(L,M,\tilde{N}_\#)$- element sequence 
\begin{equation*}
\mathscr{F}_\# = \{\phi_n \} \equiv \{ f_\ell (\{\eta_j\},  \{b_j(t)\}) \}, \; \ell\in [1,L], \, j \in [1,M],
\end{equation*}
with $\eta_i =  2 \pi k_{i}/\tilde{N}_\#,$ $k_i \in [0,\tilde{N}_\#]$, spans $\mathcal{H}$ when $\tilde{N}_\# = N_{\rm ctrl},$ and is indeed (trivially) a frame. For a different choice of parameter $\tilde{N}_\#$, e.g., if $\tilde{N}_\# < N_{\rm ctrl},$ the frame property is lost as $\mathscr{F}_\#$ no longer spans $\mathcal{H}.$ One can nevertheless proceed to build a dual sequence $\tilde{\mathscr{F}}_\# =\{ \tilde{\phi}_n (t)\}_{n=1}^{N_\#}$ via the Moore-Penrose pseudo-inverse method (see below), such that by construction the orthogonality condition $(\phi_n,\tilde{\phi}_{n'}) = \delta_{n,n'}$ is satisfied. With this one can write
\begin{equation}
\label{parsi}
\tilde{y}(t) = \sum_{n=1}^{N_\#} (y,\tilde{\phi}_n)\, \phi_n (t),
\end{equation}
and calculate the error bound
$$\max_{y(t)}\Vert y(t) - \tilde{y}(t) \Vert_2  = \max_{y(t)} \sqrt{\int_0^T |y(t)- \tilde y(t)|^2 dt} \leq \varepsilon,$$
for the candidate ${\mathscr{F}}_\#$ and $\tilde{\mathscr{F}}_\#$, as necessary to verify the FSF condition. That is, one can assess the ability of a candidate frame/dual-frame pair to approximate every $y(t) \in \mathcal{H}$ via Eq.~\eqref{parsi}, thereby verifying the parsimony of ${\mathscr{F}}_\#$.

\subsubsection{Single-qubit additive and multiplicative noise} 

For the additive dephasing noise we study in the main text (see also Appendix~\ref{oneQ} below), the five dual frame functions built for $\tilde{N}_\#=2$ and $M=1$ are depicted in Fig.~\ref{fig:dualFrame}(a). Here, the associated frame is given by $\mathscr{F}^{(a)}_\# = \{\phi_n^{(a)}(t)\} = \{1,\cos[\pi \psi^{(1)}_t ],\cos[2\pi \psi^{(1)}_t ],\sin[\pi \psi^{(1)}_t ],\sin[2\pi \psi^{(1)}_t ]\}$, and its dual can be calculated as outlined below. It is worth highlighting that when the $b_j(t)$ are non-overlapping, as in the $b_j(t)=h(t_j,t)$ case considered here, one has $N_\# =  M L ( \tilde{N}_\# +1)$, that is, the size of the frame grows \emph{linearly} with the number of pulses $M$ (Note that the $\phi_0^{(a)} = \sin [0 \,\psi_t^{(1)}] =0$ is trivial and thus excluded from the frame definition, leading to $N_\# = 5$). For any $y(t)$ one can calculate an upper bound $\varepsilon$ to the $L^2$-distance $\Vert y(t) - \tilde{y}(t) \Vert_2$ by evaluating these quantities for any $\theta_1 \seq \frac{k}{100} 2\pi $, for $k {\in}\, [0,100]$. We find that ${\varepsilon \vert_{\tilde{N}_\#=2}} =2.4\cdot 10^{-5}$ and ${\varepsilon \vert_{\tilde{N}_\#=4}} =  2.8\cdot10^{-8}$. In Fig.~\ref{fig:dualFrame}(b), we plot seven such $y(t)$ (for random $\theta_1$) and their approximated frame representation functions (\ref{parsi}) for illustration purposes.

For the case of multiplicative noise, notice that $\mathcal{H}_{\mathscr{Y}^{(m)}}={\rm span} \{ h(t_j,t)\}$, so we may simply take $\mathscr{F}^{(m)}_\#\! \equiv \! \{h(t_j,t)\}$, with $j\in[1,M]$, and trivially satisfy the FSF condition in this way. Also, $\tilde{\mathscr{F}}^{(m)}_\# ={\mathscr{F}}^{(m)}_\#/Z_m$, where $Z_m$ is a normalization factor ensuring that $\int_0^T dt \,\phi_j^{(m)}(t) {\phi^{(m)}_{j'}(t)}/{Z_{m}} =\delta_{j,j'}$.

\subsubsection{Building a dual sequence via Moore-Penrose pseudo-inverse}

In practice, given a sequence, one can build a canonical dual one via a standard Moore-Penrose pseudo-inverse construction. When the sequence under consideration is a frame, then its dual is also a frame. The starting point is a reference orthonormal basis. We first note that $\mathscr{F}_\#= \{\phi_n(t)\}_{n=1,\cdots,N_\#}$ spans a Hilbert space $\mathcal{H}_\#$ with the usual $L^2$ inner product. An orthonormal basis for $\mathcal{H}_\#$, which we denote by $\mathscr{G}_\# =\{ g_{m}\}_{m=1,\ldots, M_\#}$ with $M_\# \leq N_\#$, is then built by applying a Gram-Schmidt process to $\mathscr{F}_\#$. In turn, this means that one can write
$\phi_n = \sum_{m} ( \phi_n, g_{m} ) g_{m} ,$ which is represented in matrix language as $\vec{\phi} = \mathbf{T} \vec{g}$, with $\mathbf{T}$ generally a non-square matrix. The dual frame can then be built via Moore-Penrose pseudo-inverse. One has then that
\begin{align*}
{\phi_n} & = \sum_{m} ( \phi_n, g_{m} ) \,g_{m} 
%{\phi_n} & 
=  \sum_{n'} ( \phi_n,  {\phi}_{n'} ) \, \tilde{\phi}_{n'}  \\ 
&= \!\!\sum_{{n'},m',m,j' } \!\! \! (\phi_n,  {g}_{m}) (\phi_{n'},  g_{m} )
( \tilde{\phi}_{n'},  g_{m'} ) \, g_{m'},
\end{align*}
where we have used that $\tilde{\phi}_j = \sum_{j'} ( \tilde{\phi}_j, g_{j'} ) \,g_{j'} ,$ or, equivalently, $\vec{\tilde{\phi}} = \tilde{\mathbf{T}} \vec{g},$ and that the $\{g_{j'}\}$ are a basis. The above implies that  $\mathbf{T}^T \tilde{\mathbf{T}} = I$, which, noting that $\mathbf{T}^T \mathbf{T}$ is invertible, has the solution $\tilde{\mathbf{T}} = \mathbf{T} (\mathbf{T}^T \mathbf{T})^{-1}.$ Thus, the elements of $\tilde{\mathscr{F}}_\#$ are given by $$\vec{\tilde{\phi}}  = \mathbf{T} (\mathbf{T}^T \mathbf{T})^{-1} \vec{g}.$$

\section{Single-qubit frame-based protocols}
\label{oneQ}

\subsection{Single-qubit reduced dynamics and determination of $\bar{\mathscr{S}}\vert_{\mathscr{C}}$}

We are interested in obtaining explicit expressions for expectation values of an invertible observable $O$ at a time $T$. As mentioned in the text, this can be accomplished in a weak-coupling regime via a truncated Dyson expansion given by
\begin{align}
E[O(T)]_{\rho_S \otimes \rho_B} &\approx \big\langle \Tr[ (I_S - \mathcal{D}_{O}^{(2)}(T)) \rho_S \widetilde{O}(T)]\big\rangle  \label{subeq:D2sep}\\
&= \Tr[\rho_S \widetilde{O}(T)]  - \big\langle \Tr[\mathcal{D}_{O}^{(2)}(T) \rho_S \widetilde{O}(T)] \big\rangle, \notag
\end{align}
where the Dyson term is given by (see Eq. \eqref{eq:finaldyson})
\begin{align*}
    \mathcal{D}_{O}^{(2)}(T) &= 2!\int_0^T \!\!\!d_> \vec{t}_{[2]} \left\langle \widetilde{H}(t_{1})\widetilde{H}(t_{2})-\overline{H}(t_{2})\right.\widetilde{H}(t_1) \\ 
    & \left.-\overline{H}(t_{1})\widetilde{H}(t_{2})+\overline{H}(t_{2})\overline{H}(t_{1}) \right\rangle_c,
\end{align*}
and $\overline{H}(t)  \equiv \sum_u y^{(\alpha)}_u(t)\sum_c f_c^u \sigma_c \otimes B^{(\alpha)}(t)$, with $f_c^u=\frac{1}{2}\Tr[\widetilde{O}^{\dagger}(T)\sigma_u \widetilde{O}(T)\sigma_c]$. Assuming that there are no correlations between additive and multiplicative noise, i.e. $\langle B^{(a)}(t_1) B^{(m)}(t_2) \rangle =0,$ for all $t_1,t_2$, we have
\begin{widetext}
\begin{align}
\frac{\mathcal{D}_{O}^{(2)}(T)}{2} = \sum_{\alpha;u,v} \int_0^T d_>\vec{t}_{[2]} \Big( ~~& y^{(\alpha)}_u(t_1)y^{(\alpha)}_{v}(t_2)\sigma_u\sigma_{v}\left\langle B^{(\alpha)}(t_1)B^{(\alpha)}(t_2)\right \rangle_c \notag \\
- ~ &y^{(\alpha)}_u(t_1)y^{(\alpha)}_{v}(t_2)\Big(\sum_c f_{c}^{u}\sigma_{c}\Big)\sigma_{v}\left\langle B^{(\alpha)}(t_1)B^{(\alpha)}(t_2)\right \rangle_c \notag\\
- ~ &y^{(\alpha)}_u(t_1)y^{(\alpha)}_{v}(t_2)\Big(\sum_{c'}f_{c'}^{v}\sigma_{c'}\Big)\sigma_{u}  \left\langle B^{(\alpha)}(t_2)B^{(\alpha)}(t_1)\right \rangle_c \notag \\
+ ~ &y^{(\alpha)}_u(t_1)y^{(\alpha)}_{v}(t_2)\sum_{c,c'}f_c^{u}f_{c'}^{v}\sigma_{c'}\sigma_{c}\left\langle B^{(\alpha)}(t_2)B^{(\alpha)}(t_1)\right\rangle_c \Big)
\end{align}
which, in the frame language, reads
\begin{align*}
\frac{\mathcal{D}_{O}^{(2)}(T)}{2} 
= \frac{\sigma_0}{4} \!\!\sum_{\alpha;u;n,n'; \mu=\pm}
\hspace*{-3mm}\bar{S}^{(\mu)}_{\alpha}(n,n')   F^{(1,-)}_{\alpha;u}(n,T) F^{(1,-\mu)}_{\alpha;u}(n',T) 
+ \!\!\!\!\sum_{\alpha;u\neq v;n,n';\mu=\pm }\! 
\hspace*{-3mm} \frac{\sigma_u \sigma_{v}}{4} 
\bar{S}^{(\mu)}_{\alpha}(n,n')   F^{(1,-)}_{\alpha;u}(n,T) F^{(1,\mu)}_{\alpha;v}(n',T).
\label{eq:dysoninframe}
\end{align*}
The last equation follows after change of variables $c\leftrightarrow u$, $c'\leftrightarrow v$ and using
$F^{(1,\pm)}_{\alpha;u}(n,T)$ to denote the frame $\mathscr{F}^{(\alpha)}$-representation of $Y^{(\pm)}_{\alpha;u}(t) \equiv y^{({\alpha})}_{u}(t)\pm \sum_c f^{u}_c y^{({\alpha})}_c(t)$, with the definition for $\bar{S}^{(\pm)}_{\alpha}(n,n')$ used in the main text. Therefore, the second term in Eq.~\eq{subeq:D2sep} contains the effect of the noise that we are interested in. In the following, we isolate the quantities $\Tr [ \mathcal{D}^{(2)}_O(T) \sigma_l]$,   $l \in \{x,y,z\}$, for a given control sequence, initial system state, and measured observable, so that we can build our CA-QNS protocol by cycling over a sufficiently large set of controls and observables. First we note that
\begin{subequations}
\label{eq:combinetoaccess}
\begin{align}
    E\big[O(T)\big]_{\frac{1}{2}(I_S + \sigma_k) \otimes \rho_B} +  E\big[O(T)\big]_{\frac{1}{2}(I_S -  \sigma_k) \otimes \rho_B} &= -\left \langle \Tr\left[\mathcal{D}_{O}^{(2)}(T) \widetilde{O}(T)\right] \right\rangle, \\
    E\big[O(T)\big]_{\frac{1}{2}(I_S + \sigma_k) \otimes \rho_B} -  E\big[O(T)\big]_{\frac{1}{2}(I_S -  \sigma_k) \otimes \rho_B} &=  \Tr\left[\sigma_k \widetilde{O}(T)\right] - \left\langle \Tr\left[\mathcal{D}_{O}^{(2)}(T) \sigma_k \widetilde{O}(T)\right] \right\rangle,
\end{align}
\end{subequations}
which implies that from the (measured) expectation values we can infer the value of $\langle \Tr[\mathcal{D}_{O}^{(2)}(T) \sigma_r \widetilde{O}(T)]\rangle_c$, for $r \in \{0,x,y,z\}$. Then, we note that for a choice of control and observable, the operator $\widetilde{O}(T)=\sum_{k} \frac{\Tr[ \widetilde{O}(T) \sigma_k]}{2}\sigma_k\equiv \sum_{k}o_{k}\sigma_k$ is known and fixed, which allows us to write the system of equations
\begin{equation}
    \Big\{
    \Tr[\mathcal{D}^{(2)}_{O}(T)\sigma_r \widetilde{O}(T)] = \sum_{l=0,x,y,z}\sum_{k=x,y,z} \Tr[\mathcal{D}^{(2)}_{O}(T) \sigma_l]  o_k \frac{\Tr[\sigma_{l}\sigma_r\sigma_{k}]}{2}\Big\}_{r \in \{0,x,y,z\}},
    \end{equation}
from which the $ \{\Tr[\mathcal{D}^{(2)}_{O}(T) \sigma_l]\}$ can be inferred, as desired.

We then combine these values --  for a fixed $U_0(T)$ -- to construct $ \Tr[\mathcal{D}^{(2)}_{O}(T)\sigma_r \widetilde{O}(T)],$ for $\widetilde{O}(T) = \sigma_\gamma,$ for all $r, \gamma,$  which simplifies the $\widetilde{O}(T)$-dependent expression for $F_{\alpha,u}^{(1,\pm)}(n,T)$. A direct calculation for $r=0$ shows that
\begin{align*}
\langle \Tr[\mathcal{D}_{O}^{(2)}(T) \sigma_0 \sigma_{\gamma}]\rangle_c & =
\sum_{\alpha;n,n'} \sum_{u\neq v\neq \gamma}\sum_{\mu=\pm} i \epsilon_{uv\gamma} \bar{S}^{(\mu)}_{\alpha}(n,n') ~ F^{(1,-)}_{\alpha;u}(n,T) F^{(1,\mu)}_{\alpha;v}(n',T), \\
&= \sum_{\alpha;n,n'} \sum_{u\neq v\neq \gamma}\sum_{\mu=\pm} i \epsilon_{uv\gamma} ~ (1+\mu g^{\gamma}_{v})(1-g^{\gamma}_{u})F^{(1)}_{\alpha;u}(n,T)F^{(1)}_{\alpha;v}(n',T)\bar{S}^{(\mu)}_{\alpha}(n,n'), \\
&= 4i\sum_{\alpha;n,n'} \sum_{u\neq v\neq \gamma}\epsilon_{uv\gamma}F^{(1)}_{\alpha;u}(n,T)F^{(1)}_{\alpha;v}(n',T)\bar{S}^{(-)}_{\alpha}(n,n'), \\
&= 4i\sum_{\alpha;u< v\neq\gamma}\epsilon_{uv\gamma}\overline{\mathcal{I}}^{(2,-)}_{\alpha;u,v}(T) ,
\end{align*}
where $\epsilon_{uv\gamma}$ is the Levi-Civita symbol, $g_{u}^{v}=\frac{1}{2}\Tr[\sigma_{u}\sigma_{v}\sigma_{u}\sigma_{v}]$, and
\begin{equation*}
\overline{\mathcal{I}}_{\alpha;u,v}^{(2,-)}(T) \equiv \sum_{n,n'}\Big(\bar{S}_{\alpha}^{(-)}(n,n')-\bar{S}_{\alpha}^{(-)} (n',n)\Big)F^{(1)}_{\alpha;u}(n,T)F^{(1)}_{\alpha;v}(n',T).
\end{equation*}
Similarly, for $r\neq0$ we find
%\begin{widetext}
\begin{align*}
\langle \Tr[\mathcal{D}_{O}^{(2)}(T) \sigma_r \sigma_{\gamma}]\rangle_c
&= \sum_{\alpha;n,n'} \sum_{u}\sum_{\mu=\pm} \delta_{u,\gamma} \left(1-g^{\gamma}_{u} \right)\left(1-\mu g_{u}^{\gamma} \right)F^{(1)}_{\alpha;u}(n,T)F^{(1)}_{\alpha;u}(n',T)\bar{S}^{(\mu)}_{\alpha}(n,n')\nonumber \\
&\hspace{0.5cm} + \sum_{\alpha;n,n'}\sum_{u\neq v}\sum_{\mu=\pm} \left(\delta_{r,v}\delta_{\gamma,u}-\delta_{r,u}\delta_{\gamma,v}\right) \left(1-g_{u}^{\gamma}\right) \left(1+\mu g_{v}^{\gamma}\right) F^{(1)}_{\alpha;u}(n,T)F^{(1)}_{\alpha;v}(n',T)\bar{S}^{(\mu)}_{\alpha}(n,n')\\
&= 4\sum_{\alpha;n,n'}\sum_{u\neq\gamma}\big(\delta_{r,\gamma}F^{(1)}_{\alpha;u}(n,T)F^{(1)}_{\alpha;u}(n',T)\bar{S}^{(+)}_{\alpha}(n,n')\nonumber -
\delta_{r,u} F^{(1)}_{\alpha;u}(n,T)F^{(1)}_{\alpha;\gamma}(n',T)\bar{S}^{(+)}_{\alpha}(n,n')\big),\\
&= 4\sum_{\alpha;u\neq\gamma} \big(\delta_{r,\gamma}~ \mathcal{I}^{(2,+)}_{\alpha;u,u}(T)-
\delta_{r,u}~ \mathcal{I}^{(2,+)}_{\alpha;u,\gamma}(T)\big),
\end{align*}
\end{widetext}
where the index structure forbids the contribution from $\bar{S}^{(-)}_\alpha(n,n'),$ and we have defined
\begin{equation*}
    \mathcal{I}^{(2,+)}_{\alpha;u,v}(T) \equiv \sum_{n,n'} \bar{S}^{(+)}_{\alpha}(n,n')F^{(1)}_{\alpha;u}(n,T)F^{(1)}_{\alpha;v}(n',T).
\end{equation*}
Therefore, from the possible $r,\gamma$ configurations, and
\begin{align*}
 \langle \Tr[\mathcal{D}_{O}^{(2)}(T) \sigma_r \sigma_\gamma]\rangle_c =
\begin{cases}
4\displaystyle{\sum_{\alpha;u<v\neq\gamma}} i \epsilon_{uv\gamma} ~\overline{\mathcal{I}}^{(2,-)}_{\alpha;u,v}(T) & r=0, \\
4\displaystyle{\sum_{\alpha;u\neq\gamma}}~ \mathcal{I}^{(2,+)}_{\alpha;u,u} & r=\gamma, \\
-4 \displaystyle{\sum_{\alpha;u\neq \gamma}} \delta_{u,r} ~\mathcal{I}^{(2,+)}_{\alpha;u,\gamma} & \hspace*{-9mm} 
r \neq \gamma, r \neq 0,
\end{cases}
\end{align*}
we conclude that \emph{only} the integrals $\mathcal{I}^{(2,+)}_{\alpha;u,v}$, for all $u,v$, and $\overline{\mathcal{I}}^{(2,-)}_{\alpha;u,v}$, for all $u\neq v$, influence the reduced qubit dynamics. In turn, this implies that only the quantities $ \bar{\mathscr{S}}\vert_\mathscr{C} =\{ \bar{S}^{(+)}_\alpha (n,n'),\bar{S}^{(-)}_\alpha (n,n')-\bar{S}^{(-)}_\alpha (n',n)\}$ are relevant to the dynamics given $\mathscr{C}$. The objective of QNS is to precisely extract all the spectra in $ \bar{\mathscr{S}}\vert_\mathscr{C}$.

\subsection{Control-adapted QNS protocol with instantaneous control}
\label{App:CA-QNS}

In the case of instantaneous control, notice that each switching function is exactly expanded by appropriate digital basis as mentioned in Sec.~\ref{subsection:frame}. We thus use Walsh functions (see Appendix~\ref{app:digital}) as our frame. To perform CA-QNS for such a frame, it is enough to use rotations around the  $y$-axis. In this situation, the toggling-frame Hamiltonian specializes to $$ \widetilde{H}(t) = y_{z,z}(t) \sigma_z \otimes B(t) + y_{z,x}(t) \sigma_x \otimes B(t),$$ where the control matrix elements are such that in the $j$-th time interval they are given by
\begin{subequations}
\begin{align*}
\label{eq:SwitchSinCos}
    y_{z,z}(t) &= \cos ( \tilde \theta_j ) = \frac{1}{2}\Tr\Big[e^{i\frac{\tilde{\theta}_{j}}{2}\sigma_y}\sigma_z e^{-i\frac{\tilde{\theta}_{j}}{2}\sigma_y}\sigma_z \Big], \notag \\
    &\hspace{3cm} (j-1)\tau \le t < \tau, \\
    y_{z,x}(t) &= -\sin ( \tilde \theta_j ) = \frac{1}{2}\Tr\Big[e^{i\frac{\tilde{\theta}_{j}}{2}\sigma_y}\sigma_z e^{-i\frac{\tilde{\theta}_{j}}{2}\sigma_y}\sigma_x\Big] , \notag  \\ 
    & \hspace{3cm} (j-1)\tau \le t < \tau,
\end{align*}
\end{subequations}
with the relation between the $\tilde{\theta}_j$ and $\theta_j$ set by the equations
\begin{equation}
\label{eq:thetatildetheta}
    \begin{cases}
	0  =  \tilde\theta_1, & \\
	\theta_{k} =  \tilde\theta_{k+1} - \tilde\theta_{k}, & 1\leq k\leq M-1\notag \\
	\theta_M = -\tilde\theta_M. &
\end{cases}.
\end{equation}

The above considerations suggest that:

\smallskip

{\bf (1)} Using only $\tilde{\theta} = \pi$ pulses, one can ensure that $y_z(t)$ and $y_x(t)$ take values in $\{-1,1\}$, and equal (up to a sign) to any desired Walsh function $w_n(t)$ for $t \in [0,T]$ and $n \in [1,M].$  This implies, for example, that
\begin{equation*}
    F^{(1)}_z(n,T) = \int_0^T dt\ w_m(t) w_n(t) = T\delta_{mn},
\end{equation*}
and thus
\begin{align}
\label{eq:QWalsh}
    \mathcal{I}^{(+)}_{z,z}(T)& \equiv \mathcal{I}^{(2,+)}_{a;z,z}(T)= T^2\sum_{n,n'} \bar{S}^{(+)}(n,n') \delta_{nm} \delta_{n'm}\notag\\
    & = T^2\bar{S}^{(+)}(m,m).
\end{align}
That is, we can directly sample diagonal elements $\bar{S}^{(+)}(m,m)$.

\smallskip

{\bf (2)} Using  $\tilde{\theta} \lspace{\in} \{ \pi, \pi/2\}$, we ensure that $y_z(t)$ and $y_x(t)$ take values in $\{-1,0,1\}$, with the constraint $|y_z(t)|^2 +|y_x(t)|^2 {=} 1.$ In particular, one can choose angles such that
$y_z(t) = w_m(t) + w_{m'}(t)$, while $y_x(t) = w_m(t) - w_{m'}(t)$ and thus
\begin{align*}
    F^{(1)}_z(n,T) &= \int_0^T dt ~ \frac{1}{2} \Big( w_m(t)+w_{m'}(t) \Big) w_n(t) \notag \\
    &= \frac{T}{2} (\delta_{nm} + \delta_{nm'}), 
   \\
    F^{(1)}_x(n,T) &= \int_0^T dt ~ \frac{1}{2} \Big( w_m(t)-w_{m'}(t) \Big) w_n(t) \notag \\ 
    &= \frac{T}{2} (\delta_{nm} - \delta_{nm'}).
\end{align*}
\noindent 
When applied to our dynamical equations, the above implies that
\smallskip
\begin{widetext}
\begin{subequations}
\label{eq:QTwoWalshes}
\begin{align}
    \mathcal{I}^{(+)}_{z,z}(T) &= \frac{T^2}{4} \Big( \bar{S}^{(+)}(m,m) + \bar{S}^{(+)}(m,m') + \bar{S}^{(+)}(m',m) + \bar{S}^{(+)}(m',m') \Big), \label{eq:Qaa}\\
    \mathcal{I}^{(+)}_{z,x}(T) &= \frac{T^2}{4} \Big( \bar{S}^{(+)}(m,m) - \bar{S}^{(+)}(m,m') + \bar{S}^{(+)}(m',m) - \bar{S}^{(+)}(m',m') \Big),\\
    \overline{\mathcal{I}}^{(-)}_{z,x}(T) &= -\frac{T^2}{2} \Big( \bar{S}^{(-)}(m,m') - \bar{S}^{(-)}(m',m) \Big)
    ,\label{eq:Qaap}
\end{align}
\end{subequations}
\end{widetext}
and thus we can infer the elements $\bar{S}^{(+)}(n,n')$ and $\bar{S}^{(-)}(n,n') - \bar{S}^{(-)}(n',n)$, as desired.

\smallskip

Given access to the corresponding $\overline{\mathscr{S}}|_\mathscr{C}$ and noting that the SP and the CA pictures are related via
${S}^{(\pm)}(n,n') =\bar{S}^{(\pm)}(n,n') \pm \bar{S}^{(\pm)}(n',n),$
one can then obtain Walsh reconstructions $\hat{C}_{\pm}^{(a)}(t_1,t_2)$ of ${C}_{\pm}^{(a)}(t_1,t_2)$
given by
\begin{align*}
\hat{C}_{\pm}^{(a)}(t_1,t_2) = \!\!\sum_{n,n'=1}^{N_\#} \!\!
\Big(\bar{S}^{(\pm)}(n,n') \pm \bar{S}^{(\pm)}(n',n)\Big) w_n(t_1) w_{n'}(t_2).
\end{align*}
The reconstruction resolution will depend on the free parameters in the above protocol, namely, the total time $T$ and the minimum switching time $\tau$, which upper-bounds the value of $N_\#$. In general, a smaller $\tau$ leads to higher resolution.

\begin{figure*}
\makeatletter\def\@captype{figure}\makeatother
\begin{minipage}[t]{.5\linewidth}
\centering
\hspace{-8mm}\vspace{-6mm}\includegraphics[width=1.1\linewidth]{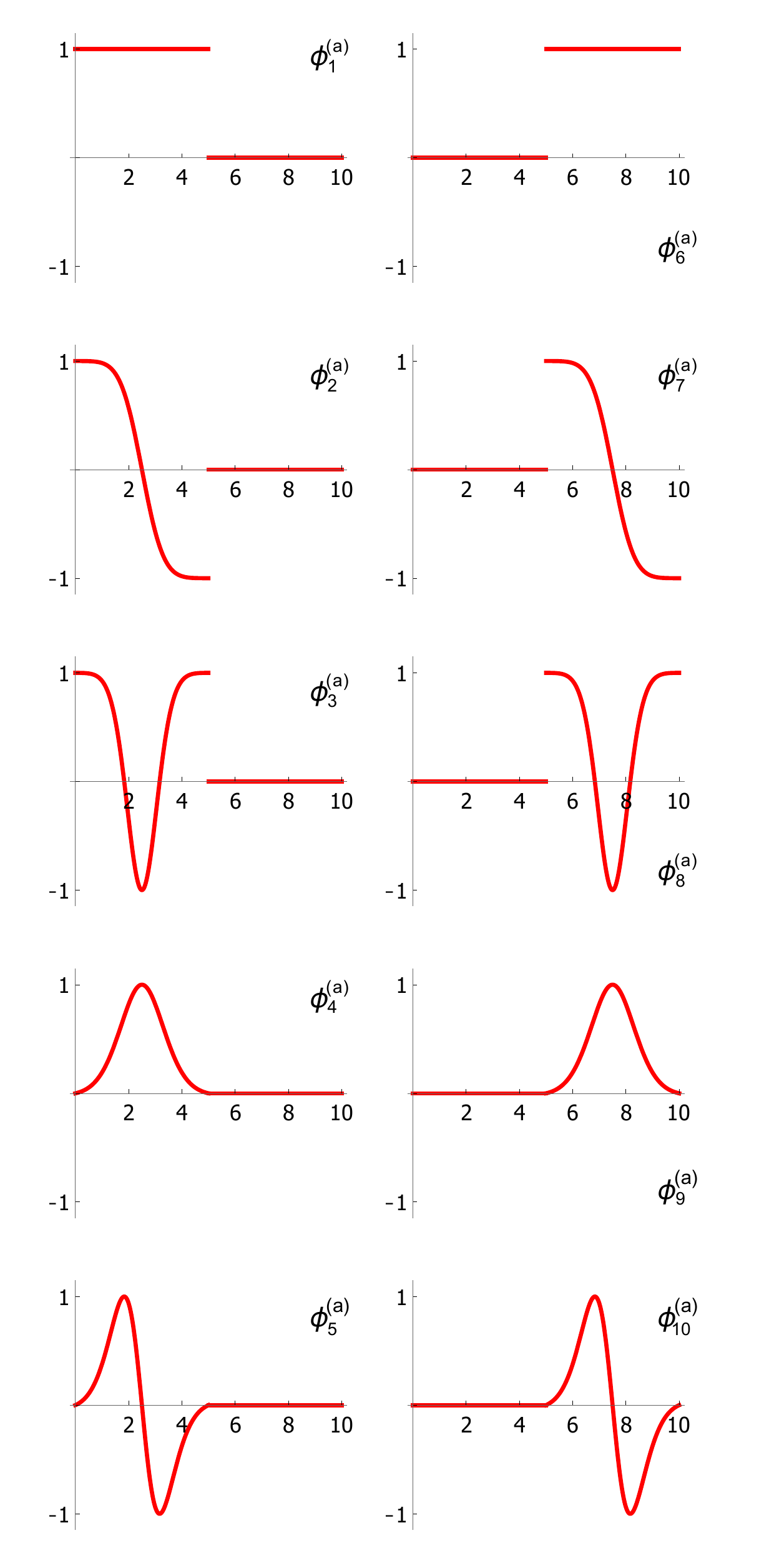}
\caption{Frame elements in $\mathscr{F}^{(a)}_\#$ for $M=2$ and $\tilde{N}_\#=2$. Horizontal axes are time, $t(\mu s)$.}
\label{frame1}
\end{minipage}
\hspace{7mm}
\makeatletter\def\@captype{table}\makeatother
\begin{minipage}[t]{.42\linewidth}
\centering
\caption{Symmetry equation list. Number 1 to 45 list all the combinations of $(j_1,k_1)$ and $(j_2,k_2)$ such that $\bar{S}_a^+(j_1,k_1)\pm\bar{S}_a^+(j_2,k_2)=0$. Number 46 to 70 list all the $(j_3,k_3)$ such that $\bar{S}_a^+(j_3,k_3)=0$.}
\renewcommand\arraystretch{1.45}
\setlength{\tabcolsep}{0.9mm}{\begin{tabular}{ |c|c||c|c| }
  \hline
1  & $(1, 1)  -  (6, 6)$  &  36  &    $(6, 2)  + ( 7,1)$   \\
 \hline
2  & $(1, 2)  +  (7, 6)$ & 37   &   $(6, 3)  - (8 ,1)$    \\
 \hline
3  & $(1, 3)  -  (8, 6)$ &  38  &    $(6, 4)  - ( 9,1)$  \\
 \hline
 4 & $(1, 4)  -  (9, 6)$ &  39  &     $(6, 5)  + (10 ,1)$   \\
 \hline
5  &  $(1, 5)  +  (10, 6)$  & 40   &  $(7, 3) +  (8 ,2)$   \\
 \hline
6  & $(6, 7)  +  (7, 6)$  &   41 &    $(7, 4) +  (9 ,2)$  \\
 \hline
7  &   $(6, 8)  -  (8, 6)$ &   42 &    $(7, 5) -  (10 ,2)$  \\
 \hline
8  & $(6, 9)  -  (9, 6)$ &  43  &   $(8,4 ) -  (9 ,3)$   \\
 \hline
9  & $(6, 10)  +  (10, 6)$ &  44  &  $(8,5 ) +  (10 ,3)$   \\
 \hline
10  & $(2, 1)  -  (7, 6)$  &   45 &    $(9, 5) +  ( 10,4)$  \\
 \hline
11  & $(2, 2)  -  (7, 7)$  &  46  &   $(1, 6)$  \\
 \hline
12  & $(2, 3)  +  (8, 7)$ &  47  &   $(1, 7)$  \\
 \hline
13  & $(2, 4)  +  (9, 7)$  &  48  &   $(1, 8)$  \\
 \hline
14  & $(2, 5)  -  (10, 7)$  & 49   &    $(1, 9)$ \\
 \hline
15  & $(7, 8)  +  (8, 7)$  &  50  &  $(1, 10)$   \\
 \hline
16  & $(7, 9)  +  (9, 7)$  &  51  &    $(2, 6)$  \\
 \hline
17  & $(7, 10)  -  (10, 7)$  & 52   &  $(2, 7)$  \\
 \hline
18  & $(3, 1)  -  (8, 6)$  &  53  &   $(2, 8)$  \\
 \hline
19  & $(3, 2)  -  (8, 7)$  &  54  &   $(2, 9)$   \\
 \hline
20  & $(3, 3)  -  (8, 8)$  &  55  &   $(2, 10)$   \\
 \hline
21  & $(3, 4)  -  (9, 8)$  &  56  &   $(3, 6)$  \\
 \hline
22  & $(3, 5)  +  (10, 8)$  &  57  &   $(3, 7)$ \\
 \hline
23  & $(8, 9)  -  (9, 8)$  &  58  &  $(3, 8)$ \\
 \hline
24  &  $(8, 10)  +  (10, 8)$  &  59  &  $(3, 9)$  \\
 \hline
25  & $(4, 1)  -  (9, 6)$  &  60  &    $(3, 10)$ \\
 \hline
26  & $(4, 2)  -  (9, 7)$  &  61  &    $(4, 6)$ \\
 \hline
27  &  $(4, 3)  -  (9, 8)$  &  62  &   $(4, 7)$  \\
 \hline
28  &  $(4, 4)  -  (9, 9)$ &  63  &    $(4, 8)$ \\
 \hline
29  &  $(4, 5)  +  (10, 9)$  &  64  &   $(4, 9)$  \\
 \hline
30  &  $(9, 10)  +  (10, 9)$ &  65  &    $(4, 10)$ \\
 \hline
31  &  $(5, 1)  -  (10, 6)$ & 66   &    $(5, 6)$ \\
 \hline
32  &  $(5, 2)  -  (10, 7)$ &  67  &    $(5, 7)$  \\
 \hline
33  &  $(5, 3)  -  (10, 8)$ &  68  &     $(5, 8)$  \\
 \hline
34  &  $(5, 4)  -  (10, 9)$  &  69  &    $(5, 9)$  \\
 \hline
35  & $(5, 5)  -  (10, 10)$  &  70  &   $(5, 10)$  \\
 \hline
\end{tabular}}
 \label{sym}
\end{minipage}
\end{figure*}

\section{Symmetry analysis for control-adapted spectra}
\label{AppSymmetry}

\begin{table*} 
\huge
 \centering
 \resizebox{\textwidth}{!}{
\begin{tabular}{c|cccc|ccc}
{\huge{Gate}}& \multicolumn{4}{c}{\huge{Model reduced, $P^\ast\vert_{\mathscr{F}}=\argmin_P \mathscr{E}^{G}_\$(P;T)\vert_\mathscr{F}$}} &  \multicolumn{3}{c}{\huge{Full knowledge, $P^\ast=\argmin_P \mathscr{E}^{G}_\$(P;T)$}}\\
\hline 
$G$ & $(\theta^\ast_1, \vec{n}^{(1)\ast})$ & $(\theta^\ast_2, \vec{n}^{(2)\ast})$ & $\mathscr{E}^{G}_\$(P^\ast\vert_{\mathscr{F}};T)\vert_\mathscr{F}$ & $\mathscr{E}^{G}_\$(P^\ast\vert_{\mathscr{F}};T)$ & $(\theta^\ast_1, \vec{n}^{(1)\ast})$ & $(\theta^\ast_2, \vec{n}^{(2)\ast})$ & $\mathscr{E}^{G}_\$(P^\ast;T)$ \\ \hline
$I$ & $(2.22,\{1,0,0\})$& $(2.22,\{-1,0,0\})$& $1.64\times 10^{-6}$ & $1.64\times 10^{-6}$& $(2.22,\{1,0,0\})$& $(2.22,\{-1,0,0\})$& $1.64\times 10^{-6}$\\
$X$ & $(1.56,\{-1,0,0\})$& $(3.14,\{1,0,0\})$& $3.14\times 10^{-3}$ &$3.21\times 10^{-3}$ &  $(1.56,\{-1,0,0\})$& $(3.14,\{1,0,0\})$& $3.21\times 10^{-3}$\\
\multirow{2}{*}{$Z$} & $(1.71,$ \vspace{-4mm}& $(1.72,$& \multirow{2}{*}{$5.26\times 10^{-3}$} & \multirow{2}{*}{$5.39\times 10^{-3}$}&  \multirow{2}{*}{$(1.71,\{0,0,1\})$}& \multirow{2}{*}{$(1.71,\{0,0,1\})$}& \multirow{2}{*}{$5.39\times 10^{-3}$}\\
 & $\{-0.99,0,-0.16\})$& $\{-0.01,0.99,-0.16\})$& & & & &\\
$e^{i{\pi}\sigma_x/8}$  & $(2.03,\{-1,0,0\})$& $(2.42,\{1,0,0\})$& $8.63\times 10^{-6}$ & $8.76\times 10^{-6}$ &  $(2.03,\{-1,0,0\})$ & $(2.42,\{1,0,0\})$& $8.77\times 10^{-6}$\\
\multirow{2}{*}{$e^{i{\pi}\sigma_z/8}$}  & $(1.57,$ \vspace{-4mm}& $(1.57,$& \multirow{2}{*}{$1.11\times 10^{-2}$} &\multirow{2}{*}{$1.13\times 10^{-2}$} & $(1.57,$ & \multirow{2}{*}{$(1.57,\{0,0,-1\})$}& \multirow{2}{*}{$1.13\times 10^{-2}$}\\
& $\{-0.27,0.96,0.05\})$& $\{-0.61,0.79,0.05\})$& &  & $\{-0.24,0.98,0.97\})$& &\\
\multirow{2}{*}{$H$} & $(1.35,$ \vspace{-4mm}& $(2.18,$& \multirow{2}{*}{$3.97\times 10^{-3}$} & \multirow{2}{*}{$4.20\times 10^{-3}$}& $(1.21,$ & $(1.81,$& \multirow{2}{*}{$4.19\times 10^{-3}$}\\
& $\{0.82,0.58,0.02\})$& $\{-0.66,0.68,0.31\})$& & & $\{0.23,0.96,0.15\})$& $\{-0.91,0.06,0.42\})$ &\\
\end{tabular}}
\caption{Optimal gate design results for the noise model in Appendix \ref{appendNoise}.}
\label{table2}
\end{table*}

Given the frame of choice as plotted in Fig.~\ref{frame1}, there are symmetries in the CA-spectra. We systematically classify any symmetries present in $\bar{S}^{+}_{a} (n,n')$ by a kernel analysis method (which also works for more general noise models) as follows.
 
\smallskip

{\bf (1)} The relevant set of parameters, $\bar{S}^{+}_{a} (n,n')$, are not linearly independent. To probe this, we decompose each of them by dividing the integration region $0\leq t_2 \leq t_1 \leq T$ into three distinct subregions, namely  $I_1 \lspace{\equiv} \{0\leq t_2\leq t_1\leq T/2\}$, $I_2 \lspace{\equiv}
\{T/2\leq t_2\leq t_1\leq T\}$ and $I_3\equiv\{0\leq t_2\leq T/2,T/2\leq t_1\leq T\},$ and thus, letting $\bar{S}^{+}_{a} (n,n')\vert_{i}$ be the component of $\bar{S}^{+}_{a} (n,n')$ in the $I_i$ integration subregion, such that
$$\bar{S}^{+}_{a} (n,n') = \bar{S}^{+}_{a} (n,n')\vert_{1}+\bar{S}^{+}_{a} (n,n')\vert_{2}+\bar{S}^{+}_{a} (n,n')\vert_{3}.$$
The key point is that this division allows the systematic study of the symmetries in the frame elements within each as well as between different integration subregions, e,g., the stationary assumption implies that $\bar{S}^{+}_{a} (n,n')\vert_{1} = \bar{S}^{+}_{a} (m,m')\vert_{2}$ when $\phi_{n}^{(a)}(t_1-T/2)=\phi_{m}^{(a)}(t_1)$ and $\phi_{n'}^{(a)}(t_2-T/2)=\phi_{m'}^{(a)}(t_2)$. Moreover, parity symmetries lead to further reduction in the free parameters, e.g., $\bar{S}^{+}_{a} (n,n')\vert_{1}=-\bar{S}^{+}_{a} (n',n)\vert_{1}$ when $\phi_{n}^{(a)}(t_1)$ is an odd function in $0\leq t_1\leq T/2$ (anti-symmetric about $t_1=T/4$) and $\phi_{n'}^{(a)}(t_2)$ is an even function in $0\leq t_2\leq T/2$ (symmetric about $t_2=T/4$). We then associate a vector $s(n,n')$ to each $\bar{S}^{+}_{a} (n,n')$ such that the $l$-th entry, $s(n,n')_l$, is non-zero if $\bar{S}^{+}_{a} (n,n')$ has a non-zero projection on the $l$-th element of the set of independent  elements $\{\bar{S}^{+}_{a} (n,n')\vert_{i}\}$ (after considering all the symmetries above), and zero otherwise. Furthermore, we use all elements of the vectors $s(n,n')$ to construct the matrix
\begin{equation*}
    S_K\equiv\left(
\begin{matrix}
{s}{(1,1)}_1 & {s}{(1,2)}_1 & \cdots\\
{s}{(1,1)}_2 & {s}{(1,2)}_2 & \cdots\\
\vdots &  \vdots &
\end{matrix}\right).
\end{equation*}
Since the kernel of $S_K$ represents the vanishing linear combinations of its columns, calculating it provides exactly the linear dependencies among $\{\bar{S}^{+}_a(n,n')\}$ we seek. For our choice of frame, we find 70 such symmetries (see Tab.~\ref{sym}), and thus the number of free parameters -- which describe the additive noise -- is now reduced from $100$ to $30$.

\smallskip 

{\bf (2)} Recalling that for multiplicative noise, the $j$-th frame element is  supported on the $j$-th interval, one then has $\bar{S}_{m}^+(j,j') =0$ for $ j<j'$ and $\bar{S}_{m}^+(j,j) =\bar{S}_{m}^+(j',j')$.

\smallskip 

{\bf (3)} Finally, we point out that the additive and multiplicative noise components can be separately inferred by first fixing $\{\theta_i\}$ and cycling over a sufficiently large set of directions $\vec{n}$, and then repeating these steps for several choices of $\{\theta_i\}.$

\section{Faster gates need not be more accurate}
\label{appendNoise}
Here we showcase another noise example where two-interval control can, somewhat surprisingly, lead to better performance in some gate design tasks than single-interval control. This noise model is the same as Sec. \ref{section:Characterization}, except that the parameter values are changed as $b^{(a)}_{0}/\hbar =2000\ {\rm kHz}$, $c_0^{(a)}=0.08\ {\rm ms}^2$, $b^{(a)}_1/\hbar= 5\cdot10^4\ {\rm kHz}$, $c^{(a)}_1=0.64\ {\rm s}^2$, $\omega_1^{(a)}=400\ {\rm kHz}$, $b^{(m)}_{0}/\hbar=0.1\ {\rm mHz}$, $c^{(m)}_0/\hbar=6\sqrt{2\pi} \ {\rm Hz}$ and $\omega_0^{(m)}=60\ {\rm Hz}$. The optimal gate design results are summarized in Tab. \ref{table2}. In comparison, the shortest implementation of the $\pi/8$ gate around $X$ allowed by $\mathscr{C}$, yields the {\it larger} error $\mathscr{E}^G_\$(\{\pi/4\};T/2) = 7.75 \cdot 10^{-3}$. As in dynamically corrected gates \cite{DCG} and composite pulses \cite{DS4}, multiple segments of evolution may be crucial to enable error cancellation, despite the gate taking longer. 


\begin{thebibliography}{52}
\bibitem [{Rel()}]{Relative}%
  \BibitemOpen
  \href@noop {} {}\bibinfo {note} {Throughout the manuscript we calculate the
  relative error between an approximated/estimated quantity $\hat{A}$ and the
  actual one$A$, by $ \mathcal{E}_{\rm relative}={\Vert \hat{A} -A
  \Vert}/{\Vert A \Vert}$.}\BibitemShut {Stop}%
\end{thebibliography}
\end{document}